\documentclass[mathpazo]{cicp}

\usepackage{color}          
\usepackage{subcaption}
\usepackage{rotating}
\usepackage{multirow, makecell}
\usepackage{booktabs}
\usepackage{lineno}
\usepackage{microtype}
\usepackage{bm}
\usepackage{algorithm}
\usepackage{algpseudocode}
\usepackage{placeins,stfloats}
\floatname{algorithm}{Algorithm}

\begin{document}
\graphicspath{{figures/}}

\title{A Conservative Multi-Level Adaptive Velocity-Space Method for the Discrete Unified Gas-Kinetic Scheme}



\author[Ren W J et~al.]{Weijie Ren\affil{1},
      Hang Yu\affil{2},
      Zhengyu Tian\affil{1},
      Wenjia Xie\affil{1}\comma\corrauth,
      and XiaoQiang Fan\affil{1}}
\address{\affilnum{1}\ College of Aerospace Science and Engineering,
         National University of Defense Technology,
         Changsha 410073, Hunan, P.R. China. \\
         \affilnum{2}\ National Key Laboratory of Aerospace Flow Physics,
         Mianyang 621000, Sichuan, P.R. China}
\emails{{\tt xiewenjia@nudt.edu.cn} (W.~Xie)}

\begin{abstract}
    Multiscale gas flows span a wide range of Knudsen numbers, from near-continuum to highly rarefied conditions, and therefore require kinetic solvers that remain accurate across these regimes. The discrete unified gas-kinetic scheme (DUGKS) provides such a unified description, but its memory and computational cost are dominated by the size of the discrete velocity space. To alleviate this bottleneck, this work develops a multi-level velocity-space adaptive DUGKS (MLVS-DUGKS), reducing velocity-space redundancy while preserving accuracy and conservation. The method identifies the velocity-space requirements of individual physical cells from their distribution functions and groups these requirements into a small number of shared representative velocity spaces, retaining sensitivity to different local flow states without constructing an independent velocity space for every cell. A linear moment-constrained correction is incorporated to maintain discrete compatibility and conservative transfer between different velocity spaces. Benchmark simulations at $\mathrm{Kn} = 0.001$--$10$ show close agreement with the results of uniform-velocity-space DUGKS (UVS-DUGKS). For the cylinder and the two-cavity problems, the average number of discrete velocity points is reduced by approximately one order of magnitude, yielding speedups of $7.00$--$7.93$ and GPU-memory reductions by factors of $6.73$--$7.49$. MLVS-DUGKS thus offers an efficient tool for simulating multiscale nonequilibrium flows.
\end{abstract}


\maketitle

\section{Introduction}\label{sec:introduction}
Multiscale flows spanning continuum and rarefied regimes arise in a wide range of engineering applications, including near-space vehicles~\cite{Li_Gas-kinetic_2025,Li_Application_2025}, satellite reaction-control systems~\cite{Glass_Numerically_2018,ChenQi_RCSDui_2018}, microelectromechanical systems (MEMS)~\cite{Karniadakis_Microflows_2005,Akhlaghi_new_2012,Li_Lattice_2011}, and shale-gas extraction~\cite{Sharma_Molecular_2015,Li_Current_2014,Ozdemir_Modeling_2009}. In such flows, the local Knudsen number may vary by several orders of magnitude across the computational domain, and the flow can range from near-equilibrium hydrodynamic behavior to strongly nonequilibrium molecular transport. Conventional continuum solvers based on the Navier-Stokes equations are no longer valid under such conditions. Instead, the Boltzmann equation and its model equations provide a kinetic description that remains valid across the entire Knudsen-number range.

Compared with the direct simulation Monte Carlo (DSMC)~\cite{Bird_DSMC_2013,Bird_Approach_1963} and unified gas-kinetic wave-particle (UGKWP)~\cite{Liu_Unified_2020,Zhu_Unified_2019} methods, which rely on stochastic particles, deterministic kinetic solvers are free of statistical noise. They therefore provide smooth numerical solutions in the near-continuum, low-speed, and microscale regimes. Among various deterministic kinetic solvers, the unified gas-kinetic scheme (UGKS)~\cite{Xu_unified_2010,Huang_Unified_2012} and the discrete unified gas-kinetic scheme (DUGKS)~\cite{Guo_Discrete_2013,Guo_Discrete_2015} have emerged as highly efficient multiscale methods. By coupling particle transport and collisions, these methods evaluate the evolution of the distribution function at the numerical cell and time-step scales. DUGKS can be viewed as a simplified variant of UGKS, in which the coupling of free transport and collision is achieved by numerically integrating the kinetic equation along the characteristic line. Extensive numerical evidence has shown that DUGKS and UGKS produce virtually identical results, while the former reportedly offers a $10\%$--$20\%$ improvement in computational efficiency~\cite{Wang_Unified_2015,Guo_Progress_2021}. With these advantages, DUGKS has been applied to a broad range of flow problems spanning wide ranges of Knudsen and Mach numbers~\cite{Liu_Discrete_2026,Zhang_Large-eddy_2020,Bo_DUGKS_2017,Wang_Comparison_2016,Zhang_Implicit_2025,Xin_discrete_2023,Zhang_Discrete_2018a}. Nevertheless, as a deterministic discrete-velocity method, DUGKS still requires the evolution and storage of distribution functions over the velocity space. Its computational cost and memory consumption are therefore related to the number of discrete velocity points. In a uniform velocity space (UVS), the same velocity discretization is applied throughout the computational domain, regardless of the local characteristics of the distribution function. Consequently, many velocity points may be redundant in regions where the distribution is close to equilibrium or has a narrow effective support.

To alleviate this redundancy, adaptive velocity-space (AVS) strategies have been investigated in recent years. These methods adjust the velocity-space discretization according to the local characteristics of the distribution function, reducing unnecessary velocity points while retaining adequate resolution in dynamically important regions. Based on how the adapted space is organized among physical cells, existing AVS methods can be broadly classified into local and global approaches. In local methods, each cell independently adapts its own velocity discretization~\cite{Aristov_method_1977,Chen_unified_2012,Kolobov_adaptive_2012,Arslanbekov_Kinetic_2012,Brull_Local_2014,Filbet_Rescaling_2013,Bernard_Local_2014}, which offers the highest degree of flexibility and the maximum achievable compression. However, neighboring cells may employ non-aligned velocity grids, requiring interpolation or remapping to transfer distribution information across cell interfaces. Since these operations do not automatically preserve the discrete moments of mass, momentum, and energy, additional conservative corrections are required~\cite{Brull_Local_2014}. The resulting grid-to-grid transfers, conservative corrections, and irregular data structures increase algorithmic complexity, reduce numerical robustness, and limit the efficiency of large-scale parallel implementations~\cite{Chen_global_2024a,Brull_Two_2016,Arslanbekov_Kinetic_2012}. In global methods, by contrast, a single velocity space is shared by all cells and adapted to the evolving flow field as a whole~\cite{Yuan_conservative_2020,Chen_Conserved_2018,Baranger_Locally_2014,Chen_global_2024a}. This shared representation avoids cell-to-cell interpolation and remapping, thereby simplifying interface flux evaluation and data management, improving numerical robustness, and facilitating efficient parallel implementation. Substantial computational acceleration has been demonstrated in~\cite{Chen_global_2024a}. Its compression capability, however, is limited because the single shared space must accommodate the most demanding local requirement among all physical cells.

A complementary strategy, instead of adapting the velocity space, is to reduce the portion of the domain that requires a kinetic description by employing hybrid continuum-kinetic solvers~\cite{Yang_Adaptive_2023,Xiao_velocity-space_2020,Wei_Adaptive_2024,Wu_Hybrid_2026,Long_implicit_2024}. In such solvers, the Navier-Stokes equations or the gas-kinetic scheme (GKS) is used in near-equilibrium regions, while kinetic solvers, including UGKS, DUGKS, or other discrete-velocity methods, are retained in regions where nonequilibrium effects are important. In this way, such hybrid methods can reduce computational cost while preserving kinetic resolution in rarefied or strongly nonequilibrium regions. Their overall efficiency, however, depends on the portion of the domain that can be treated accurately by the continuum solver. If kinetic effects extend over most of the computational domain, the potential savings become limited.

The above studies indicate that the local and global families embody two complementary design principles: local AVS emphasizes cell-level flexibility, whereas global AVS emphasizes a shared velocity-space discretization and simplified inter-cell couplings. In practical flows, the local flow states often take a limited number of characteristic forms, so that their velocity-space requirements can be grouped into a small set of representative types. Motivated by this observation, this paper develops the multi-level velocity-space adaptive DUGKS (MLVS-DUGKS). Instead of assigning an independent velocity space for every physical cell, the proposed method first identifies the local velocity-space requirements and clusters similar requirements into a small number of shared representative velocity spaces. In this way, the MLVS-DUGKS combines the shared-space organization of global AVS with a controlled degree of local adaptivity, providing an intermediate organization between fully local and single-space global adaptation. A linear correction is further developed to conserve the discrete moments of the representative distributions throughout both the collision step and the inter-space projection.

The remainder of this paper is organized as follows. Section~\ref{sec: model and method} briefly introduces the BGK model equation and the UVS-DUGKS. Section~\ref{sec: MLVS-DUGKS} presents the proposed MLVS-DUGKS in detail. Section~\ref{sec:numerical_results} evaluates the accuracy and efficiency of the present method through a series of benchmark test cases. Section~\ref{sec: Conclusions} concludes the paper and outlines directions for future work.

\section{Kinetic model equation and discrete unified gas-kinetic scheme}
\label{sec: model and method}

\subsection{Kinetic model equation}

The Shakhov model modifies the BGK equilibrium distribution to recover the Prandtl number of a monatomic gas. Following Guo et al.~\cite{Guo_Discrete_2015}, we use the reduced distribution functions
\begin{equation}
	\begin{aligned}
		g(\boldsymbol{x},\boldsymbol{\xi},t)
		 & = \int f(\boldsymbol{x},\boldsymbol{\xi},\boldsymbol{\eta},\boldsymbol{\zeta},t)
		\mathrm{d}\boldsymbol{\eta}\mathrm{d}\boldsymbol{\zeta},                            \\
		h(\boldsymbol{x},\boldsymbol{\xi},t)
		 & = \int (\eta^{2}+\zeta^{2})
		f(\boldsymbol{x},\boldsymbol{\xi},\boldsymbol{\eta},\boldsymbol{\zeta},t)
		\mathrm{d}\boldsymbol{\eta}\mathrm{d}\boldsymbol{\zeta},
	\end{aligned}
	\label{eq:reduced_dist}
\end{equation}
where $f(\boldsymbol{x},\boldsymbol{\xi},\boldsymbol{\eta},\boldsymbol{\zeta},t)$ is the velocity distribution function, $\boldsymbol{\xi}=(\xi_x,\xi_y)$ is the particle velocity in the two-dimensional physical space, $\boldsymbol{\eta}$ denotes the vector of the remaining $L=3-D$ components of the three-dimensional molecular velocity, and $\boldsymbol{\zeta}$ is a vector of length $K$ representing the internal degrees of freedom of the molecules. The two-dimensional monatomic flows considered in this work correspond to $D=2$ and $K=0$, for which $\boldsymbol{\eta}=\xi_z$ reduces to the single velocity component in the third direction, $\boldsymbol{\zeta}$ is absent, and Eq.~\eqref{eq:reduced_dist} reduces to $g=\int f\mathrm{d}\xi_z$ and $h=\int \xi_z^2 f\mathrm{d}\xi_z$. The conservative variables and heat flux are
\begin{equation}
	\begin{aligned}
		 & \rho = \int g\mathrm{d}\boldsymbol{\xi},
		\quad
		\rho\boldsymbol{u} = \int \boldsymbol{\xi}g\mathrm{d}\boldsymbol{\xi}, \\
		 & \rho E = \frac{1}{2}\int
		(\xi^2 g+h)\mathrm{d}\boldsymbol{\xi},
		\quad
		\boldsymbol{q} = \frac{1}{2}\int
		\boldsymbol{c}(c^2g+h)\mathrm{d}\boldsymbol{\xi},
	\end{aligned}
	\label{eq:reduced_moments}
\end{equation}
with $p=\rho RT$ and $\rho E=\tfrac{1}{2}\rho u^2+\tfrac{3}{2}\rho RT$, where $\boldsymbol{u}$ is the flow velocity, $\boldsymbol{c}=\boldsymbol{\xi}-\boldsymbol{u}$ is the peculiar velocity.

The reduced distributions satisfy
\begin{equation}
	\begin{aligned}
		\frac{\partial g}{\partial t} + \boldsymbol{\xi}\cdot\nabla g
		 & = -\frac{1}{\tau}(g-g^{S}), \\
		\frac{\partial h}{\partial t} + \boldsymbol{\xi}\cdot\nabla h
		 & = -\frac{1}{\tau}(h-h^{S}),
	\end{aligned}
	\label{eq:reduced_gh_eq}
\end{equation}
where $\tau=\mu/p$, and $g^{S}$ and $h^{S}$ are the reduced Shakhov distributions. For a monatomic gas with $\mathrm{Pr}=2/3$, they are
\begin{equation}
	\begin{aligned}
		g^{S} & = g^{eq}\left[
			1+(1-\mathrm{Pr})\frac{\boldsymbol{c}\cdot\boldsymbol{q}}{5pRT}
		\left(\frac{c^{2}}{RT}-4\right)\right], \\
		h^{S} & = RTg^{eq}\left[
			1+(1-\mathrm{Pr})\frac{\boldsymbol{c}\cdot\boldsymbol{q}}{5pRT}
			\left(\frac{c^{2}}{RT}-2\right)\right],
	\end{aligned}
	\label{eq:g_h_eq_Pr}
\end{equation}
with
\begin{equation}
	g^{eq}=\frac{\rho}{2\pi RT}
	\exp\!\left[-\frac{(\boldsymbol{\xi}-\boldsymbol{u})^{2}}{2RT}\right].
\end{equation}

\subsection{Discrete unified gas-kinetic scheme}

The DUGKS applies a finite-volume discretization to the reduced kinetic equations~\cite{Guo_Discrete_2013,Guo_Discrete_2015}. Let $\varphi$ denote either $g$ or $h$, with collision term $\Omega=-(\varphi-\varphi^{S})/\tau$. Integrating over a control volume $V_j$ and one time step gives
\begin{equation}
	\varphi_j^{n+1}-\varphi_j^{n}
	+\frac{\Delta t}{|V_j|}F_j^{n+1/2}
	=\frac{\Delta t}{2}
	\left(\Omega_j^{n+1}+\Omega_j^{n}\right),
\end{equation}
where the microscopic flux through the cell faces is
\begin{equation}
	F_j^{n+1/2}
	=\sum_{f\in\partial V_j}\int_{A_f}
	(\boldsymbol{\xi}\cdot\boldsymbol{n}_f)
	\varphi(\boldsymbol{x}_f,\boldsymbol{\xi},t_{n+1/2})
	\mathrm{d}S.
\end{equation}
To make the update explicit, define
\begin{equation}
	\tilde{\varphi} = \varphi-\frac{\Delta t}{2}\Omega,
	\quad
	\tilde{\varphi}^{+} = \varphi+\frac{\Delta t}{2}\Omega.
\end{equation}
The cell update becomes
\begin{equation}
	\tilde{\varphi}_j^{n+1}
	=\tilde{\varphi}_j^{+,n}
	-\frac{\Delta t}{|V_j|}F_j^{n+1/2}.
	\label{eq:explicit_update}
\end{equation}
The macroscopic variables are recovered from the transformed distributions as
\begin{equation}
	\begin{aligned}
		\rho           & = \int\tilde{g}\mathrm{d}\boldsymbol{\xi},
		\quad
		\rho\boldsymbol{u} = \int\boldsymbol{\xi}\tilde{g}\mathrm{d}\boldsymbol{\xi},
		\quad
		\rho E = \frac{1}{2}\int(\xi^{2}\tilde{g}+\tilde{h})\mathrm{d}\boldsymbol{\xi},   \\
		\boldsymbol{q} & = \frac{2\tau}{2\tau+\Delta t\mathrm{Pr}}\tilde{\boldsymbol{q}},
		\quad\text{with}\quad
		\tilde{\boldsymbol{q}} = \frac{1}{2}\int \boldsymbol{c}(c^{2}\tilde{g}+\tilde{h})\mathrm{d}\boldsymbol{\xi}.
	\end{aligned}
	\label{eq:W_and_flux_from_tilde}
\end{equation}
Using the definitions above, $\tilde{\varphi}^{+}$ can be obtained from $\tilde{\varphi}$ and $\varphi^{S}$:
\begin{equation}
	\tilde{\varphi}^{+}
	=\frac{2\tau-\Delta t}{2\tau+\Delta t}\tilde{\varphi}
	+\frac{2\Delta t}{2\tau+\Delta t}\varphi^{S}.
\end{equation}

The interface distribution is evaluated along a characteristic over the half time step $s=\Delta t/2$. Define
\begin{equation}
	\bar{\varphi} = \varphi-\frac{s}{2}\Omega,
	\quad
	\bar{\varphi}^{+} = \varphi+\frac{s}{2}\Omega.
\end{equation}
Integration along the characteristic gives
\begin{equation}
	\bar{\varphi}(\boldsymbol{x}_b,\boldsymbol{\xi},t_n+s)
	=\bar{\varphi}^{+}
	(\boldsymbol{x}_b-\boldsymbol{\xi}s,\boldsymbol{\xi},t_n),
\end{equation}
where $\boldsymbol{x}_b$ is the interface center. The upstream value is reconstructed from the cell-centered value using a limited linear profile,
\begin{equation}
	\bar{\varphi}^{+}
	(\boldsymbol{x}_b-\boldsymbol{\xi}s,\boldsymbol{\xi},t_n)
	=
	\bar{\varphi}^{+}(\boldsymbol{x}_j,\boldsymbol{\xi},t_n)
	+(\boldsymbol{x}_b-\boldsymbol{x}_j-\boldsymbol{\xi}s)
	\cdot\boldsymbol{\sigma}_j,
\end{equation}
where $\boldsymbol{\sigma}_j$ is the limited cell gradient. Its cell-centered value follows from
\begin{equation}
	\bar{\varphi}^{+}
	=\frac{2\tau-s}{2\tau+\Delta t}\tilde{\varphi}
	+\frac{3s}{2\tau+\Delta t}\varphi^{S}.
\end{equation}
At the interface, the conservative variables are computed from $\bar{g}$ and $\bar{h}$ via the moment relations in Eq.~\eqref{eq:reduced_moments}, and the interface heat flux follows as $\boldsymbol{q} = 2\tau / (2\tau + s\mathrm{Pr}) \bar{\boldsymbol{q}}$, where $\bar{\boldsymbol{q}}$ is given by the heat-flux integral in Eq.~\eqref{eq:reduced_moments} with $g,h$ replaced by $\bar{g},\bar{h}$. The physical interface distribution is then recovered as
\begin{equation}
	\varphi(\boldsymbol{x}_b,\boldsymbol{\xi},t_n+s)
	=\frac{2\tau}{2\tau+s}\bar{\varphi}
	+\frac{s}{2\tau+s}\varphi^{S}.
\end{equation}
Substitution into the face integral gives $F_j^{n+1/2}$ in Eq.~\eqref{eq:explicit_update}.

\subsection{Discrete velocity formulation}
In the DUGKS formulation described above, the distribution functions are continuous functions of the particle velocity $\boldsymbol{\xi}$. In practical computations, the velocity space must be discretized into a finite set of discrete velocity points. Let $\{\boldsymbol{\xi}_i\}_{i=1}^{N_{\xi}}$ be the discrete velocity set and $\{w_i\}_{i=1}^{N_{\xi}}$ the corresponding quadrature weights, with which all velocity space integrals are evaluated by numerical quadrature. For example, the conservative variables obtained from Eq.~\eqref{eq:W_and_flux_from_tilde} become
\begin{equation}\label{eq:discrete_W_and_q}
	\begin{aligned}
		\rho           & = \sum_{i=1}^{N_{\xi}} w_i\tilde{g}(\boldsymbol{\xi}_i),\quad
		\rho\boldsymbol{u} = \sum_{i=1}^{N_{\xi}} w_i\boldsymbol{\xi}_i\tilde{g}(\boldsymbol{\xi}_i),\quad
		\rho E = \frac{1}{2}\sum_{i=1}^{N_{\xi}} w_i\bigl[\xi_i^{2}\tilde{g}(\boldsymbol{\xi}_i) + \tilde{h}(\boldsymbol{\xi}_i)\bigr], \\
		\boldsymbol{q} & = \frac{2\tau}{2\tau+\Delta t\,\mathrm{Pr}}\tilde{\boldsymbol{q}},\quad\text{with}\quad
		\tilde{\boldsymbol{q}} = \frac{1}{2}\sum_{i=1}^{N_{\xi}} w_i\boldsymbol{c}_i\bigl[c_i^{2}\tilde{g}(\boldsymbol{\xi}_i) + \tilde{h}(\boldsymbol{\xi}_i)\bigr].
	\end{aligned}
\end{equation}

The discrete velocity set can be chosen based on the Newton-Cotes quadrature with a uniform Cartesian velocity grid, which is widely used in DUGKS because of its simplicity and robustness for strongly nonequilibrium flows.  A simpler alternative is to take the cell centers of the velocity mesh as the discrete velocity points and the cell areas as the quadrature weights. Such a low-order rule is less accurate in the integration, but it removes the restriction to uniform Cartesian grids, and is therefore widely adopted for unstructured and quadtree/octree discretizations of the velocity space.

\section{Multi-level velocity space adaptive DUGKS}\label{sec: MLVS-DUGKS}

\subsection{Overview of the multi-level velocity space adaptation}
The uniform-velocity-space DUGKS (UVS-DUGKS) applies a single velocity space to all physical cells, which becomes redundant when the local distribution functions vary strongly across the flow field. To reduce this redundancy, the present method first identifies the velocity-space requirements of individual cells from their local distribution functions and groups similar requirements. The DUGKS evolution for each group is then performed on a dynamically constructed representative velocity space, thereby retaining local adaptivity without assigning a distinct velocity space to every cell. Throughout this paper, a ``level'' denotes such a representative space rather than a nested refinement depth. Interfaces between cells at the same level retain the standard UVS-DUGKS formulation, whereas interfaces between different levels are treated using the conservative inter-level coupling described in Section~\ref{subsection:inter_level_coupling}. This coupling relies on an auxiliary interface velocity representation, $\mathcal{V}^{\mathrm{int}}$, constructed from the velocity cells required by the participating levels, which is detailed in Section~\ref{subsection:inter_level_coupling}.

\subsection{Adaptive construction of representative velocity spaces}
This subsection first describes the criterion used to determine the velocity-space resolution required by each physical cell. The criterion is based primarily on the local contributions of the discrete velocity points to the distribution function, as commonly adopted in adaptive velocity-space methods~\cite{Chen_unified_2012,Chen_global_2024a}. For a given physical cell $j$, the normalized mass and energy contributions associated with the discrete velocity point $\boldsymbol{\xi}_i$ are defined as
\begin{equation}
	\begin{aligned}
		M_1(\boldsymbol{\xi}_i)
		 =
		\frac{w_i g_j(\boldsymbol{\xi}_i)}
		{\displaystyle\sum_k w_k g_j(\boldsymbol{\xi}_k)}, \quad
		M_2(\boldsymbol{\xi}_i)
		=
		\frac{
			\frac{1}{2}w_i
			\left[
				h_j(\boldsymbol{\xi}_i)
				+
				|\boldsymbol{\xi}_i|^2 g_j(\boldsymbol{\xi}_i)
				\right]}
		{\displaystyle
			\sum_k
			\frac{1}{2}w_k
			\left[
				h_j(\boldsymbol{\xi}_k)
				+
				|\boldsymbol{\xi}_k|^2 g_j(\boldsymbol{\xi}_k)
				\right]}.
	\end{aligned}
	\label{eq:M2}
\end{equation}
For a candidate velocity block $B$, the contributions of all velocity points within the block are accumulated. The block indicator is defined as
\begin{equation}
	I_B
	=
	\max
	\left(
	\sum_{\boldsymbol{\xi}_i\in B}M_1(\boldsymbol{\xi}_i),
	\sum_{\boldsymbol{\xi}_i\in B}M_2(\boldsymbol{\xi}_i)
	\right).
	\label{eq:block-contribution-indicator}
\end{equation}
A three-regime decision rule is then applied:
\begin{equation}
	\text{block status}
	=
	\begin{cases}
		\text{candidate for merging},
		 & I_B<C_{\mathrm{merge}},
		\\
		\text{retain the previous status},
		 & C_{\mathrm{merge}}\leq I_B\leq C_{\mathrm{split}},
		\\
		\text{split (keep fine)},
		 & I_B>C_{\mathrm{split}},
	\end{cases}
	\label{eq:block-hysteresis-rule}
\end{equation}
where $C_{\mathrm{merge}}<C_{\mathrm{split}}$ are prescribed thresholds. This hysteresis treatment prevents frequent switching of the velocity-space configuration between consecutive time steps.

For flows with large Knudsen numbers, the contribution-based criterion is supplemented by a transport-based criterion to further control the accuracy of velocity-space coarsening. Let
\begin{equation}
    \boldsymbol{m}(\boldsymbol{\xi}) = \left( g,\ \xi_x g,\ \xi_y g,\ \frac{1}{2}\left( h+|\boldsymbol{\xi}|^{2}g \right) \right)
\end{equation}
denote the velocity-point contributions to mass, momentum, and energy. The fine-grid directional flux of the $r$-th component of a block $B$ is
\begin{equation}
    F_{r,d}^{\pm}(B) = \sum_{\substack{\boldsymbol{\xi}_i\in B,\; \pm\xi_{i,d}\geq 0}} w_i\,\xi_{i,d}\,m_r(\boldsymbol{\xi}_i), \quad d\in\{x,y\},
\end{equation}
and the normalized merging error is defined as
\begin{equation}
    \varepsilon_B = \max_{r,d,\pm} \left| F_{r,d}^{\pm}(B)-\tilde{F}_{r,d}^{\pm}(B) \right| / S_r, \label{eq:merge-transport-error}
\end{equation}
where $\tilde{F}_{r,d}^{\pm}$ is evaluated directly from the merged block representation without correction, and the characteristic scale $S_r=(\rho c_s,\ \rho c_s^{2},\ \rho c_s^{2},\ \rho\,c_s E)$ is chosen as the local scale of the corresponding conserved quantity, with $c_s$ the speed of sound. A block is allowed to merge only if $I_B<C_{\mathrm{merge}}$ and $\varepsilon_B\leq\varepsilon_{\mathrm{merge}}$, with the latter serving primarily as a supplement for large Knudsen number cases. 

Through the above coarsening procedure, the local velocity-space requirements of each physical cell are determined, indicating which regions of $\mathcal{V}_{\mathrm{base}}$ (the base velocity grid, from which all refinement and coarsening proceed) must be retained and which may be coarsened. To compare and cluster these local requirements, each requirement is decoded into a blockwise coarsening pattern $\mathcal{K}$ over the $2\times2$ blocks of $\mathcal{V}_{\mathrm{base}}$. For example, if a three-stage merging strategy with $2\times2$, $4\times4$, and $8\times8$ levels is employed, each block $b$ can be assigned an index $\kappa_{\mathcal{K}}(b)\in\{0,1,2,3\}$, corresponding respectively to retaining the original $1\times1$ points and to merging at these three levels. The discrepancy between two patterns is defined as
\begin{equation}
	D(\mathcal{K}_a,\mathcal{K}_b) = \frac{1}{3N_b} \sum_{b=1}^{N_b} \left| \kappa_{\mathcal{K}_a}(b) - \kappa_{\mathcal{K}_b}(b) \right|,
\end{equation}
where $N_b$ is the number of $2\times2$ blocks; since the indices span $\{0,1,2,3\}$, $D\in[0,1]$. Using this distance metric, the patterns are greedily grouped into a limited number of clusters, from which the active representative velocity spaces are constructed. The cellwise patterns are first deduplicated, and the occurrence frequency of each unique pattern is recorded. The unique patterns are then processed in descending order of frequency. The first pattern initializes the first cluster. For each subsequent pattern, its distance to the existing cluster centers is evaluated. If the minimum distance satisfies $D\leq\tau_{\mathrm{clust}}$, the pattern is assigned to the nearest cluster; otherwise, a new cluster is initialized. The clustering threshold $\tau_{\mathrm{clust}}\in[0,1]$ controls the degree of velocity-space sharing among physical cells. As $\tau_{\mathrm{clust}}\to0$, only nearly identical requirements are grouped and the configuration approaches a local AVS strategy, whereas $\tau_{\mathrm{clust}}=1$ assigns all requirements to a single cluster whose enclosing representative space is shared by the entire domain, corresponding to a global AVS strategy. Intermediate values yield a finite number of representative spaces between these two limits.

For each cluster, an enclosing representative pattern is constructed. Since a smaller $\kappa$ corresponds to a finer velocity-space resolution, the representative coarsening index of each block is defined as the minimum over the cluster members,
\begin{equation}
	\kappa_{\mathrm{rep}}(b)
	=\min_{\mathcal{K}_m\in\mathcal{C}}\,\kappa_{\mathcal{K}_m}(b),
	\label{eq:kappa_rep}
\end{equation}
where $\mathcal{C}$ denotes the patterns belonging to the cluster. The representative index array is then converted into velocity points and quadrature weights according to the merging rules, with the parent-child consistency of the quadtree hierarchy enforced.

\subsection{Linear moment-constrained conservation correction} \label{sec: Linear correction}
The conservation and accuracy of discrete velocity methods rely on moment consistency, as both macroscopic variables and nonequilibrium transport quantities are evaluated via discrete velocity-space quadrature. In the MLVS-DUGKS, discrete moment discrepancies arise primarily from the quadrature of the collision-model distribution and from the transfer of distribution functions between different velocity spaces. Despite their different origins, both can be formulated as the recovery of prescribed target moments from a provisional distribution. To address this, a unified linear moment-constrained correction is introduced below.

Let $f_i^p$ denote the provisional distribution at the velocity $\boldsymbol{\xi}_i$, and $f_i^*$ the corrected distribution. The linear correction is written as
\begin{equation}
	f_i^*=f_i^p+\chi_i\sum_{m=1}^{N_m}a_m\,\phi_m(\boldsymbol{\xi}_i),
	\label{eq:linear_correction}
\end{equation}
where $\phi_m$ are the velocity-space basis functions, $\chi_i$ is the weighting function, $a_m$ are the unknown coefficients, and $N_m$ is the number of imposed moment constraints. The corrected distribution must satisfy the moment constraints
\begin{equation}
	\sum_i w_i\,\psi_k(\boldsymbol{\xi}_i)\,f_i^*=M_k^{\mathrm{tar}},
	\quad k=1,\ldots,N_m,
	\label{eq:linear_moment_constraint}
\end{equation}
where $w_i$ is the quadrature weight, $\psi_k$ is the integration kernel of the $k$th moment, and $M_k^{\mathrm{tar}}$ is the prescribed target value. Substituting Eq.~\eqref{eq:linear_correction} into Eq.~\eqref{eq:linear_moment_constraint} gives
\begin{equation}
	\begin{aligned}
		 & \sum_{m=1}^{N_m}A_{km}\,a_m=\Delta M_k,                                            \\
		 & A_{km}=\sum_i w_i\,\psi_k(\boldsymbol{\xi}_i)\,\chi_i\,\phi_m(\boldsymbol{\xi}_i), \\
		 & \Delta M_k=M_k^{\mathrm{tar}}-\sum_i w_i\,\psi_k(\boldsymbol{\xi}_i)\,f_i^p.
	\end{aligned}
	\label{eq:linear_correction_system}
\end{equation}
The provisional distribution, target moments, weighting function, and basis functions can be chosen according to the specific application, while the algebraic form of the correction and the resulting small $N_m\times N_m$ system remain unchanged.

\subsubsection{Discrete compatibility correction of the model distribution}
In continuous velocity space, the reduced collision-model distributions satisfy
\begin{equation}
	\int
	\begin{bmatrix}
		g^{S}-g                   \\
		\xi_x\left(g^{S}-g\right) \\
		\xi_y\left(g^{S}-g\right) \\
		\dfrac{1}{2}
		\left[
			h^{S}-h
			+
			\xi^2\left(g^{S}-g\right)
			\right]
	\end{bmatrix}
	\,\mathrm{d}\boldsymbol{\xi}
	=
	\boldsymbol{0}.
	\label{eq:continuous_collision_compatibility}
\end{equation}
However, due to numerical integration errors, this relation no longer holds exactly in the discrete velocity space. The resulting compatibility error enters the time evolution through the collision term. To address this problem, Mieussens~\cite{MIEUSSENS_DISCRETE_2003} developed a conservative discrete velocity ordinate method that constructs a discrete Maxwellian distribution satisfying the compatibility relation exactly on a given velocity grid. Related iterative corrections have since been developed in various kinetic solvers~\cite{Dzanic_positivity-preserving_2023,Johnson_moment-conserving_2025a}. Within the DUGKS framework, Zhang et al.~\cite{Zhang_microscopically_2026} proposed a microscopically conservation-enforced scheme (MicroC-DUGKS) that restores the discrete compatibility through such an iterative construction. However, iterative corrections of the discrete equilibrium require nonlinear solvers at every update, which introduces appreciable computational cost, particularly in GPU implementations.

Let $g_i^{S}\equiv g^{S}(\boldsymbol{\xi}_i)$ and $h_i^{S}\equiv h^{S}(\boldsymbol{\xi}_i)$ denote the uncorrected discrete Shakhov model distributions. The linear correction takes the form
\begin{equation}
	\varphi_i^* = \varphi_i^{S} + \chi_i^{\varphi} \sum_{m=1}^{N_m} a_m\phi_m(\boldsymbol{\xi}_i),
	\quad
	\varphi\in\{g,h\},
	\label{eq:gh_model_correction}
\end{equation}
with the weighting functions set to their respective uncorrected model distributions. The corresponding correction matrix for the joint moment constraints is
\begin{equation}
	A_{km}	=
	\sum_i w_i
	\left[
		\psi_k^g(\boldsymbol{\xi}_i)\chi_i^g
		+
		\psi_k^h(\boldsymbol{\xi}_i)\chi_i^h
		\right]
	\phi_m(\boldsymbol{\xi}_i),
	\label{eq:joint_gh_correction_matrix}
\end{equation}
where $\psi_k^g$ and $\psi_k^h$ are the $g$- and $h$-distribution kernels associated with the $k$th constrained moment.

When only the discrete compatibility associated with the collision invariants needs to be enforced, the correction uses the following four basis functions
\begin{equation}
	\boldsymbol{\phi}^{eq}_4 =
	\left[1,\, \xi_x,\, \xi_y,\, \mathcal{E}_{\xi} \right]^T,
	\quad 
	\mathcal{E}_{\xi} = \frac{1}{2} \left[ (3-D+K)RT+\xi^2 \right].
	\label{eq:eq_four_basis}
\end{equation}
The correction coefficients are then determined by requiring the discrete moments of the corrected reduced distributions to satisfy
\begin{equation}
	\sum_i w_i
	\begin{bmatrix}
		g_i^*          \\
		\xi_{x,i}g_i^* \\
		\xi_{y,i}g_i^* \\
		\dfrac{1}{2}
		\left(
		h_i^*+\xi_i^2g_i^*
		\right)
	\end{bmatrix}
	=
	\begin{bmatrix}
		\rho   \\
		\rho u \\
		\rho v \\
		\rho E
	\end{bmatrix}.
	\label{eq:eq_four_moment_constraint}
\end{equation}

When the heat-flux and stress moments of the collision-model distribution are also required to be represented consistently, the basis is augmented by four additional functions,
\begin{equation}
	\boldsymbol{\phi}^{eq}_{4\rightarrow8} = \left[\, \mathcal{E}_c c_x,\; \mathcal{E}_c c_y,\; c_x^2-c_y^2,\; c_xc_y \,\right]^T, \quad \mathcal{E}_c = \tfrac{1}{2}\left[(3-D+K)RT+c^2\right],
	\label{eq:eq_eight_extra_basis}
\end{equation}
where $\mathcal{E}_c\,g^{eq}=\tfrac{1}{2}(c^2g^{eq}+h^{eq})$. The first four constraints are unchanged from Eq.~\eqref{eq:eq_four_moment_constraint}, and the four additional constraints are imposed as
\begin{equation}
	\sum_i w_i
	\begin{bmatrix}
		\dfrac{1}{2}\left( h_i^*+c_i^2g_i^* \right)c_{x,i} \\
		\dfrac{1}{2}\left( h_i^*+c_i^2g_i^* \right)c_{y,i} \\
		\left(c_{x,i}^2-c_{y,i}^2\right)g_i^* \\
		c_{x,i}c_{y,i}g_i^*
	\end{bmatrix}
	=
	\begin{bmatrix}
		q_x^{\mathrm{tar}} \\ q_y^{\mathrm{tar}} \\ 0 \\ 0
	\end{bmatrix},
	\label{eq:eq_eight_extra_constraints}
\end{equation}
where $\boldsymbol{q}^{\mathrm{tar}}=(1-\mathrm{Pr})\boldsymbol{q}$, and the vanishing targets of the last two constraints reflect the isotropy of the continuous Shakhov stress tensor. The coefficients follow from the joint correction matrix in Eq.~\eqref{eq:joint_gh_correction_matrix} with the moment kernels of the present constraints.

On a fixed discrete velocity quadrature, replacing $(g^S,h^S)$ in Eq.~\eqref{eq:reduced_gh_eq} with the eight-moment corrected distributions constructed above gives the local collision equations. For every input where the correction is well defined, the eight-moment constraints enforce $\boldsymbol W(g^*,h^*)=\boldsymbol W$ and $\boldsymbol q(g^*,h^*)=(1-\mathrm{Pr})\,\boldsymbol q$. The conservative variables, the bulk velocity, and the peculiar velocities entering the heat flux therefore remain constant during local collision, yielding the exact moment relaxation relations $\mathrm{d}\boldsymbol W/\mathrm{d}t=\boldsymbol 0$ and $\mathrm{d}\boldsymbol q/\mathrm{d}t=-(\mathrm{Pr}/\tau)\,\boldsymbol q$. Assuming the corrected model is differentiable in $(\boldsymbol W,\boldsymbol q)$, the dependence on only six input moments decomposes the linearized collision dynamics into four conservative modes, two heat-flux modes, and the remaining moment-free modes. For $\mathrm{Pr}=2/3$, any fixed $\Delta t>0$, and equilibrium relaxation time $\tau_0>0$, the DUGKS trapezoidal collision update has an amplification factor of $1$ for the conservative modes, and
\begin{equation}
	r_q=\frac{1-\Delta t/(3\tau_0)}{1+\Delta t/(3\tau_0)},
	\qquad
	r_\perp=\frac{1-\Delta t/(2\tau_0)}{1+\Delta t/(2\tau_0)}
	\label{eq:local_collision_factors}
\end{equation}
for the heat-flux and remaining modes, respectively. Both factors have modulus smaller than $1$, yielding a full update spectral radius of $1$. On the invariant subspace of zero-conservative-moment perturbations, the spectral radius is $r_b=\max\{|r_q|,|r_\perp|\}<1$, ensuring asymptotic decay under repeated local updates. However, in the stiff limit $\Delta t/\tau_0\to\infty$, both factors approach $-1$, degrading the effective damping despite unconditional linear stability. Detailed derivations and implementation verifications are provided in Appendices~\ref{app:stability} and~\ref{app:response_protocol}.

\subsubsection{Moment-consistent mapping between different velocity spaces}
\label{subsubsection:mapping}
Conservative distribution-function mapping is required when evaluating microscopic fluxes at cross-level interfaces and when a physical cell is reassigned to a different velocity space. Let $\mathcal{V}^{s}=\{(\boldsymbol{\xi}_j^{s},w_j^{s})\}$ and $\mathcal{V}^{t}=\{(\boldsymbol{\xi}_i^{t},w_i^{t})\}$ denote the source and target velocity spaces, with the provisional projection
\begin{equation}
	(g^{p,t},h^{p,t})=\mathcal{P}_{s\rightarrow t}(g^{s},h^{s}),
	\label{eq:hierarchical_projection_operator}
\end{equation}
where the superscript $p$ denotes the provisional target-space representation. Since both grids are quadtrees, the projection follows the parent-child relations in two directions. For coarsening, the target cell $i$ aggregates the source leaf cells in $\mathcal{C}(i)$, i.e., those whose velocity domains are contained in that of cell $i$. Its weight and velocity coordinate, together with $g$ and the combined energy density $\mathcal{H}=h+|\boldsymbol{\xi}|^{2}g$, are aggregated as
\begin{equation}
	\begin{aligned}
		w_i^{t}   & =\sum_{j\in\mathcal{C}(i)}w_j^{s},
		          & \boldsymbol{\xi}_i^{t}                                            & =\Bigl(\sum_{j\in\mathcal{C}(i)}w_j^{s}\boldsymbol{\xi}_j^{s}\Bigr)\Big/w_i^{t}, \\[2pt]
		g_i^{p,t} & =\Bigl(\sum_{j\in\mathcal{C}(i)}w_j^{s}g_j^{s}\Bigr)\Big/w_i^{t},
		          & \mathcal{H}_i^{p,t}                                               & =\Bigl(\sum_{j\in\mathcal{C}(i)}w_j^{s}\mathcal{H}_j^{s}\Bigr)\Big/w_i^{t}.
	\end{aligned}
	\label{eq:mapping_g_projection}
\end{equation}
$\mathcal{H}$ is projected first, and $h$ is recovered as $h_i^{p,t}=\mathcal{H}_i^{p,t}-|\boldsymbol{\xi}_i^{t}|^{2}g_i^{p,t}$. Refinement reverses the traversal. Each target child inherits $g_i^{p,t}=g_{p(i)}^{s}$ and $\mathcal{H}_i^{p,t}=\mathcal{H}_{p(i)}^{s}$ from its source parent $p(i)$ and recovers $h$ by the same formula with its own velocity coordinate.

The projection above preserves mass and total energy, but since the source and target quadratures differ, errors can remain in other moments such as momentum, stress, and heat flux. In the global adaptive velocity-space method~\cite{Chen_global_2024a}, all cells share one velocity space, so cross-interface mapping is avoided. When the shared space is re-adapted, the projected distributions are corrected by the forced-conservation method
\begin{equation}
	\varphi_i^{\mathrm{FC}}
	=\varphi_i^{p}
	+\varphi_i^{eq}\left(\boldsymbol{W}^{s}\right)
	-\varphi_i^{eq}\left(\boldsymbol{W}^{p}\right),
	\label{eq:forced_conservation_mapping}
\end{equation}
The correction is designed to recover the conserved moments but leaves the higher-order nonequilibrium moments unconstrained. Moreover, the recovery is exact only when the discrete equilibrium itself satisfies the compatibility conditions on the target quadrature, and otherwise the errors accumulate over repeated mappings. In the present work, mapping in both situations employs the linear moment-constrained conservation correction introduced above.

When only the basic conserved quantities are to be preserved, the target moments are
\begin{equation}
	\boldsymbol{M}_4^{\mathrm{tar}}
	= \left[ \rho,\, \rho u_x,\, \rho u_y,\, \rho E \right]^T.
	\label{eq:mapping_four_targets}
\end{equation}
The mapping correction uses unit weighting ($\chi_i=1$), and the $g$-correction employs the raw-velocity basis
\begin{equation}
	\boldsymbol{\phi}_4^g
	= \left[ 1,\, \xi_x,\, \xi_y,\, \xi_x^2+\xi_y^2 \right]^T,
	\label{eq:raw_four_g_basis}
\end{equation}
whose first three functions correspond to mass and the two momentum components, and the fourth controls the radial second-order moment of $g$. Because total energy involves both reduced distributions, their corrections are coupled through the equilibrium-inspired relation $\delta h_i = (3-D+K)RT^{\mathrm{tar}}\delta g_i$. The coefficients are then determined jointly from the four moment constraints. The corrected distributions then satisfy the same four physical moment constraints as in Eq.~\eqref{eq:eq_four_moment_constraint}.

To ensure the consistency of the stress and heat flux, the correction is therefore extended to the nine target moments
\begin{equation}
	\boldsymbol{M}_9^{\mathrm{tar}}
	= \left[ \rho,\, \rho u_x,\, \rho u_y,\, \rho E,\,
	G_{xx}^{s},\, G_{yy}^{s},\, G_{xy}^{s},\, J_x^{s},\, J_y^{s} \right]^T,
	\label{eq:mapping_nine_targets}
\end{equation}
with the first four targets taken from the same stored macroscopic state as in Eq.~\eqref{eq:mapping_four_targets}. The complete second-order raw moments and total-energy-flux moments are evaluated from the source-space distribution as
\begin{equation}
	G_{\alpha\beta}^{s}
	= \sum_j w_j^{s} \xi_{\alpha,j}^{s}\xi_{\beta,j}^{s}g_j^{s},
	\quad
	J_{\alpha}^{s} = \frac{1}{2}\sum_j w_j^{s}\xi_{\alpha,j}^{s}
	\left[ h_j^{s}+(\xi_j^{s})^2g_j^{s} \right],
	\quad \mathrm{with} \quad
	\alpha,\beta\in\{x,y\}.
	\label{eq:source_mapping_moments}
\end{equation}
These raw moments are related to the central pressure tensor and heat flux through
\begin{equation}
	\begin{aligned}
		P_{\alpha\beta} = G_{\alpha\beta}-\rho u_\alpha u_\beta, \quad
		\Pi_{\alpha\beta} = P_{\alpha\beta} - p\,\delta_{\alpha\beta},\quad
		J_\alpha = \rho E u_\alpha+P_{\alpha\beta}u_\beta+q_\alpha,
	\end{aligned}
	\label{eq:raw_central_moment_relations}
\end{equation}
where $\Pi_{\alpha\beta}$ is the viscous stress tensor examined in Section~\ref{subsec:projection_test}, and summation over the repeated index $\beta$ is implied. Preserving the complete second-order raw moments and total-energy-flux moments together with the conserved quantities is therefore equivalent to preserving the central pressure tensor and heat flux.

The nine-moment correction is implemented in two successive blocks
\begin{equation}
	\begin{aligned}
		\delta g_i & =\sum_{m=1}^{6}a_m\phi_{m,i}^{g},
		           & \boldsymbol{\phi}_6^g             & =\left[ 1,\xi_x,\xi_y,\xi_x^2,\xi_y^2,\xi_x\xi_y \right]^T, \\
		\delta h_i & =\sum_{n=1}^{3}b_n\phi_{n,i}^{h},
		           & \boldsymbol{\phi}_3^h             & =\left[ 1,\xi_x,\xi_y \right]^T.
	\end{aligned}
	\label{eq:mapping_block_bases}
\end{equation}
The six coefficients $a_m$ are determined first to recover the prescribed mass, two momentum components, and three second-order raw moments. Since this correction also changes the total energy and its fluxes, the coefficients $b_n$ are determined from the remaining defects
\begin{equation}
	\frac{1}{2}\sum_i w_i
	\begin{bmatrix}
		1 \\ \xi_{x,i}\\ \xi_{y,i}
	\end{bmatrix}
	\delta h_i
	=
	\begin{bmatrix}
		\Delta(\rho E) \\ \Delta J_x\\ \Delta J_y
	\end{bmatrix}
	-
	\frac{1}{2}\sum_i w_i
	\begin{bmatrix}
		\xi_i^2 \\ \xi_{x,i}\xi_i^2\\ \xi_{y,i}\xi_i^2
	\end{bmatrix}
	\delta g_i.
	\label{eq:mapping_h_residual_correction}
\end{equation}
Because the $g$-block is solved first, its contribution to the combined moments is subtracted on the right-hand side. The $h$-block then removes only the remaining defects, completing the energy and energy-flux corrections without altering the six moments recovered by the $g$-block. The zeroth- and first-order moments of $h$ are not separately preserved, but rather are adjusted only to meet the combined targets.

Because unit weighting and raw-velocity bases are used, the matrices of the four- and nine-moment corrections depend only on the discrete velocities, weights, and basis functions of the target representative space, and are independent of the local flow state. Therefore, they are assembled and factorized once when each representative space is constructed. Subsequent mappings require only the state-dependent moment defects and the thermodynamic coupling terms on the right-hand sides, with no matrix reassembly.

The linear correction recovers the prescribed discrete moments but does not guarantee pointwise nonnegativity of the corrected distributions. A positivity-preserving fallback can be obtained by replacing the additive linear correction with an exponential moment-constrained correction,
\begin{equation}
	f_i^* = f_i^p \exp\!\left[\sum_{m=1}^{N_m} a_m \phi_m(\boldsymbol{\xi}_i)\right], \label{eq:exponential_moment_correction}
\end{equation}
where the coefficients are determined by enforcing the prescribed discrete moment constraints. Whenever a finite solution exists, the strictly positive exponential factor preserves the pointwise sign of the provisional distribution.

\subsection{Conservative inter-level coupling in the MLVS-DUGKS}
\label{subsection:inter_level_coupling}
In the MLVS-DUGKS, the linear correction of Section~\ref{subsubsection:mapping} is invoked in two situations: when a cell is reassigned to a different velocity space during adaptation, and when adjacent cells with different representative spaces are coupled across a physical interface.

Upon reassignment, hierarchical projection first provides a provisional representation on the new space. The linear correction is then applied with the pre-mapping moments as targets, so that
\begin{equation}
	\boldsymbol{M}_m^{\mathrm{new}} = \boldsymbol{M}_m^{\mathrm{old}}.
	\label{eq:level_switch_conservation}
\end{equation}
The corrected distributions are stored as the new cellwise representation. For interfaces between cells sharing one representative space, the standard DUGKS interface reconstruction and evolution apply directly. For a cross-level interface, an auxiliary velocity space $\mathcal{V}^{\mathrm{int}}$ is constructed as the minimal union of the velocity cells required by the participating velocity-space quadtrees. The post-collision distributions of the adjacent cells, along with the neighboring data required by the reconstruction stencil, are mapped to this space using the moment-constrained hierarchical projection of Section~\ref{subsubsection:mapping}. Spatial reconstruction, characteristic tracing, and interface-distribution recovery then follow the standard DUGKS procedure. With the unit normal $\boldsymbol{n}_f$ pointing from the left cell to the right cell, the half-time-step microscopic interface flux is
\begin{equation}
	\boldsymbol{\mathcal{F}}_{\mathrm{int},i} =
	\left(
	\boldsymbol{\xi}_i^{\mathrm{int}} \cdot \boldsymbol{n}_f
	\right)
	\begin{bmatrix}
		g_{\mathrm{int},i}^{n+1/2} , \;h_{\mathrm{int},i}^{n+1/2}
	\end{bmatrix}^T.
	\label{eq:interface_reduced_fluxes}
\end{equation}
The flux is then projected back to each native space and corrected by the same framework, acting directly on the microscopic flux, which already includes the normal-velocity factor. The common target is $\boldsymbol{F}_m^{\mathrm{int}} = \boldsymbol{M}_m^{\mathrm{int}}(\boldsymbol{\mathcal{F}}_{\mathrm{int}})$, and the corrected native-space fluxes satisfy
\begin{equation}
	\boldsymbol{F}_m^{L,*} = \boldsymbol{F}_m^{R,*} = \boldsymbol{F}_m^{\mathrm{int}},
	\quad m\in\{4,9\}.
	\label{eq:native_flux_moment_matching}
\end{equation}
Both sides define these targets with the same normal but update with opposite outward normals.

\section{Numerical results}\label{sec:numerical_results}
This section evaluates the performance of the proposed MLVS-DUGKS through benchmark problems covering a broad range of rarefaction regimes. The transport coefficients follow the variable hard-sphere (VHS) model~\cite{Bird_DSMC_2013}, characterized by the viscosity exponent $\omega = 0.81$. The dynamic viscosity follows the power law $\mu = \mu_{\mathrm{ref}} (T/T_{\mathrm{ref}})^{\omega}$ and the Knudsen number of each case is defined as $\mathrm{Kn}=\lambda/L_{\mathrm{ref}}$, with the VHS mean free path
\begin{equation}
	\lambda
	=
	\frac{16}{5}\,
	\frac{(5-2\omega)(7-2\omega)}{24}\,
	\frac{\mu}{\rho\sqrt{2\pi RT}} .
	\label{eq:vhs-mfp}
\end{equation}
Here $L_{\mathrm{ref}}$ is the reference length of the case. For $\omega=1/2$, Eq.~\eqref{eq:vhs-mfp} reduces to the classical hard-sphere mean free path $16\mu/(5\rho\sqrt{2\pi RT})$. For the dimensional cases, $T_{\mathrm{ref}}=273.15~\mathrm{K}$ and $\mu_{\mathrm{ref}}=2.125\times10^{-5}~\mathrm{Pa\cdot s}$.

The global time step corresponds to $\mathrm{CFL}=0.5$. For velocity-space adaptation, blocks of size $2\times2$, $4\times4$, and $8\times8$ are considered as candidates for merging. The velocity-space range, base resolution, adaptation interval, and criterion parameters for each benchmark are specified in the corresponding subsection. All computations are performed on an NVIDIA L40 GPU.

\subsection{Assessment of moment-constrained corrections}\label{sec: conservation}
This section evaluates the linear moment-constrained conservation correction introduced in Section~\ref{sec: Linear correction}, with the assessment proceeding from isolated operator-level tests to complete DUGKS simulations. These tests not only verify the effectiveness of the correction but also provide guidance for its adoption in the subsequent test cases.

\subsubsection{Discrete compatibility recovery of the model distribution}
\label{subsec:compatibility_test}
\begin{figure}[htbp]
	\centering
	\includegraphics[width=0.25\textwidth]{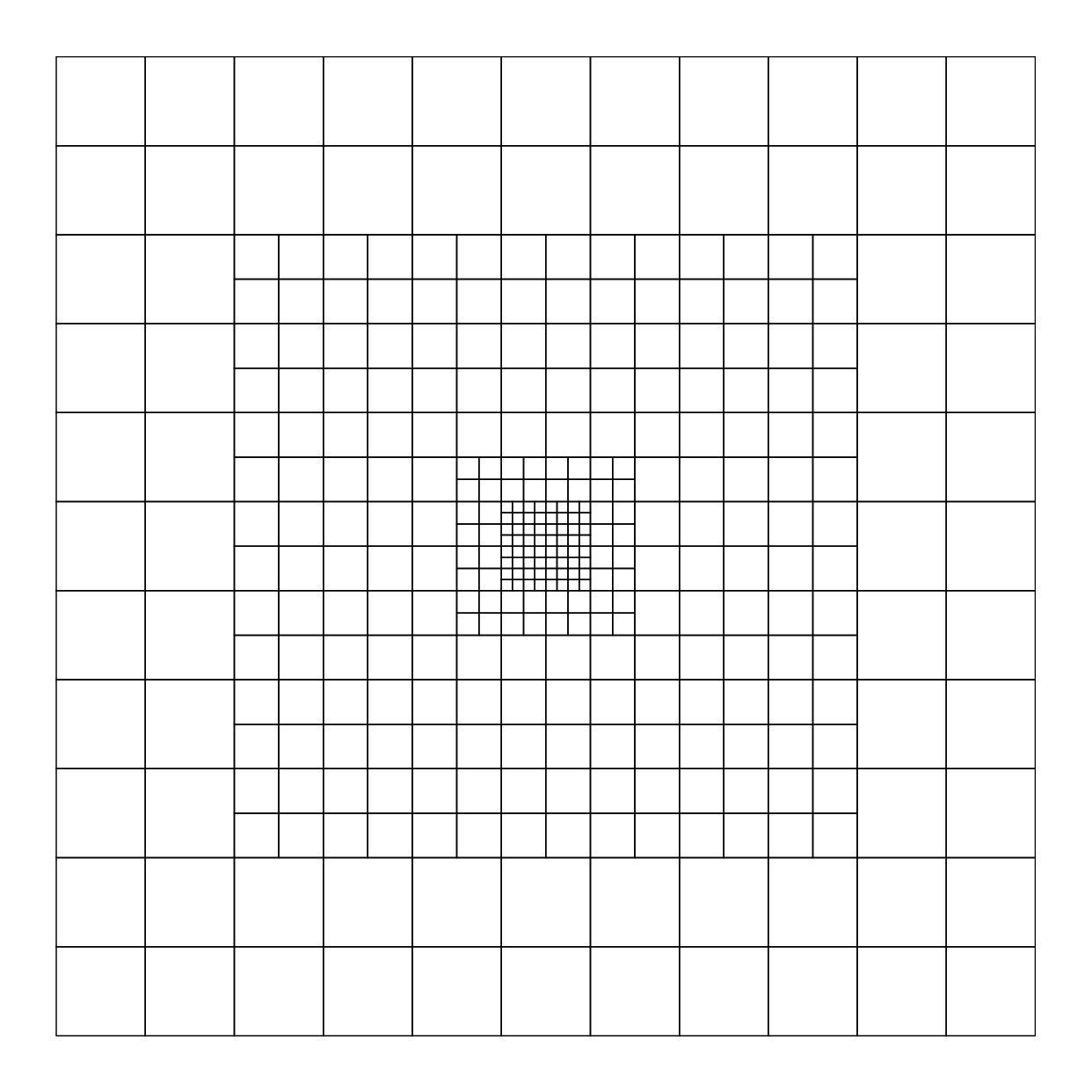}
	\caption{Velocity grid for the compatibility test.}
	\label{fig:velocity_grid}
\end{figure}

To verify the linear correction for enforcing the discrete compatibility of the model distribution, a single-step recovery test is conducted on the quadtree velocity grid shown in Fig.~\ref{fig:velocity_grid}. The nonuniform grid is intended to represent the irregular velocity-space resolution encountered in adaptive computations, where discrete quadrature errors are generally more pronounced than those on a sufficiently resolved uniform grid. A prescribed macroscopic state, $(\rho,u_x,u_y,T,R)=(1,1,0.2,1,0.5)$, is used to construct the reduced Maxwellian distributions $(g^{\mathrm{eq}},h^{\mathrm{eq}})$. The maximum relative error $\varepsilon_M 	= \max_k\left\{\left|M_k-M_k^{\mathrm{tar}}\right| / \left|M_k^{\mathrm{tar}}\right|\right\}$ is summarized in Table~\ref{tab:compatibility_correction}. Without correction, the maximum relative moment error is $\varepsilon_M=2.63\times10^{-3}$. The four-moment linear correction reduces the error to $1.11\times10^{-16}$, while the Newton iterative correction gives $2.78\times10^{-16}$. Both values are at the level of double-precision round-off error, indicating that the discrete compatibility defect has been effectively eliminated. As for computational efficiency, the proposed linear correction achieves an approximately 3.5-fold speedup over the Newton iterative method. Overall, the linear correction attains accuracy comparable to that of the Newton iterative method at markedly higher efficiency, and is therefore adopted in the remainder of this work.

\begin{table}[htbp]
	\centering
	\caption{Comparison of correction methods in the compatibility recovery test.}
	\label{tab:compatibility_correction}
	\setlength{\tabcolsep}{6pt}
	\small
	\begin{tabular}{l c c c}
		\toprule
		Method           & $\varepsilon_M$      & Avg. time per run (ms) & Ratio \\
		\midrule
		No correction    & $2.63\times10^{-3}$  & ---                    & ---   \\
		4-moment linear  & $1.11\times10^{-16}$ & 5.65                  & 1.00  \\
		Newton iterative & $2.78\times10^{-16}$ & 19.96                 & 3.53  \\
		\bottomrule
	\end{tabular}
\end{table}

\subsubsection{Repeated velocity-space projection}
\label{subsec:projection_test}
To assess the moment-consistent velocity-space mapping described in Section~\ref{subsubsection:mapping}, a repeated-projection test is performed. The initial state is $(\rho,u_x,u_y,T,R)=(1,1,0,1,0.5)$, and the reduced distribution is initialized as $g=g^{\mathrm{eq}}(1+\varepsilon\delta)$, where $\varepsilon$ controls the perturbation amplitude and $\delta$ is discretely orthogonalized against the basic moment basis, so that it alters the prescribed higher-order nonequilibrium moments while leaving the initial mass, momentum, and total energy unchanged. Perturbations are imposed independently on the heat-flux and stress moments to assess the preservation of $q_x$ and $\Pi_{xx}-\Pi_{yy}$, respectively. In each case, the velocity space is coarsened from $128\times128$ to $8\times8$ cells by halving the resolution in alternating directions, and is finally projected back to the original $128\times128$ resolution.

\begin{table}[htbp]
	\centering
	\caption{Final moments after the repeated projection sequence. For the heat-flux perturbation, $(\Pi_{xx}-\Pi_{yy})_0=0$ and $(q_x)_0=0.7071\varepsilon$; for the stress perturbation, $(\Pi_{xx}-\Pi_{yy})_0=2\varepsilon$ and $(q_x)_0=0$.}
	\label{tab:projection_merged}
	\scriptsize
	\begin{tabular}{c c l c c c c c}
		\toprule
		$\varepsilon$ & Perturbation & Method & $\rho$ & $\rho u_x$ & $\rho E$ & $\Pi_{xx}-\Pi_{yy}$ & $q_x$ \\
		\midrule
		\multirow{8}{*}{$10^{-3}$} & \multirow{4}{*}{Heat flux} & Projection only   & 1.000 & 1.000 & 1.250 & $-4.097\times10^{-1}$ & $-5.437\times10^{-1}$ \\
		                           &                            & Forced correction & 1.000 & 1.000 & 1.250 & $-4.094\times10^{-1}$ & $-5.439\times10^{-1}$ \\
		                           &                            & 4-moment linear   & 1.000 & 1.000 & 1.250 & $-4.091\times10^{-1}$ & $-5.529\times10^{-1}$ \\
		                           &                            & 9-moment linear   & 1.000 & 1.000 & 1.250 & $4.700\times10^{-14}$ & $7.071\times10^{-4}$  \\
		\cmidrule(l){2-8}
		                           & \multirow{4}{*}{Stress}    & Projection only   & 1.000 & 1.000 & 1.250 & $-4.077\times10^{-1}$ & $-5.450\times10^{-1}$ \\
		                           &                            & Forced correction & 1.000 & 1.000 & 1.250 & $-4.077\times10^{-1}$ & $-5.450\times10^{-1}$ \\
		                           &                            & 4-moment linear   & 1.000 & 1.000 & 1.250 & $-4.077\times10^{-1}$ & $-5.450\times10^{-1}$ \\
		                           &                            & 9-moment linear   & 1.000 & 1.000 & 1.250 & $2.000\times10^{-3}$  & $-1.210\times10^{-13}$ \\
		\midrule
		\multirow{8}{*}{$10^{-1}$} & \multirow{4}{*}{Heat flux} & Projection only   & 1.000 & 1.030 & 1.250 & $-4.097\times10^{-1}$ & $-4.555\times10^{-1}$ \\
		                           &                            & Forced correction & 0.983 & 1.004 & 1.259 & $-3.754\times10^{-1}$ & $-4.718\times10^{-1}$ \\
		                           &                            & 4-moment linear   & 1.000 & 1.000 & 1.250 & $-3.503\times10^{-1}$ & $-1.373$ \\
		                           &                            & 9-moment linear   & 1.000 & 1.000 & 1.250 & $1.241\times10^{-13}$ & $7.071\times10^{-2}$  \\
		\cmidrule(l){2-8}
		                           & \multirow{4}{*}{Stress}    & Projection only   & 1.000 & 1.000 & 1.250 & $-2.101\times10^{-1}$ & $-5.864\times10^{-1}$ \\
		                           &                            & Forced correction & 1.000 & 1.000 & 1.250 & $-2.101\times10^{-1}$ & $-5.864\times10^{-1}$ \\
		                           &                            & 4-moment linear   & 1.000 & 1.000 & 1.250 & $-2.101\times10^{-1}$ & $-5.864\times10^{-1}$ \\
		                           &                            & 9-moment linear   & 1.000 & 1.000 & 1.250 & $2.000\times10^{-1}$  & $-1.891\times10^{-15}$ \\
		\bottomrule
	\end{tabular}
\end{table}

Table~\ref{tab:projection_merged} presents the results with $\varepsilon=10^{-3}$ and $\varepsilon=10^{-1}$. At the smaller amplitude, all four methods preserve the basic conserved moments. At the larger amplitude, the distinction emerges for the heat-flux perturbation, where projection only distorts the streamwise momentum to $\rho u_x=1.030$, and the forced correction distorts the density and total energy to $0.983$ and $1.259$, respectively. In all cases, the two linear corrections maintain the basic conserved moments. Regarding the higher-order moments, only the nine-moment linear correction preserves them at their prescribed target values, with deviations ranging from $10^{-15}$ to $10^{-13}$ regardless of the perturbation amplitude. For the other methods, the higher-order moments deviate significantly from their targets. Since these methods impose no constraints on the nonequilibrium moments, the resulting distortion reflects both the coarse-grid quadrature error and the perturbation itself.

\subsubsection{Uniform flow}
\label{subsec:uniform_conservation}
The preceding tests assessed the correction in individual operations. Here, the uniform-flow simulation evaluates its performance within the complete DUGKS solving procedure. Because the distribution remains uniform, velocity-space adaptation assigns the same velocity grid to all physical cells, meaning no interfaces between different velocity levels are present. This primarily isolates the influence of velocity grid coarsening on discrete compatibility and the accumulation of resulting errors during DUGKS updates. The periodic domain $[0,1]\times[0,0.1]$ is discretized into $100\times10$ uniform cells with $L_{\mathrm{ref}}=1$, and the initial state is $(\rho,u_x,u_y,T)=(1,\,0.1,\,0,\,1)$. Three Knudsen numbers, $\mathrm{Kn}=10^{-3}$, $0.1$, and $10$, are advanced for 1000 steps with both the UVS- and MLVS-DUGKS. The UVS-DUGKS employs a uniform $89\times89$ velocity grid over $\left[-4\sqrt{2RT},4\sqrt{2RT}\right]^2$, whereas the MLVS-DUGKS updates its representative velocity spaces every 10 steps with $C_{\mathrm{split}}=2.0\times10^{-3}$, $C_{\mathrm{merge}}=1\times10^{-3}$, $\varepsilon_{\mathrm{merge}}=2.0\times10^{-5}$, and a clustering threshold of $0.05$.

\begin{table*}[htbp] \centering
    \caption{Errors in the monitored moments for the uniform-flow test.}
    \label{tab:uniform_correction_summary}
    \renewcommand{\arraystretch}{1.15}
    \scriptsize
    \begin{tabular}{c l l c c c c c c}
        \toprule
        $\mathrm{Kn}$ & Method & Correction & $\varepsilon_{\rho}$ & $\varepsilon_{\rho u_x}$ & $\varepsilon_E$     & $\varepsilon_{\Pi_{xx}}$ & $\varepsilon_{\Pi_{yy}}$ & $\varepsilon_{q_x}$ \\
        \midrule
        \multirow{5}{*}{$10^{-3}$}
                      & UVS    & None       & $1.25\times10^{-5}$  & $2.05\times10^{-5}$      & $1.12\times10^{-4}$ & $4.27\times10^{-7}$      & $4.26\times10^{-7}$      & $1.88\times10^{-7}$ \\
                      & UVS    & 4-moment   & $2.37\times10^{-8}$  & $3.89\times10^{-8}$      & $2.13\times10^{-7}$ & $6.21\times10^{-7}$      & $6.21\times10^{-7}$      & $1.91\times10^{-7}$ \\
                      & MLVS   & None       & $1.94\times10^{-1}$  & $7.13\times10^{-3}$      & $3.32\times10^{-1}$ & $3.13\times10^{-4}$      & $3.35\times10^{-4}$      & $1.89\times10^{-4}$ \\
                      & MLVS   & 4-moment   & $2.32\times10^{-8}$  & $3.82\times10^{-8}$      & $2.09\times10^{-7}$ & $5.64\times10^{-5}$      & $5.76\times10^{-5}$      & $1.35\times10^{-4}$ \\
                      & MLVS   & 8-moment   & $2.32\times10^{-8}$  & $3.82\times10^{-8}$      & $2.09\times10^{-7}$ & $6.84\times10^{-7}$      & $5.33\times10^{-7}$      & $9.22\times10^{-7}$ \\
        \midrule
        \multirow{5}{*}{$0.1$}
                      & UVS    & None       & $1.87\times10^{-7}$  & $3.07\times10^{-7}$      & $1.68\times10^{-6}$ & $4.26\times10^{-7}$      & $4.26\times10^{-7}$      & $2.34\times10^{-7}$ \\
                      & UVS    & 4-moment   & $3.60\times10^{-8}$  & $5.92\times10^{-8}$      & $3.24\times10^{-7}$ & $6.18\times10^{-7}$      & $6.18\times10^{-7}$      & $2.34\times10^{-7}$ \\
                      & MLVS   & None       & $3.01\times10^{-3}$  & $2.81\times10^{-4}$      & $5.89\times10^{-3}$ & $6.33\times10^{-4}$      & $5.24\times10^{-4}$      & $1.26\times10^{-4}$ \\
                      & MLVS   & 4-moment   & $3.52\times10^{-8}$  & $5.80\times10^{-8}$      & $3.18\times10^{-7}$ & $5.48\times10^{-5}$      & $5.60\times10^{-5}$      & $1.27\times10^{-4}$ \\
                      & MLVS   & 8-moment   & $3.52\times10^{-8}$  & $5.80\times10^{-8}$      & $3.18\times10^{-7}$ & $6.80\times10^{-7}$      & $5.31\times10^{-7}$      & $8.74\times10^{-7}$ \\
        \midrule
        \multirow{5}{*}{$10$}
                      & UVS    & None       & $3.77\times10^{-8}$  & $6.19\times10^{-8}$      & $3.39\times10^{-7}$ & $4.26\times10^{-7}$      & $4.26\times10^{-7}$      & $1.29\times10^{-7}$ \\
                      & UVS    & 4-moment   & $3.62\times10^{-8}$  & $5.94\times10^{-8}$      & $3.25\times10^{-7}$ & $4.34\times10^{-7}$      & $4.34\times10^{-7}$      & $1.29\times10^{-7}$ \\
                      & MLVS   & None       & $3.01\times10^{-5}$  & $1.98\times10^{-6}$      & $5.84\times10^{-5}$ & $2.32\times10^{-5}$      & $2.23\times10^{-5}$      & $1.88\times10^{-6}$ \\
                      & MLVS   & 4-moment   & $3.54\times10^{-8}$  & $5.83\times10^{-8}$      & $3.19\times10^{-7}$ & $2.35\times10^{-9}$      & $8.48\times10^{-7}$      & $1.93\times10^{-6}$ \\
                      & MLVS   & 8-moment   & $3.54\times10^{-8}$  & $5.83\times10^{-8}$      & $3.19\times10^{-7}$ & $4.28\times10^{-7}$      & $4.22\times10^{-7}$      & $1.58\times10^{-7}$ \\
        \bottomrule
    \end{tabular}
\end{table*}

Table~\ref{tab:uniform_correction_summary} shows that conservation errors arise even in the UVS-DUGKS due to discrete quadrature imbalance, and that velocity-space coarsening significantly amplifies these errors in the MLVS-DUGKS. As $\mathrm{Kn}$ increases, the conservation errors decrease and the gap between the two methods narrows, consistent with the diminished role of conservation correction in highly rarefied regimes~\cite{Zhang_microscopically_2026}. With the collision-model correction activated, the four-moment variant suppresses the basic-moment errors to the round-off level for both methods. However, in the MLVS-DUGKS, the higher-order moments remain unconstrained, so errors in $\Pi_{xx}$, $\Pi_{yy}$, and $q_x$ persist at $10^{-5}$--$10^{-4}$. Extending the correction to eight moments reduces these higher-order errors, while leaving the basic-moment accuracy essentially unchanged. Accordingly, the eight-moment collision-model correction, which enforces the heat flux in addition to the basic conserved moments, is adopted for the remaining benchmarks.

\subsubsection{Sinusoidal wave flow}
\label{subsec:sinusoidal_wave}
Following the uniform-flow assessment, this section examines the sinusoidal-wave problem, in which velocity-space adaptation, inter-level distribution mapping, and cross-level flux transfer are all active. The flow field is initialized as 
\begin{equation}
    \rho(x) = 1 + 0.2\sin\left(\frac{2\pi x}{L}\right), 
    \quad (u_x, u_y) = (0.2, 0), 
    \quad p = 1.0, 
    \quad T(x) = \frac{p}{\rho(x) R}.
    \label{eq:sinusoidal_field}
\end{equation}
The velocity-space setup is the same as that in Section~\ref{subsec:uniform_conservation}.

\paragraph{Perturbation decay response} \leavevmode\par
To assess the perturbation decay, paired runs are performed on this isobaric background on a $16\times2$ uniform mesh over $[0,1]\times[0,0.125]$ at $\mathrm{Kn}=10^{-3}$, comparing the UVS-DUGKS with the MLVS-DUGKS using the four- and nine-moment inter-level corrections. The imposed perturbation consists of a small isobaric disturbance of the conserved moments that drives the macroscopic response, and a random kinetic component that carries no conserved moments and excites only nonequilibrium modes. This combination is normalized by the discrete distribution norm and scaled to $\varepsilon_0\in\{10^{-5},\,3\times10^{-6},\,10^{-6},\,3\times10^{-7}\}$. Each perturbed run is paired with a zero-perturbation baseline. All runs are advanced with $\Delta t=0.005$ to $t=25$.

\begin{figure}[htbp]
	\centering
	\includegraphics[width=0.4\linewidth]{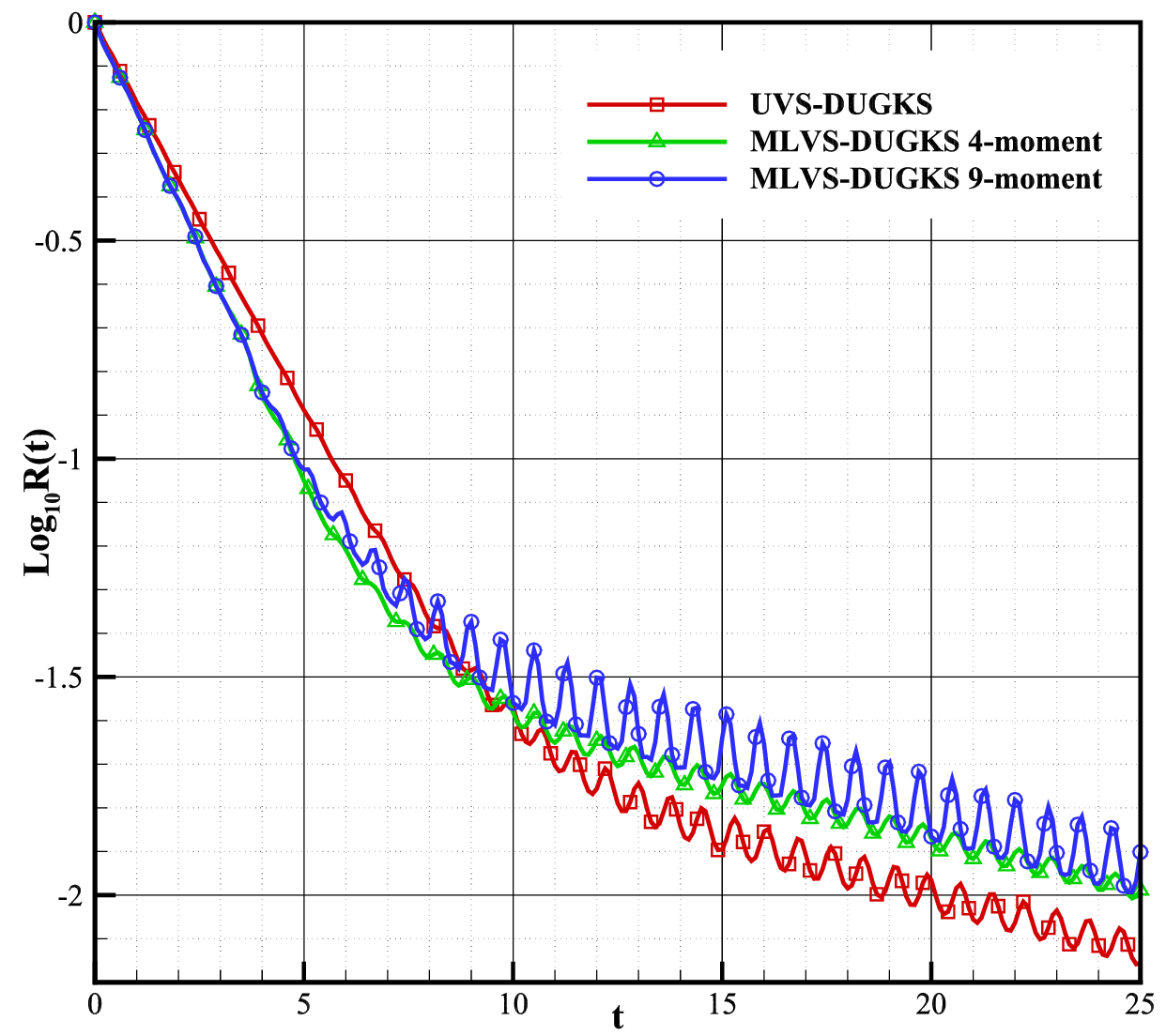}
	\caption{Normalized macroscopic perturbation response at $\mathrm{Kn}=10^{-3}$ for $\varepsilon_0=10^{-5}$.}
	\label{fig:sine_decay}
\end{figure}

The response is measured by the volume-weighted RMS relative difference of the conserved variables between each perturbed run and its baseline. The response is then normalized by its initial amplitude as $R(t)=\varepsilon(t)/\varepsilon(0)$. Since the computations at different amplitudes produce nearly identical responses, only the result for $\varepsilon_0=10^{-5}$ is plotted in Fig.~\ref{fig:sine_decay}. All three curves exhibit the same two-stage decay. At early times ($t\lesssim 8$), the decay is exponential, consistent with the local linear analysis in Appendix~\ref{app:stability}. During this stage, the MLVS-DUGKS calculations decay slightly faster than the UVS-DUGKS. For $t\gtrsim 8$, the curves transition to a slower oscillatory decay whose envelope decreases gradually, and the three curves separate, with the UVS-DUGKS decaying fastest, the four-moment result following, and the nine-moment result decaying slowest. The four- and nine-moment results nearly coincide during the exponential stage, their difference remaining within about half an order of magnitude thereafter. By $t=25$, all three responses have decayed to about the one-percent level of the initial amplitude. Overall, these results indicate that the MLVS-DUGKS possesses a perturbation damping capability comparable to that of the UVS-DUGKS, demonstrating good numerical stability.

\paragraph{Global conservation and final-field accuracy} \leavevmode\par

The companion calculations are performed on a $100\times10$ physical mesh covering $[0,1]\times[0,0.1]$, with a CFL number of $0.5$ and a final time of $t=5$, for $\mathrm{Kn}\in\{10^{-3},\,0.1,\,10\}$. Table~\ref{tab:sine_conservation_errors} summarizes the resulting global conservation errors. As shown, without inter-level correction, the errors are significantly larger than those of the UVS-DUGKS. Both the four- and nine-moment corrections reduce these errors to the same order of magnitude as the UVS-DUGKS values across all three Knudsen numbers.

In addition to global conservation, the final moment fields are compared with the UVS-DUGKS solution using the volume-weighted $L_2$ error (Table~\ref{tab:sine_field_errors}). All basic-field errors remain below $2\%$, with only the $x$-momentum error at $\mathrm{Kn}=10$ exceeding $1\%$. Neither correction shows a systematic advantage. This arises because the inter-level correction only acts on coarsened regions that are identified as low-importance by the adaptation criterion. Since these points contribute little to the recovered macroscopic fields, the higher-order constraints cannot yield a significant improvement in the overall macroscopic accuracy. Nevertheless, the influence of these constraints is evident in the nonequilibrium fields. Based on this, the four-moment correction is adopted for subsequent distribution mapping and flux transfer due to its lower algebraic cost and comparable conserved-field errors, combined with the eight-moment collision-model correction.

\begin{table}[htbp]
	\centering
	\caption{Global conservation errors obtained with the UVS-DUGKS and the MLVS-DUGKS.}
	\label{tab:sine_conservation_errors}
	\renewcommand{\arraystretch}{1.15}
	\setlength{\tabcolsep}{5pt}
	\small
	\begin{tabular}{c l c c c}
		\toprule
		$\mathrm{Kn}$ & Method and correction strategy
		              & $\rho$                          & $\rho u_x$           & $\rho E$             \\
		\midrule
		\multirow{4}{*}{$10^{-3}$}
		              & UVS-DUGKS
		              & $1.401\times10^{-7}$            & $5.135\times10^{-8}$ & $1.909\times10^{-6}$ \\
		              & MLVS-DUGKS, no correction
		              & $9.209\times10^{-4}$            & $3.429\times10^{-4}$ & $2.823\times10^{-3}$ \\
		              & MLVS-DUGKS, 4-moment correction
		              & $1.379\times10^{-7}$            & $3.901\times10^{-7}$ & $1.674\times10^{-6}$ \\
		              & MLVS-DUGKS, 9-moment correction
		              & $1.379\times10^{-7}$            & $3.904\times10^{-7}$ & $1.674\times10^{-6}$ \\
		\midrule
		\multirow{4}{*}{$0.1$}
		              & UVS-DUGKS
		              & $2.004\times10^{-7}$            & $5.084\times10^{-7}$ & $2.455\times10^{-6}$ \\
		              & MLVS-DUGKS, no correction
		              & $1.195\times10^{-4}$            & $1.582\times10^{-4}$ & $3.002\times10^{-4}$ \\
		              & MLVS-DUGKS, 4-moment correction
		              & $1.973\times10^{-7}$            & $5.395\times10^{-7}$ & $2.391\times10^{-6}$ \\
		              & MLVS-DUGKS, 9-moment correction
		              & $1.973\times10^{-7}$            & $5.395\times10^{-7}$ & $2.391\times10^{-6}$ \\
		\midrule
		\multirow{4}{*}{$10$}
		              & UVS-DUGKS
		              & $2.011\times10^{-7}$            & $5.439\times10^{-7}$ & $2.436\times10^{-6}$ \\
		              & MLVS-DUGKS, no correction
		              & $4.362\times10^{-5}$            & $2.395\times10^{-7}$ & $1.315\times10^{-4}$ \\
		              & MLVS-DUGKS, 4-moment correction
		              & $1.980\times10^{-7}$            & $5.411\times10^{-7}$ & $2.399\times10^{-6}$ \\
		              & MLVS-DUGKS, 9-moment correction
		              & $1.980\times10^{-7}$            & $5.411\times10^{-7}$ & $2.399\times10^{-6}$ \\
		\bottomrule
	\end{tabular}
\end{table}

\begin{table}[htbp]
	\centering
	\caption{Final-field $L_2$ errors of MLVS-DUGKS relative to the UVS-DUGKS.}
	\label{tab:sine_field_errors}
	\renewcommand{\arraystretch}{1.15}
	\setlength{\tabcolsep}{5pt}
	\small
	\begin{tabular}{c l c c c c c}
		\toprule
		$\mathrm{Kn}$ & Correction
		              & $\rho$                 & $\rho u_x$            & $\rho E$            & $\Pi_{xx}$ & $q_x$ \\
		\midrule
		\multirow{2}{*}{$10^{-3}$}
		              & 4-moment
		              & $1.665\times10^{-4}$    & $1.376\times10^{-4}$ & $2.299\times10^{-5}$
		              & $4.017\times10^{-6}$    & $5.135\times10^{-5}$                                               \\
		              & 9-moment
		              & $1.644\times10^{-4}$    & $1.111\times10^{-4}$ & $2.073\times10^{-5}$
		              & $1.480\times10^{-6}$    & $5.203\times10^{-5}$                                               \\
		\midrule
		\multirow{2}{*}{$0.1$}
		              & 4-moment
		              & $4.137\times10^{-5}$    & $8.364\times10^{-6}$ & $4.313\times10^{-5}$
		              & $1.053\times10^{-4}$    & $3.827\times10^{-5}$                                               \\
		              & 9-moment
		              & $5.349\times10^{-5}$    & $2.050\times10^{-5}$ & $6.168\times10^{-5}$
		              & $6.119\times10^{-5}$    & $1.021\times10^{-5}$                                               \\
		\midrule
		\multirow{2}{*}{$10$}
		              & 4-moment
		              & $5.832\times10^{-4}$    & $1.638\times10^{-2}$ & $2.466\times10^{-3}$
		              & $4.078\times10^{-3}$    & $5.924\times10^{-3}$                                               \\
		              & 9-moment
		              & $4.415\times10^{-4}$    & $1.400\times10^{-2}$ & $2.324\times10^{-3}$
		              & $2.218\times10^{-3}$    & $4.944\times10^{-3}$                                               \\
		\bottomrule
	\end{tabular}
\end{table}

\subsection{Lid-driven cavity flow}
The lid-driven cavity flow is a standard benchmark for assessing numerical methods across different rarefaction regimes. The present configuration follows Ref.~\cite{Zhu_Discrete_2016}. Calculations are performed at $\mathrm{Kn}=0.075$, $1.0$, and $10$ with $40\times40$ uniform cells. The UVS-DUGKS employs a uniform $101\times101$ velocity grid over $[-4\sqrt{2RT_w},\,4\sqrt{2RT_w}]^2$, and the MLVS-DUGKS shares the physical discretization, updating the representative velocity spaces every 100 time steps with $C_{\mathrm{split}}=5.0\times10^{-3}$, $C_{\mathrm{merge}}=2.5\times10^{-3}$, $\varepsilon_{\mathrm{merge}}=5.0\times10^{-5}$, and a clustering threshold of $0.05$. All cases are advanced for 10,000 time steps, and UGKS solutions~\cite{Xu_unified_2010} are included as qualitative references.

\begin{figure}[htbp]
	\centering
	\begin{subfigure}[t]{0.32\textwidth}
		\centering
		\includegraphics[width=0.9\textwidth]{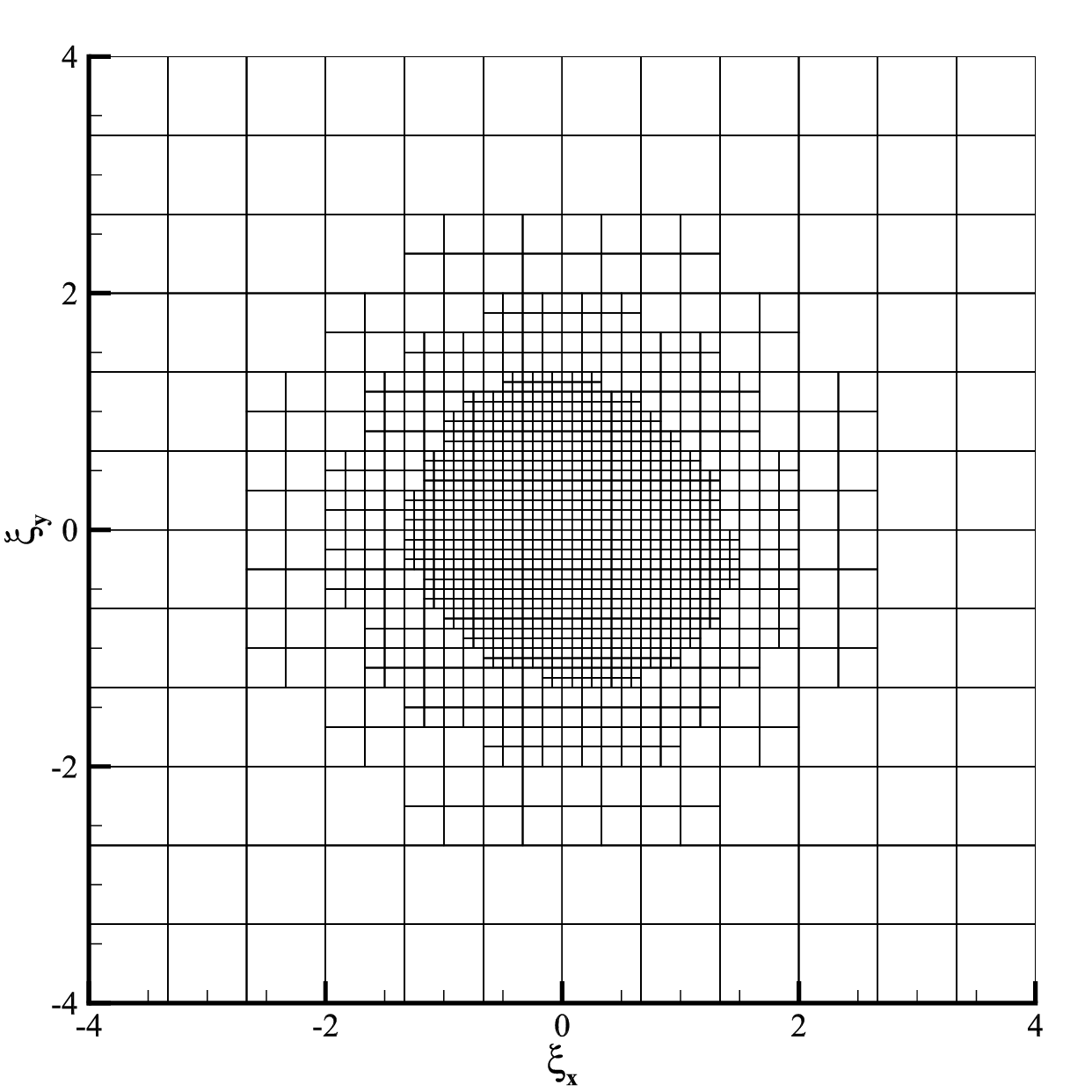}
		\caption{$\mathrm{Kn} = 0.075$, 1,254 points}
	\end{subfigure}
	\begin{subfigure}[t]{0.32\textwidth}
		\centering
		\includegraphics[width=0.9\textwidth]{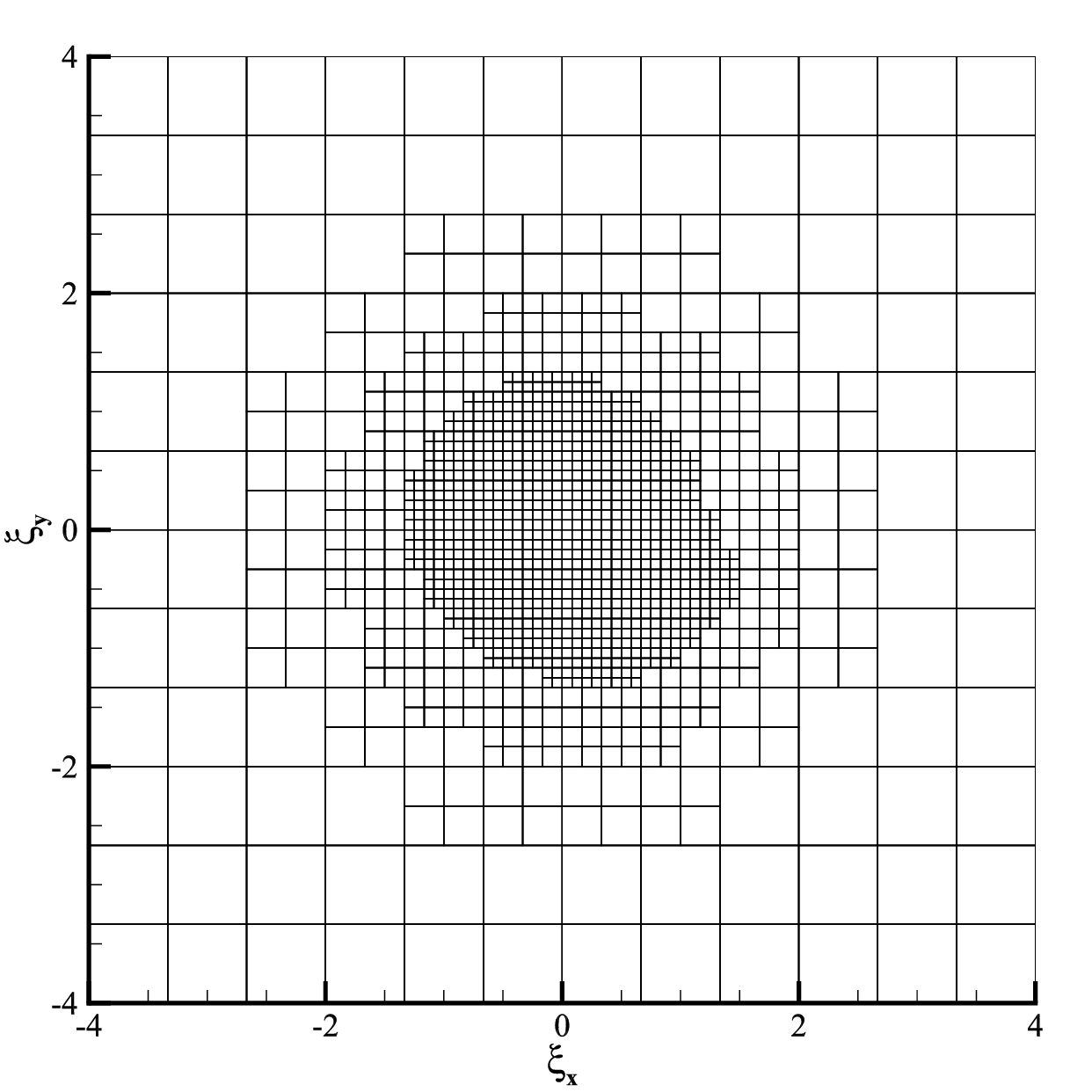}
		\caption{$\mathrm{Kn} = 1.0$, 1,254 points}
	\end{subfigure}
	\begin{subfigure}[t]{0.32\textwidth}
		\centering
		\includegraphics[width=0.9\textwidth]{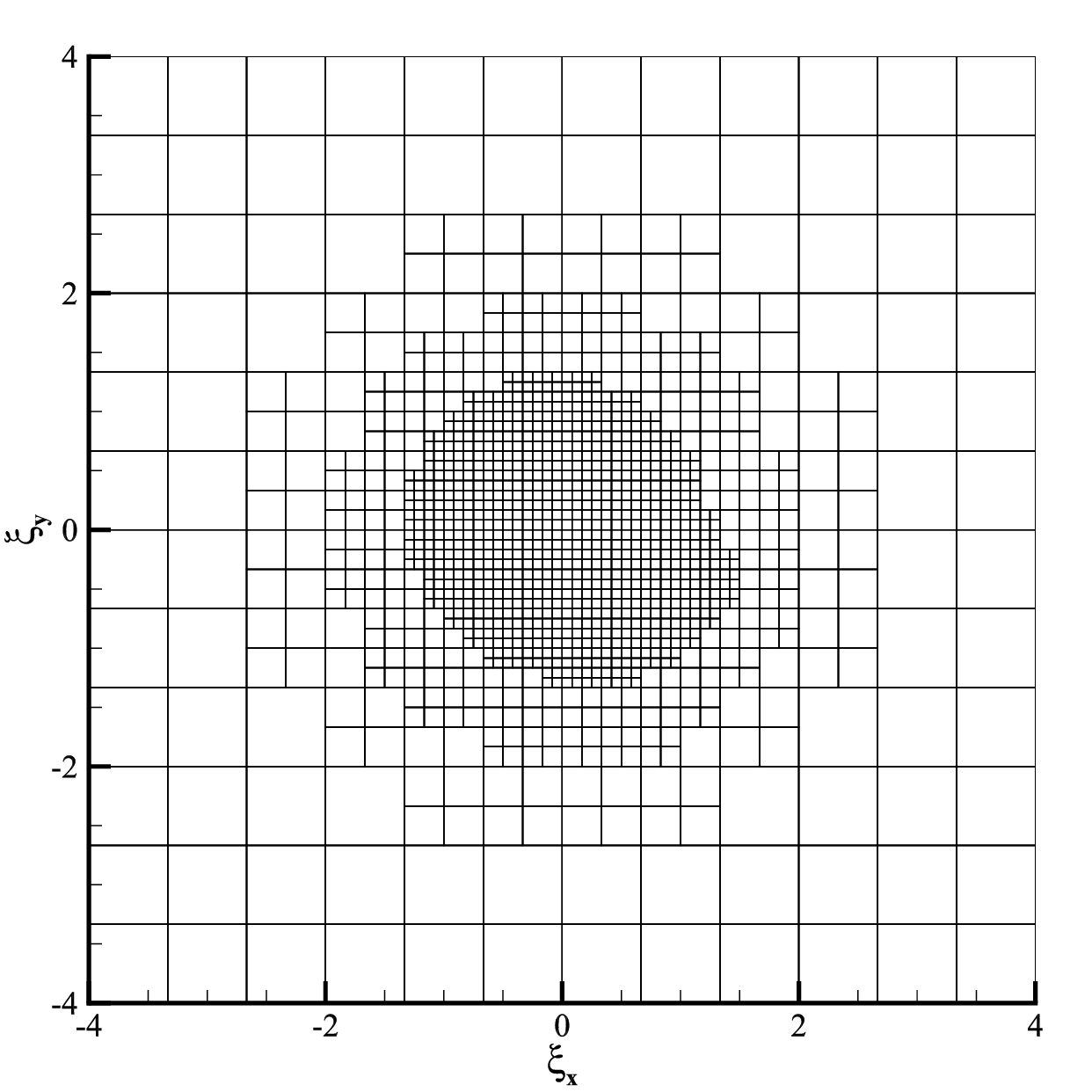}
		\caption{$\mathrm{Kn} = 10$, 1,254 points}
	\end{subfigure}

	\caption{Representative adaptive velocity spaces for the lid-driven cavity flow.}
	\label{fig:cavity_velocityspace}
\end{figure}

Since the lid-driven cavity exhibits weak variation of the macroscopic flow across the domain, the distribution functions vary correspondingly little. Consequently, at each Knudsen number all physical cells are assigned to a single active representative velocity space, as shown in Fig.~\ref{fig:cavity_velocityspace}. The three adapted spaces contain the same 1,254 velocity points and are nearly identical. This behavior reflects the inherent flow configuration rather than a limitation of the method. If a finer velocity-space separation is desired, multiple representative spaces can be generated by reducing the clustering threshold.

\begin{figure}[!htbp] \centering
    \begin{subfigure}[b]{0.9\textwidth} \centering
        \includegraphics[width=0.33\textwidth]{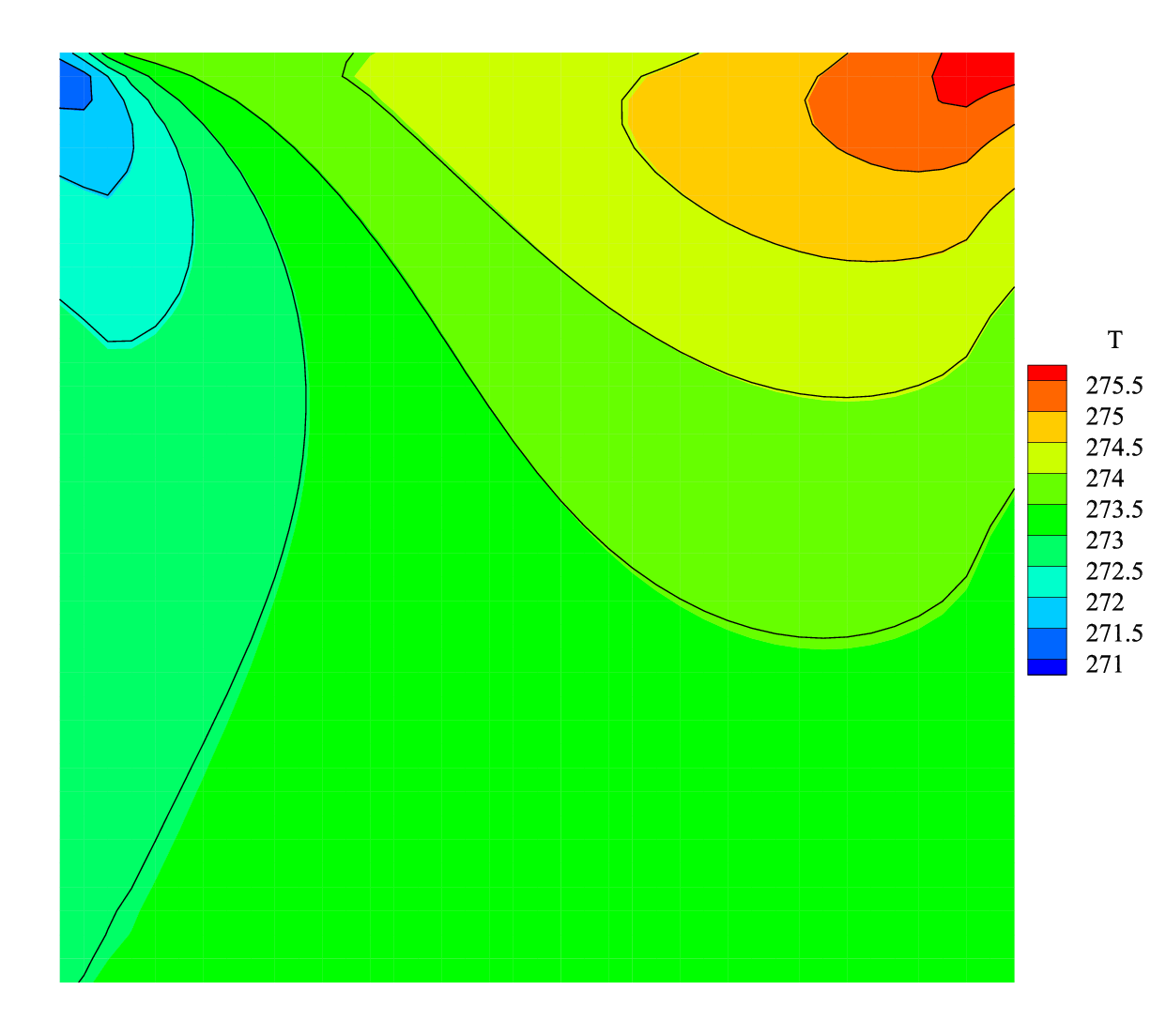}\hfill
        \includegraphics[width=0.30\textwidth]{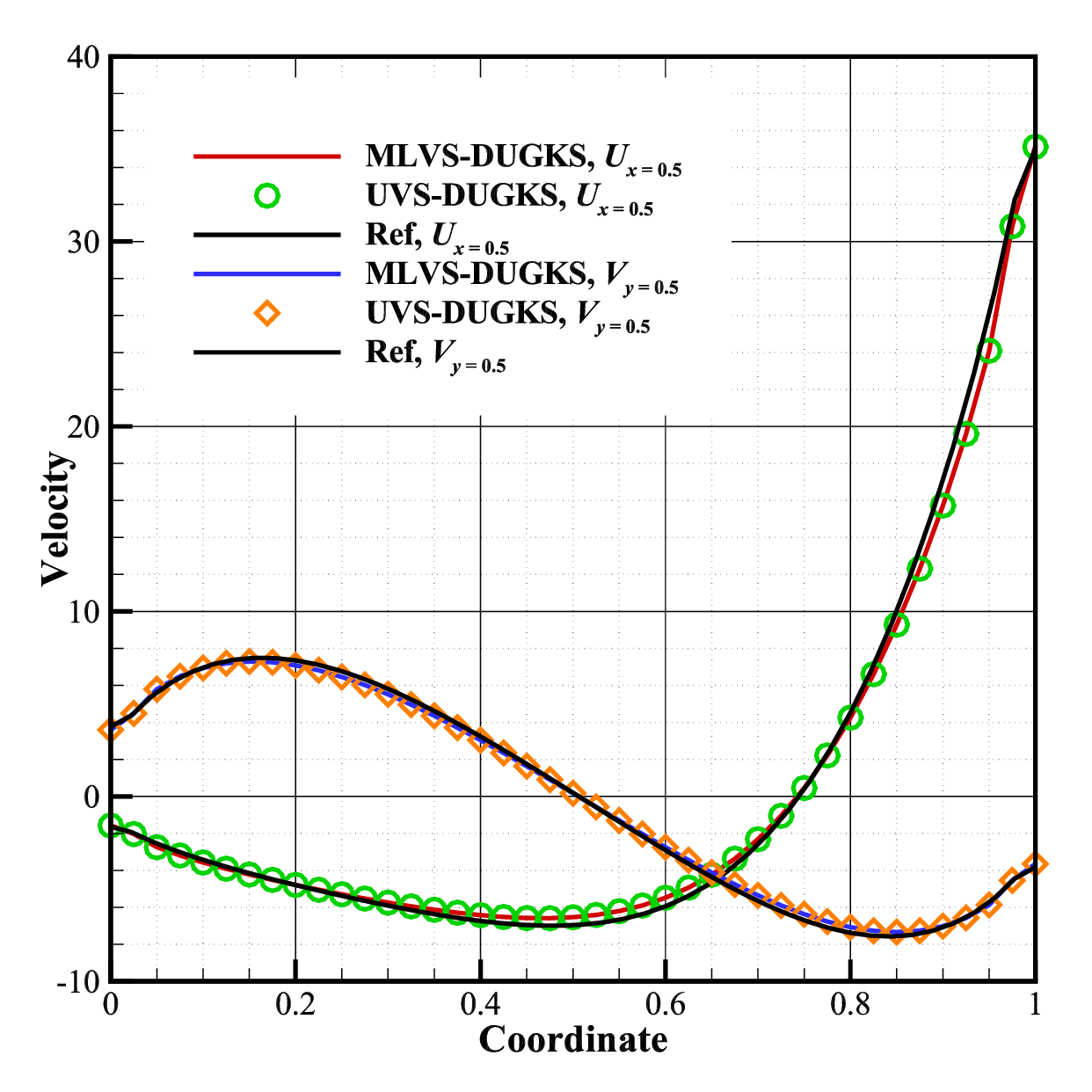}\hfill
        \includegraphics[width=0.30\textwidth]{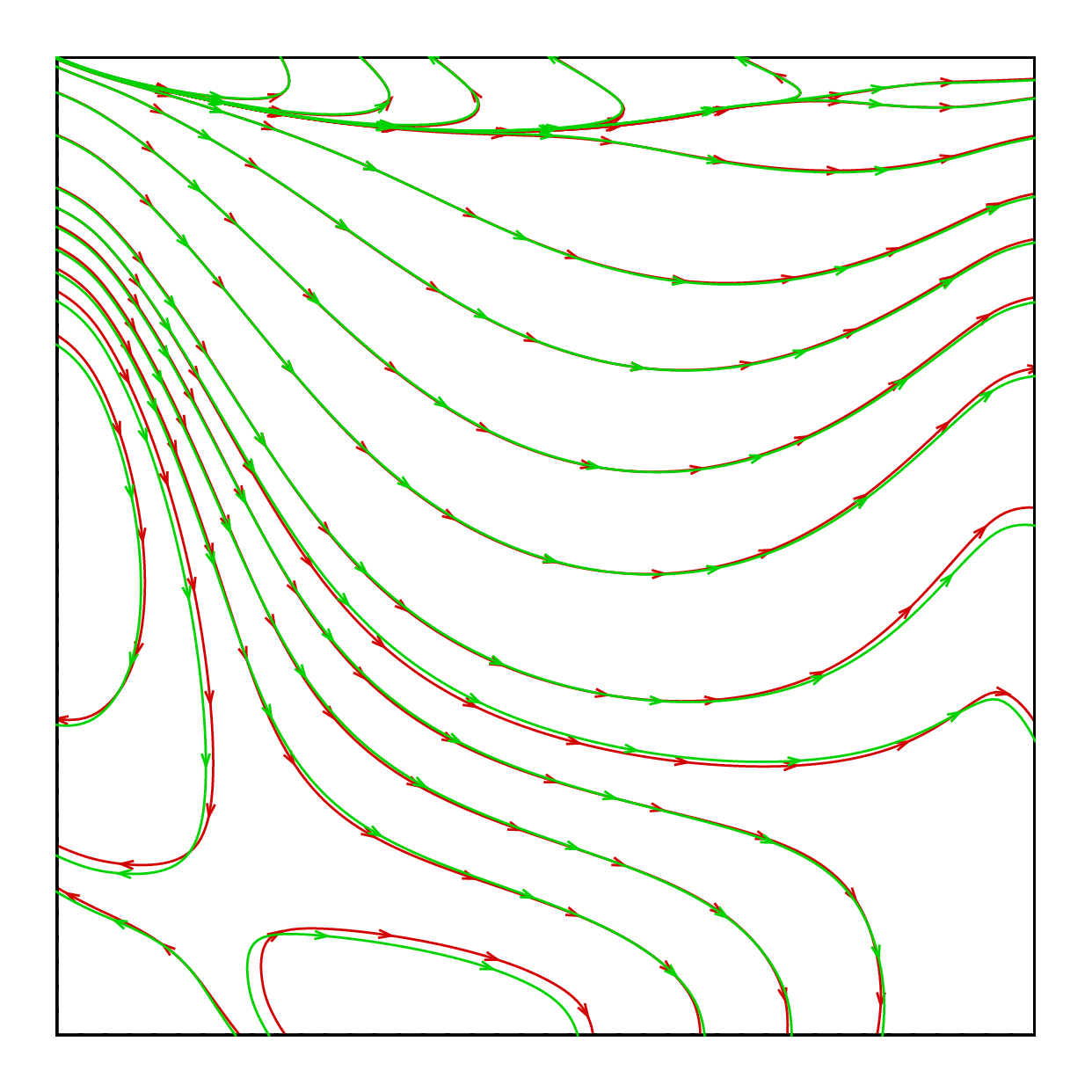}
        \caption{$\mathrm{Kn}=0.075$}
    \end{subfigure}

    \begin{subfigure}[b]{0.9\textwidth} \centering
        \includegraphics[width=0.33\textwidth]{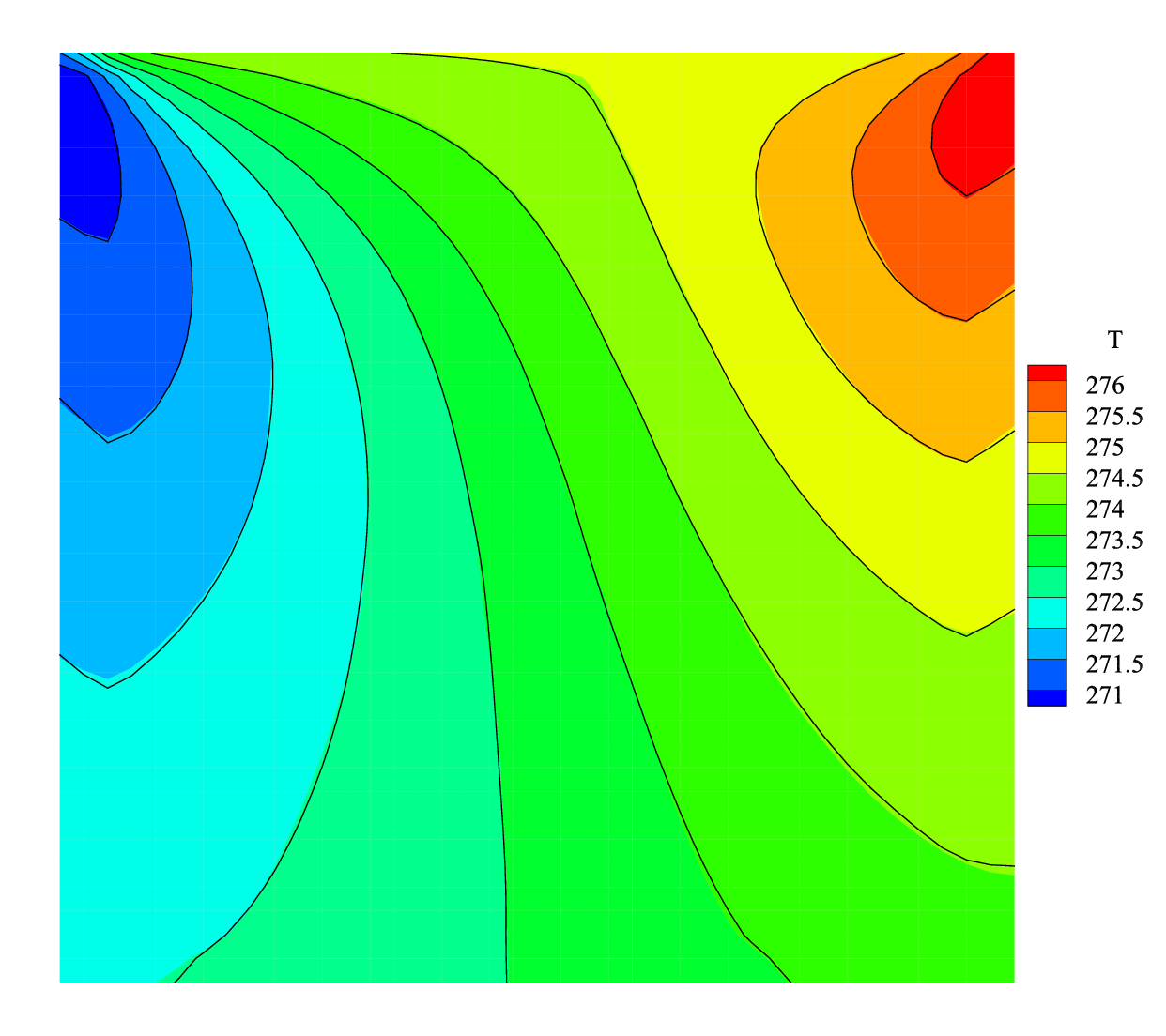}\hfill
        \includegraphics[width=0.30\textwidth]{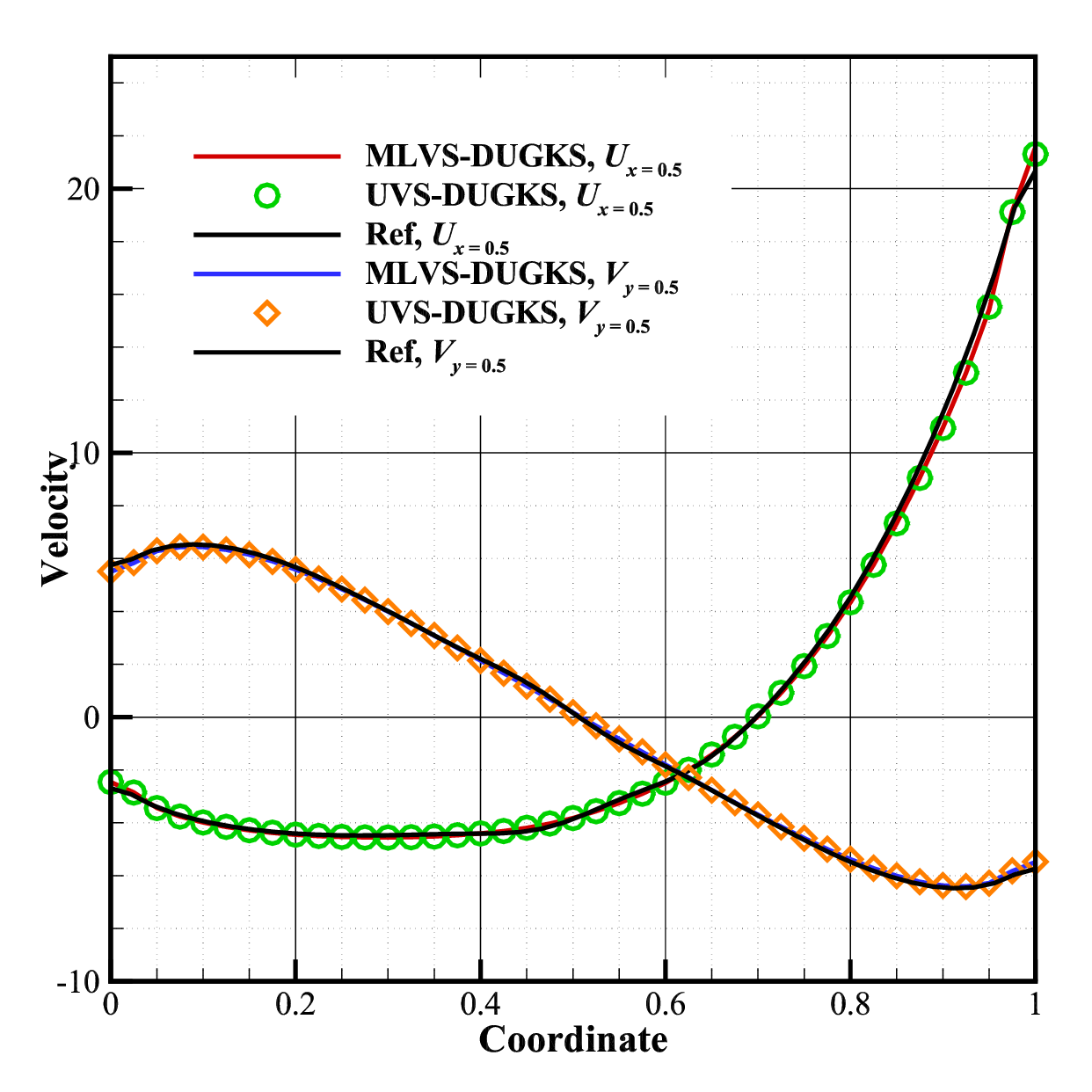}\hfill
        \includegraphics[width=0.30\textwidth]{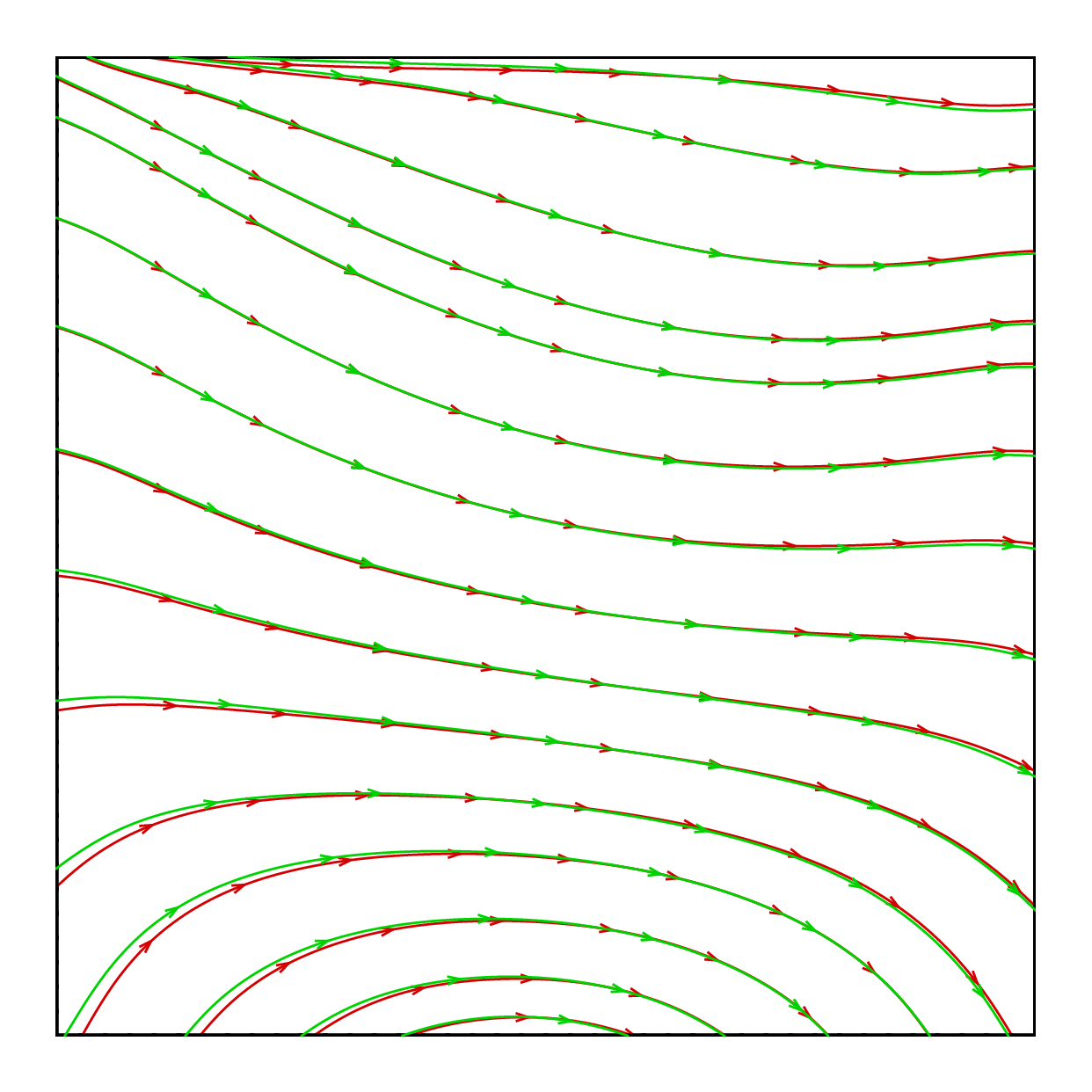}
        \caption{$\mathrm{Kn}=1.0$}
    \end{subfigure}

    \medskip
    \begin{subfigure}[b]{0.9\textwidth} \centering
        \includegraphics[width=0.33\textwidth]{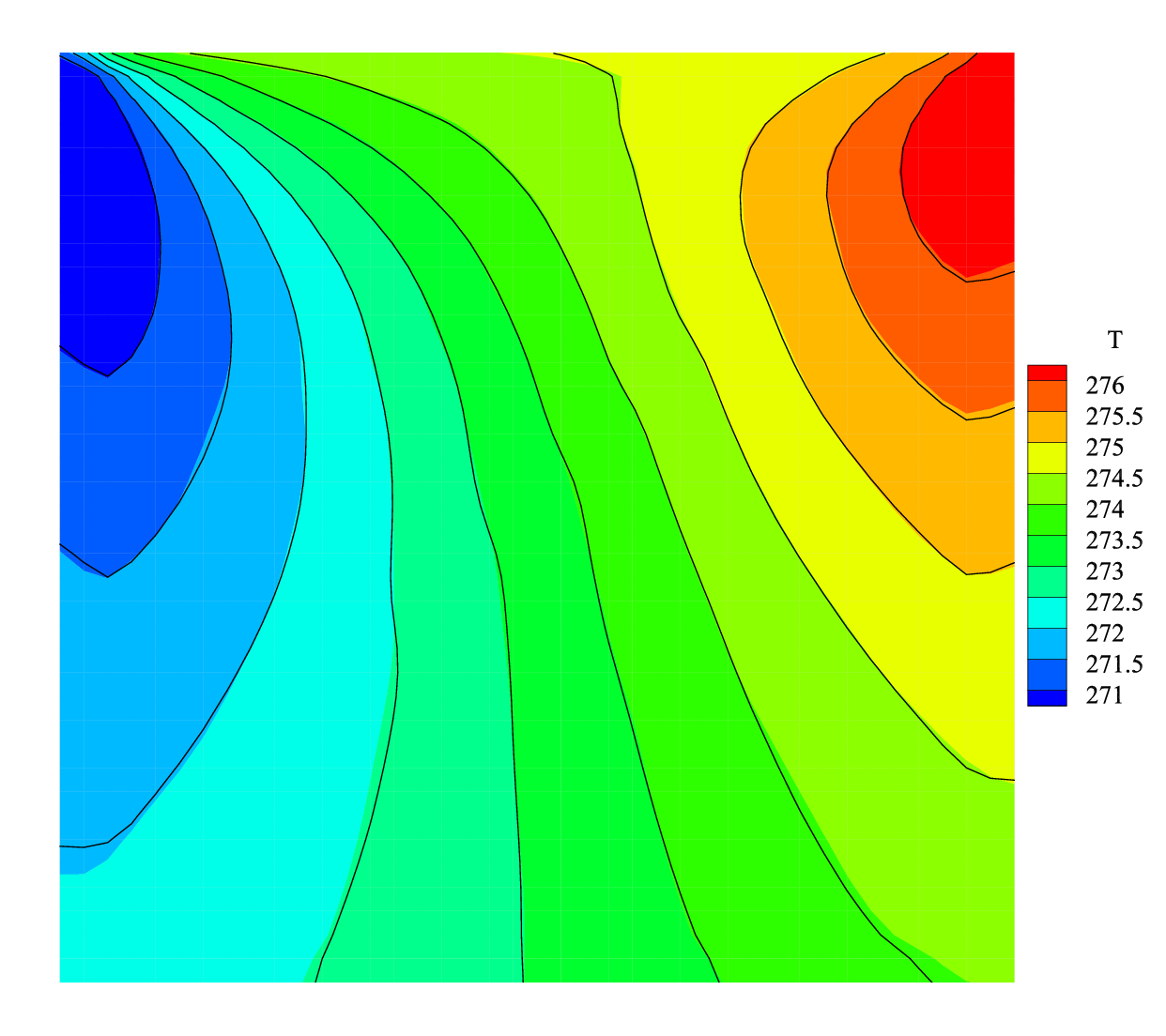}\hfill
        \includegraphics[width=0.30\textwidth]{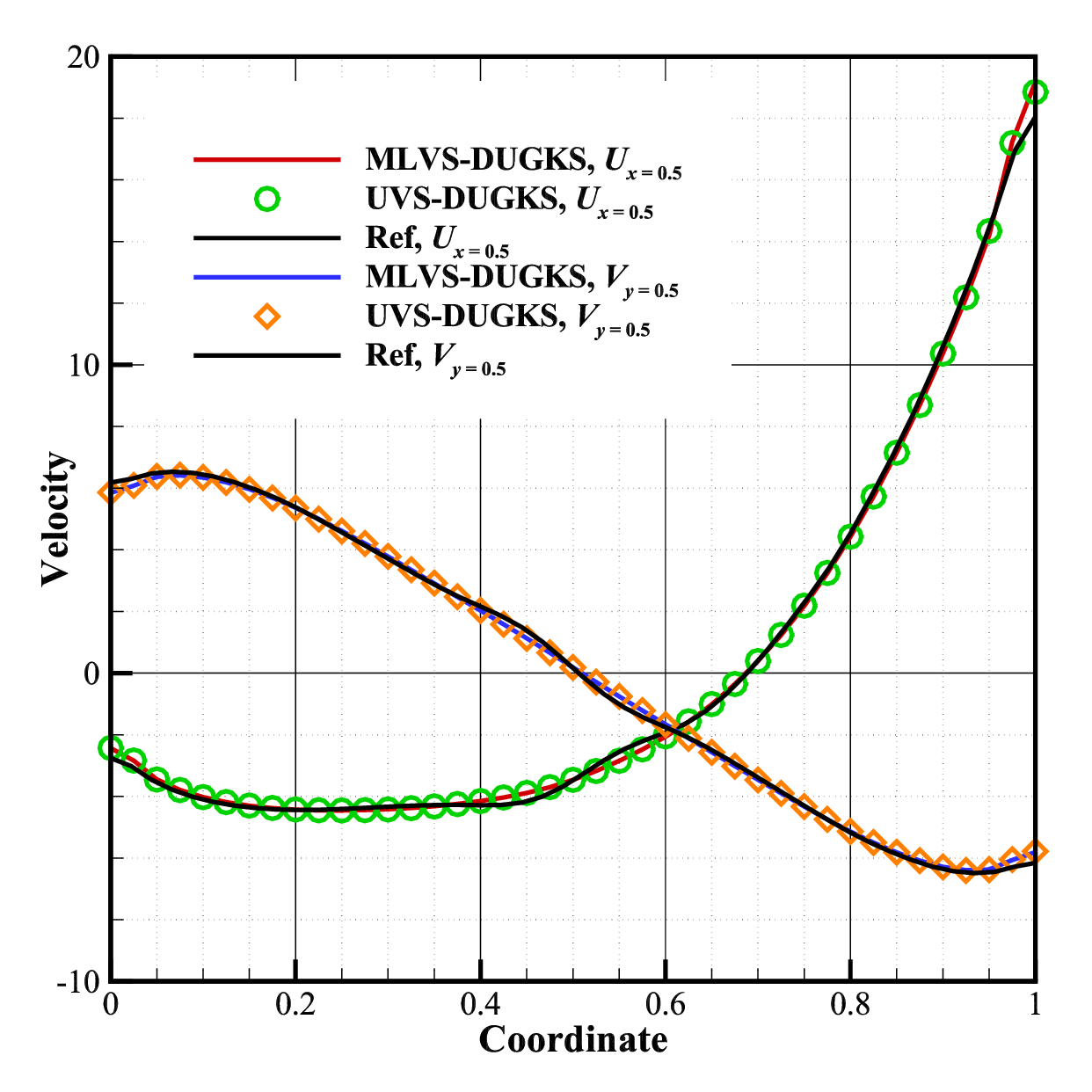}\hfill
        \includegraphics[width=0.30\textwidth]{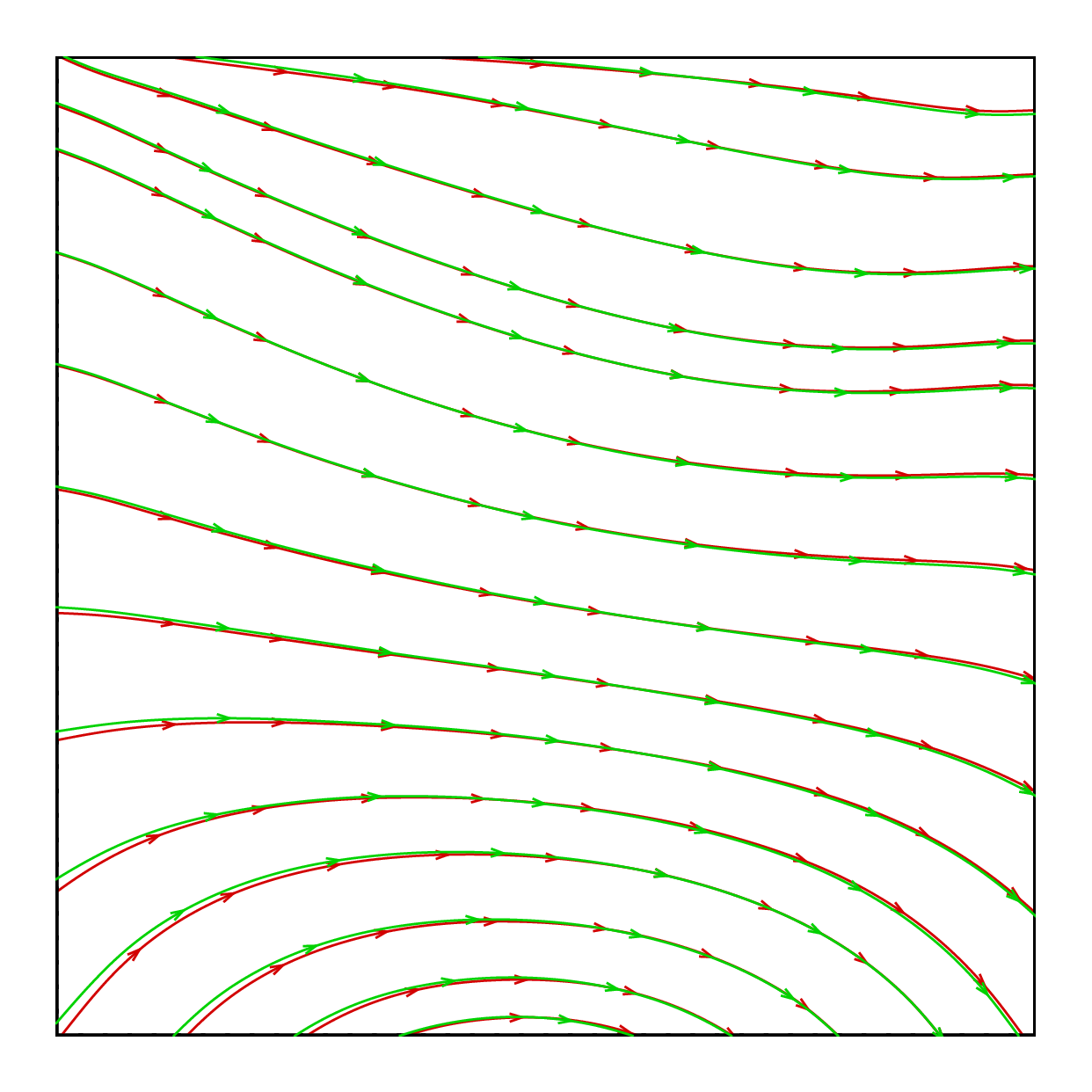}
        \caption{$\mathrm{Kn}=10$}
    \end{subfigure}
    \caption{Temperature contours (left), centerline velocities (middle), and heat-flux lines (right) for the lid-driven cavity flow. In the temperature panels, contours represent the MLVS-DUGKS solution and lines are the UVS-DUGKS solution. In the heat-flux panels, the red and green lines correspond to the MLVS and UVS-DUGKS results, respectively.}
    \label{fig:cavity_result}
\end{figure}

\begin{table}[htbp]
	\centering
	\caption{$L_2$ differences between the MLVS and UVS-DUGKS for the lid-driven cavity flow.}
	\label{tab:cavity_field_error}
	\renewcommand{\arraystretch}{1.15}
	\begin{tabular}{c c c c c}
		\toprule
		$\mathrm{Kn}$ & $\varepsilon_{2,\rho}^{\mathrm{rel}}$ & $\varepsilon_{2,\boldsymbol{u}}^{\mathrm{rel}}$ & $\varepsilon_{2,T}^{\mathrm{rel}}$ & $\varepsilon_{2,\boldsymbol{q}}^{\mathrm{rel}}$ \\
		\midrule
		0.075         & $6.75\times10^{-5}$                   & $6.91\times10^{-4}$                             & $4.42\times10^{-5}$                & $1.78\times10^{-2}$                             \\
		1.0           & $1.69\times10^{-4}$                   & $7.17\times10^{-3}$                             & $3.81\times10^{-5}$                & $1.96\times10^{-2}$                             \\
		10            & $2.87\times10^{-4}$                   & $1.03\times10^{-2}$                             & $7.46\times10^{-5}$                & $1.92\times10^{-2}$                             \\
		\bottomrule
	\end{tabular}
\end{table}

\begin{figure}[htbp]
    \centering
    \includegraphics[width=0.4\textwidth]{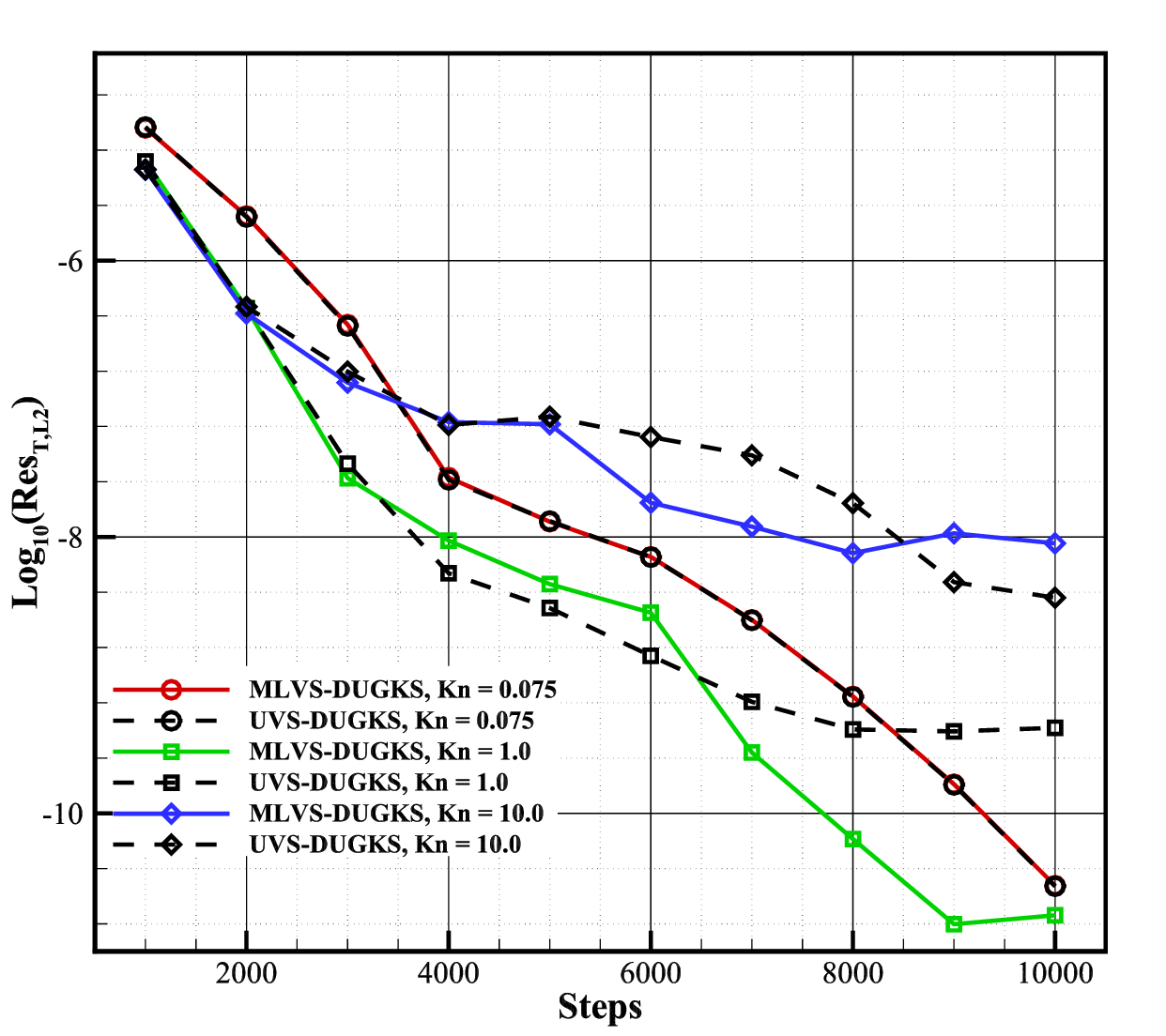}
    \caption{Comparison of the residual histories for the lid-driven cavity flow.}
    \label{fig:cavity_residual}
\end{figure}

Fig.~\ref{fig:cavity_result} presents the results across the investigated Knudsen number ranges. As shown, both methods agree closely in the temperature and heat-flux distributions, including the counter-gradient heat transport observed at $\mathrm{Kn}=0.075$. The centerline velocity profiles, compared with the UGKS and UVS-DUGKS references, likewise remain consistent across all three regimes. Table~\ref{tab:cavity_field_error} subsequently quantifies the differences between the MLVS-DUGKS and the UVS-DUGKS. The relative $L_2$ differences in the basic flow variables remain below $1.03\%$, while those in the heat flux range from $1.78\%$ to $1.96\%$. Meanwhile, the residual histories in Fig.~\ref{fig:cavity_residual} indicate that the MLVS-DUGKS and the UVS-DUGKS exhibit similar overall decay trends. At $\mathrm{Kn}=0.075$, the two curves nearly overlap throughout the recorded interval. At $\mathrm{Kn}=1.0$ and $10$, although their late-stage residual levels differ slightly, all residuals drop below $10^{-8}$ within 10,000 time steps. These results indicate that the MLVS-DUGKS retains convergence characteristics broadly similar to those of the UVS-DUGKS. These results are achieved at a substantially lower computational cost. As summarized in Table~\ref{tab:cavity_performance}, the MLVS-DUGKS achieves speedups of $6.83$--$6.91$ and reduces GPU-memory consumption by a factor of $6.70$. These ratios remain nearly invariant across the three regimes, indicating that the efficiency gain is insensitive to the Knudsen number.

\begin{table}[htbp]
	\centering
	\caption{Computational performance of the cavity calculations.}
	\label{tab:cavity_performance}
	\renewcommand{\arraystretch}{1.15}
	\small
	\begin{tabular}{l ccc ccc}
		\toprule
		& \multicolumn{3}{c}{Time (s)} & \multicolumn{3}{c}{GPU memory (MB)} \\
		\cmidrule(lr){2-4} \cmidrule(lr){5-7}
		$\mathrm{Kn}$ & UVS & MLVS & Speedup & UVS & MLVS & Ratio \\
		\midrule
		$0.075$ & 460.12 & 67.41 & 6.83 & 1068 & 159.31 & 6.70 \\
		$1.0$   & 438.07 & 63.44 & 6.91 & 1068 & 159.31 & 6.70 \\
		$10$    & 441.36 & 64.27 & 6.87 & 1068 & 159.31 & 6.70 \\
		\bottomrule
	\end{tabular}
\end{table}


\begin{figure}[htbp]
	\centering
	\begin{subfigure}[t]{0.32\textwidth}
		\centering
		\includegraphics[width=0.9\textwidth]{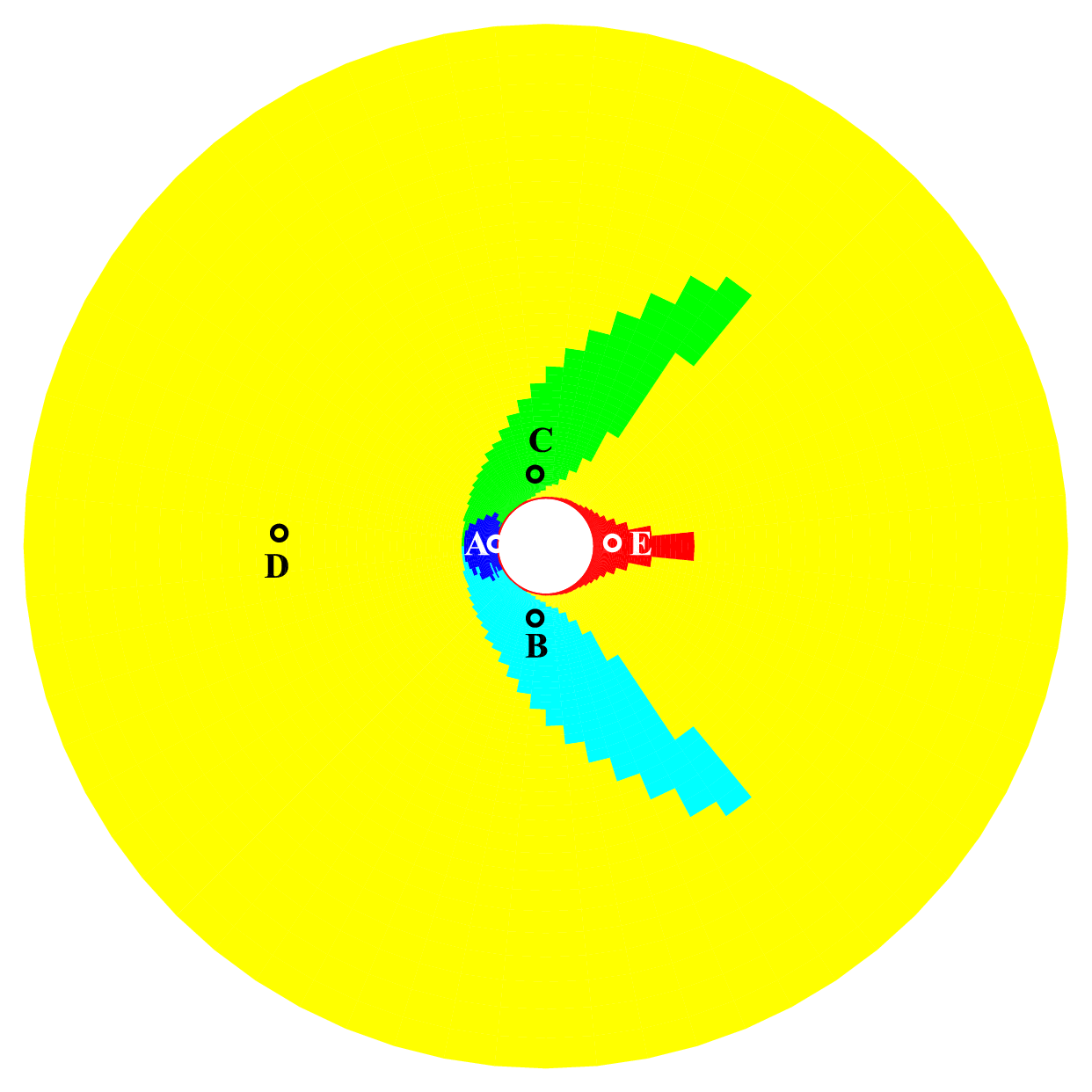}
		\caption{Velocity-space levels}
	\end{subfigure}
	\begin{subfigure}[t]{0.32\textwidth}
		\centering
		\includegraphics[width=0.9\textwidth]{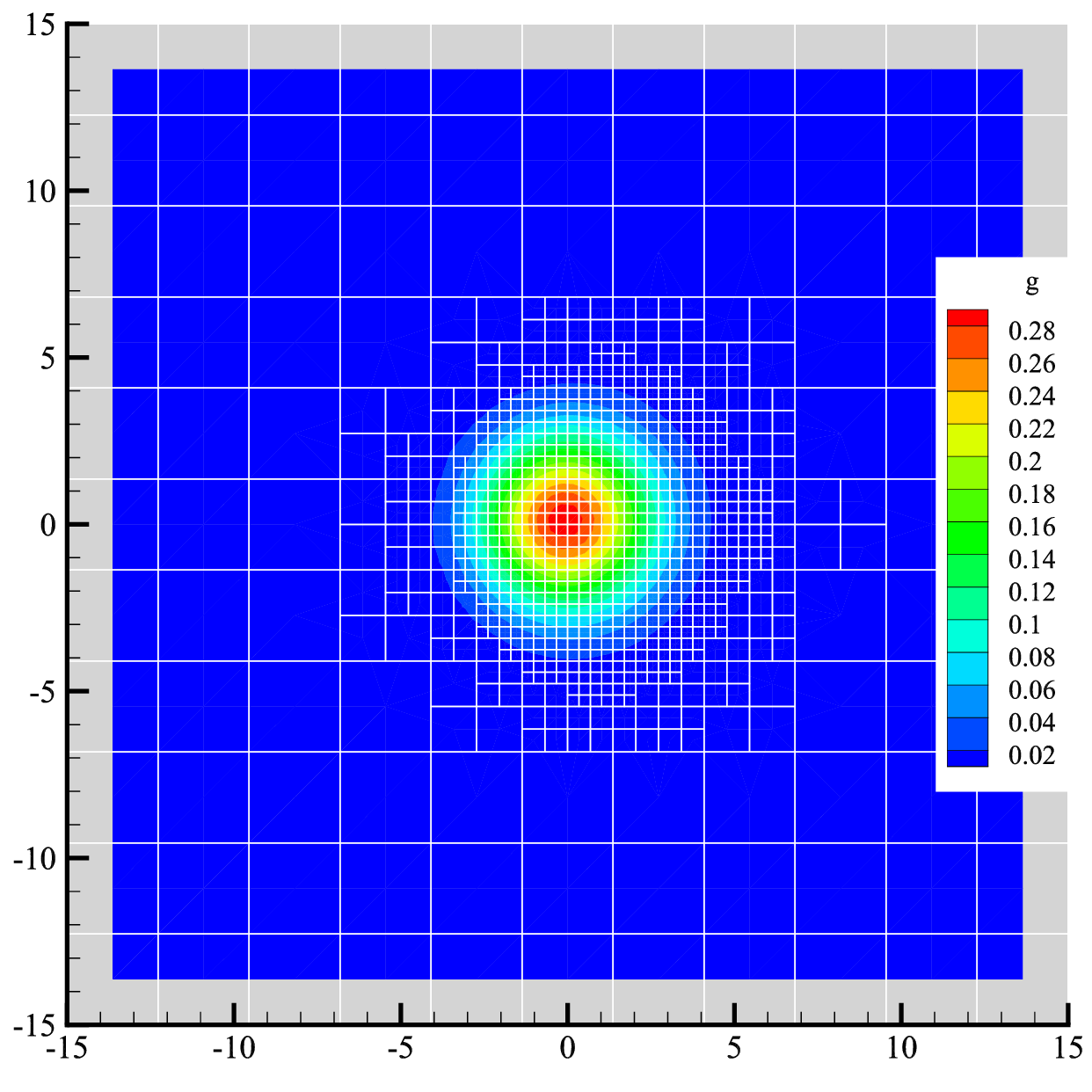}
		\caption{Post-shock region (point A)}
	\end{subfigure}
	\begin{subfigure}[t]{0.32\textwidth}
		\centering
		\includegraphics[width=0.9\textwidth]{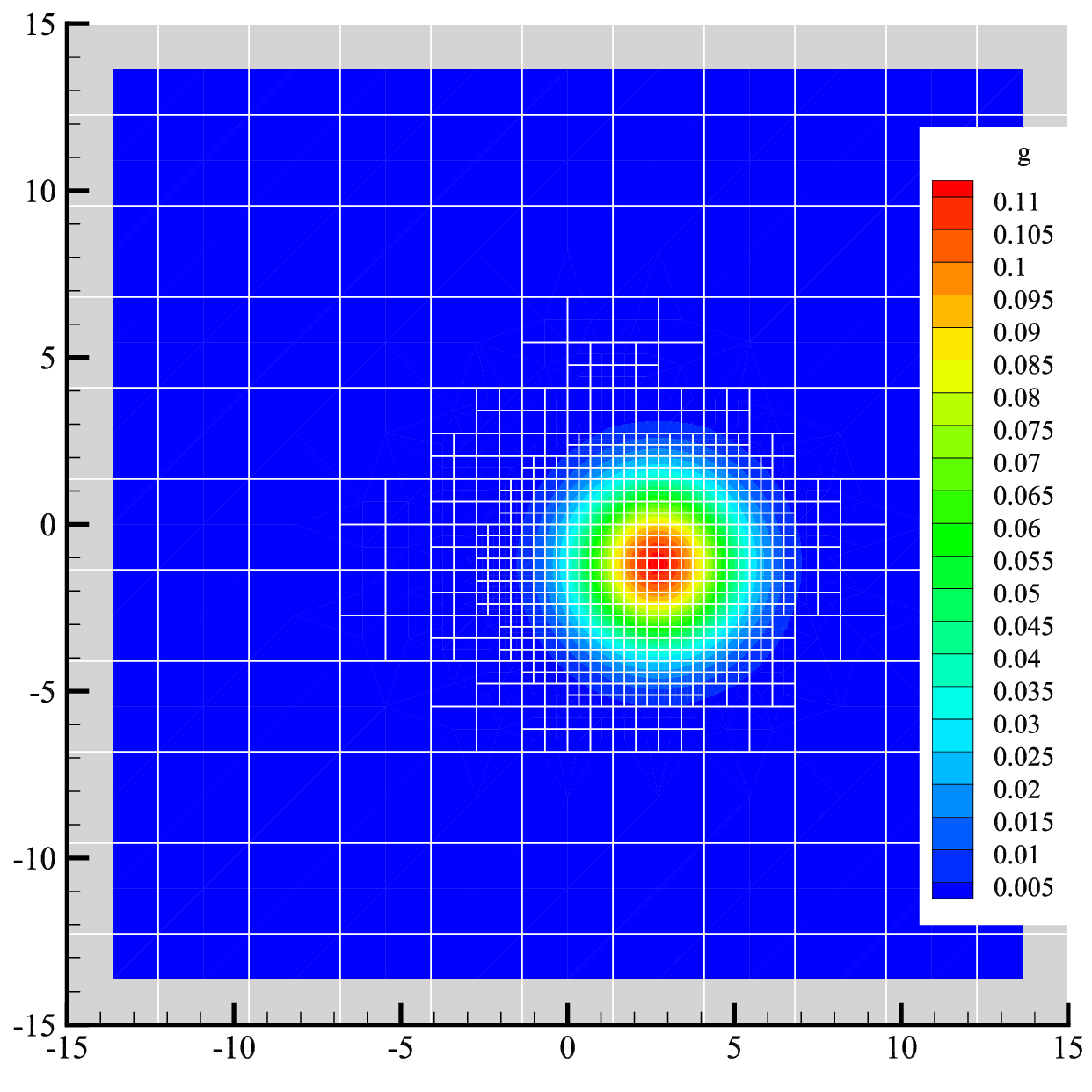}
		\caption{Lower side (point B)}
	\end{subfigure}

	\begin{subfigure}[t]{0.32\textwidth}
		\centering
		\includegraphics[width=0.9\textwidth]{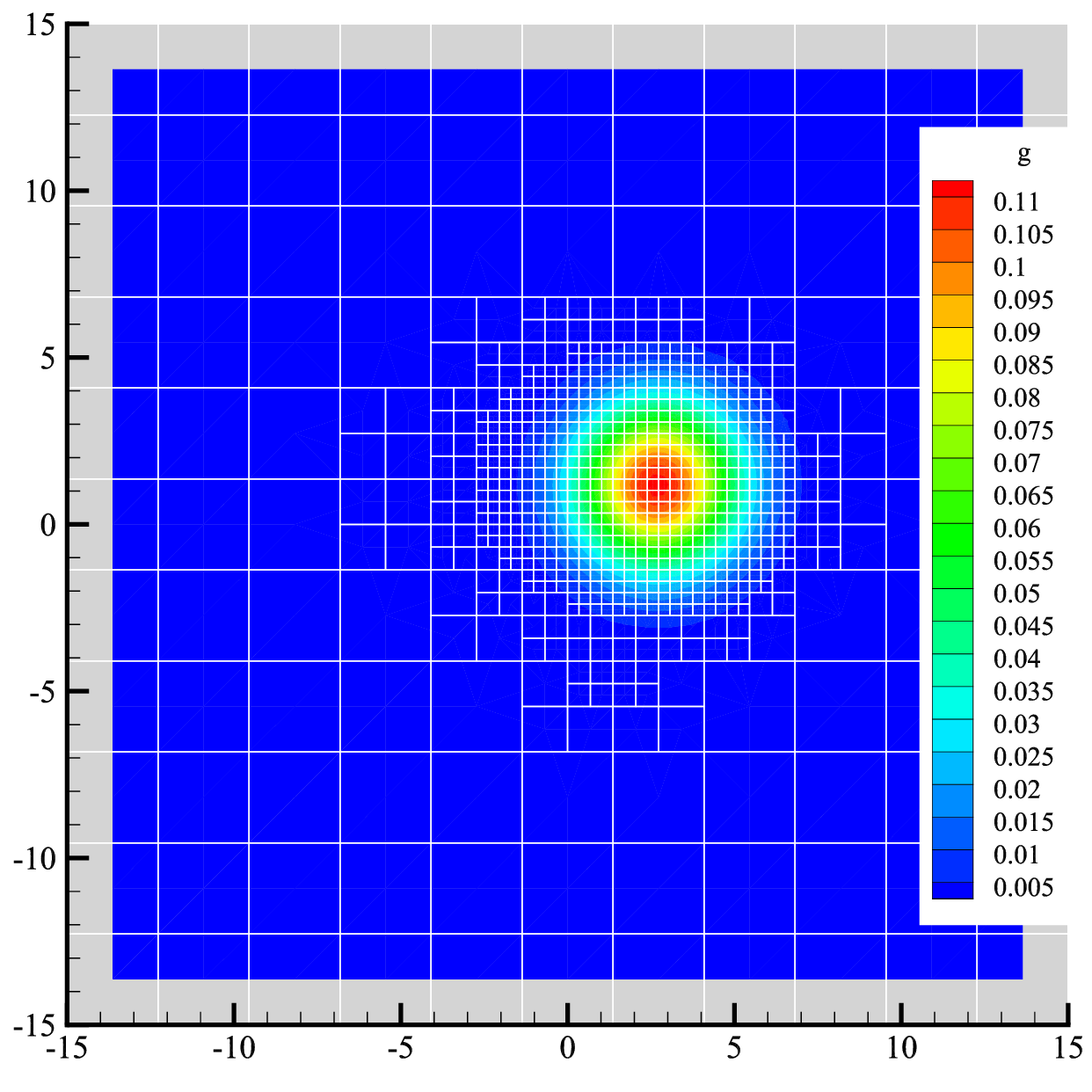}
		\caption{Upper side (point C)}
	\end{subfigure}
	\begin{subfigure}[t]{0.32\textwidth}
		\centering
		\includegraphics[width=0.9\textwidth]{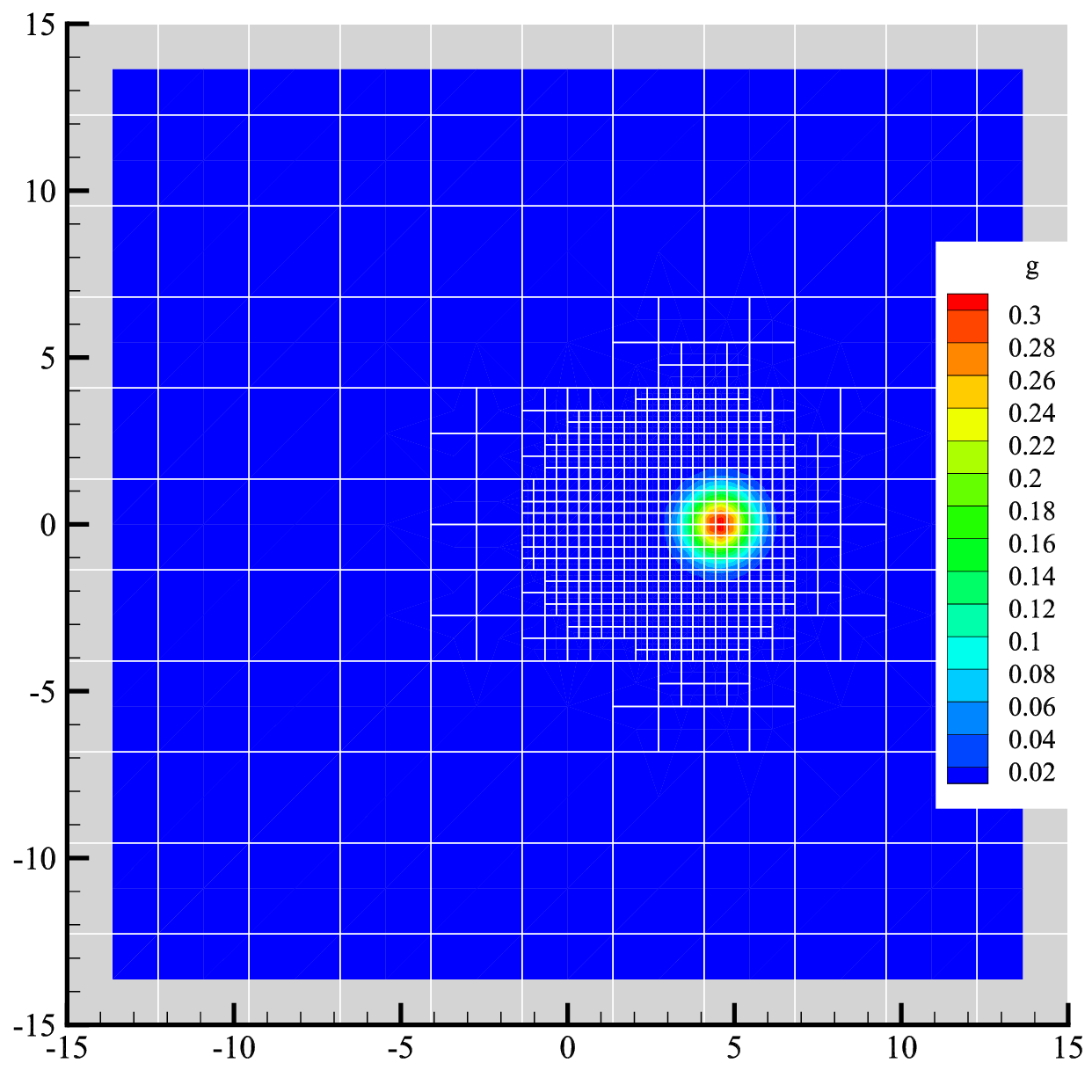}
		\caption{Freestream (point D)}
	\end{subfigure}
	\begin{subfigure}[t]{0.32\textwidth}
		\centering
		\includegraphics[width=0.9\textwidth]{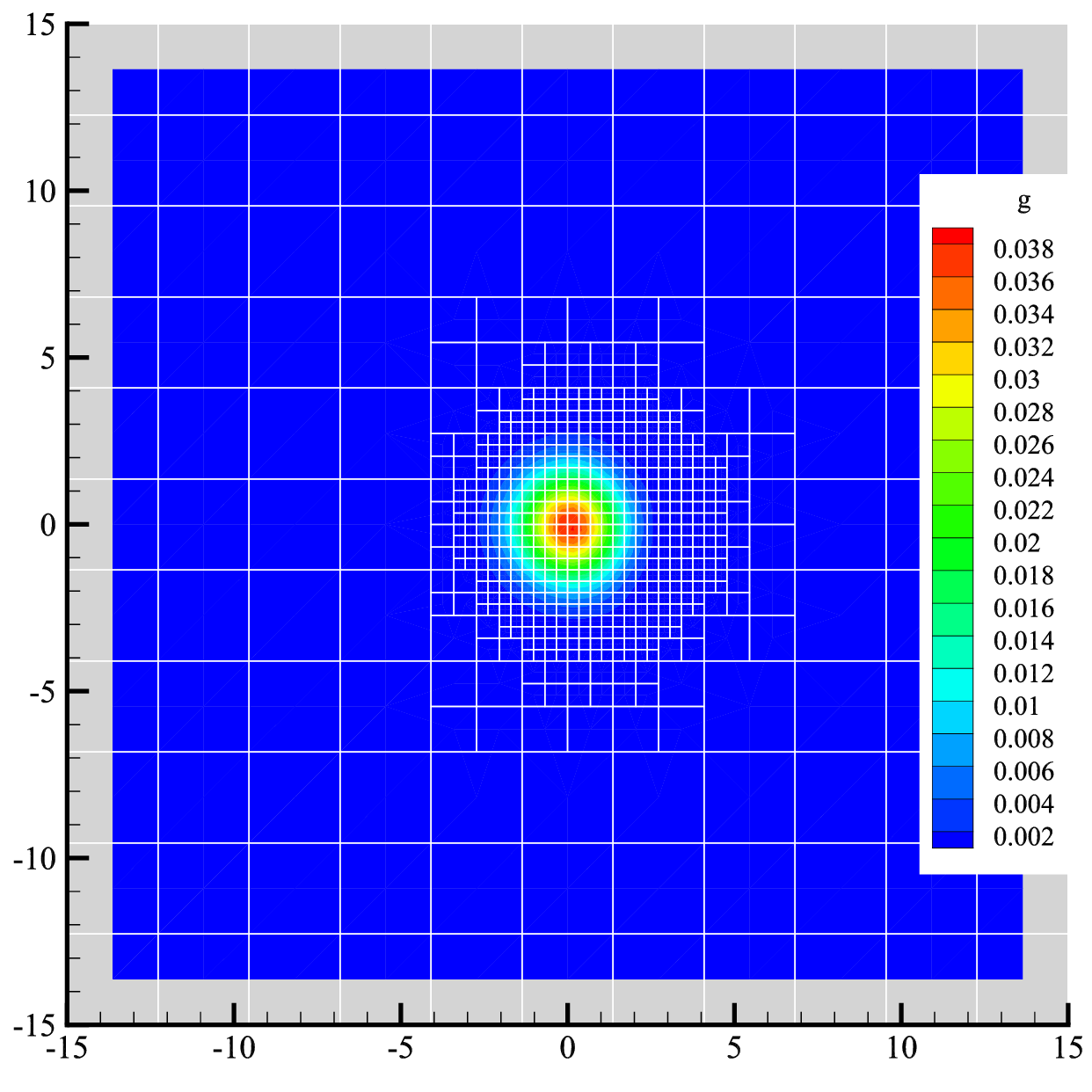}
		\caption{Behind the cylinder (point E)}
	\end{subfigure}

	\caption{Velocity-space level and distribution functions at $\mathrm{Kn}=0.01$.}
	\label{fig:cylinder_velocityspace_Kn0.01}
\end{figure}

\subsection{Mach-5 rarefied gas flow past a circular cylinder}
To evaluate the performance of the MLVS-DUGKS for high-speed rarefied flows, Mach 5 flow past a circular cylinder is simulated at $\mathrm{Kn}=0.01$, $0.1$, and $1.0$. The setup follows the configuration of Ref.~\cite{Zhu_Discrete_2016} with 5,120 structured cells. For reference, the UVS-DUGKS employs an $89\times89$ uniform velocity grid over $\left[-15\sqrt{2RT_\infty},\,15\sqrt{2RT_\infty}\right]^2$. The MLVS-DUGKS uses the same velocity-space range, where representative velocity spaces are updated every 100 time steps using $C_{\mathrm{split}}=0.01$, $C_{\mathrm{merge}}=0.005$, and a clustering threshold of $0.05$. The computations are advanced for $10^5$ time steps.

Figure~\ref{fig:cylinder_velocityspace_Kn0.01} presents the local distribution functions at the five sampling locations for $\mathrm{Kn}=0.01$; the results at $\mathrm{Kn}=0.1$ and $1.0$ exhibit the same qualitative features and are omitted for brevity. Five representative velocity spaces are generated, each adapting to the local distribution by retaining fine resolution where it is appreciable and remaining sparse in the low-contribution tails. The configurations at the three Knudsen numbers are summarized in Table~\ref{tab:cylinder_velocity_config}: as $\mathrm{Kn}$ increases, the spatial variation of the local distributions becomes more diffuse, allowing cells with similar velocity-space requirements to be grouped into fewer representative spaces. The cell-weighted average numbers of velocity points are approximately 715, 701, and 780 at $\mathrm{Kn}=0.01$, $0.1$, and $1.0$, respectively—about one tenth of the 7,921 points per physical cell used by the UVS-DUGKS.

\begin{figure}[htbp]
	\centering
	\begin{subfigure}[t]{0.3\textwidth}
		\centering
		\includegraphics[width=\textwidth]{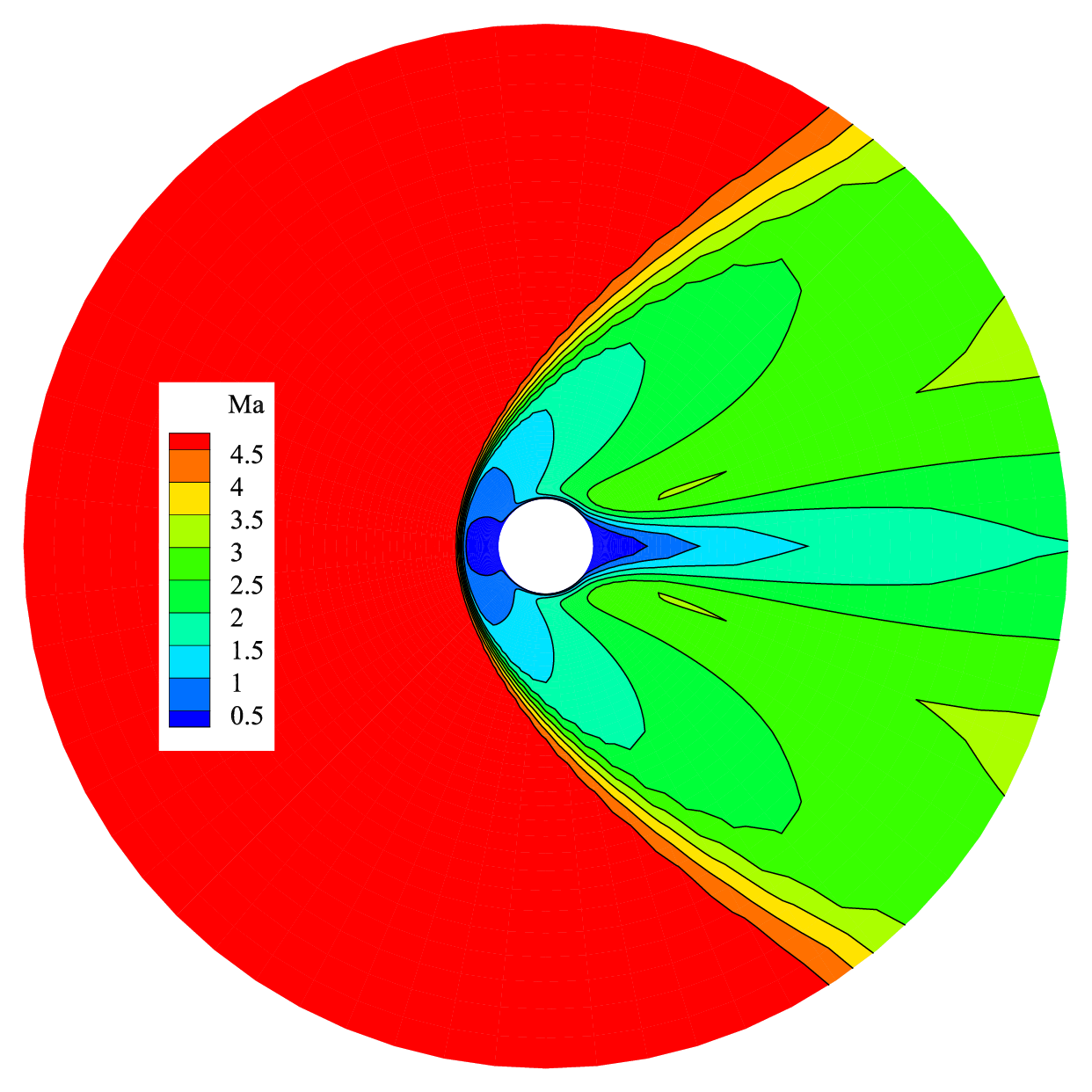}
		\caption{Mach number of $\mathrm{Kn} = 0.01$}
	\end{subfigure}
	\begin{subfigure}[t]{0.3\textwidth}
		\centering
		\includegraphics[width=\textwidth]{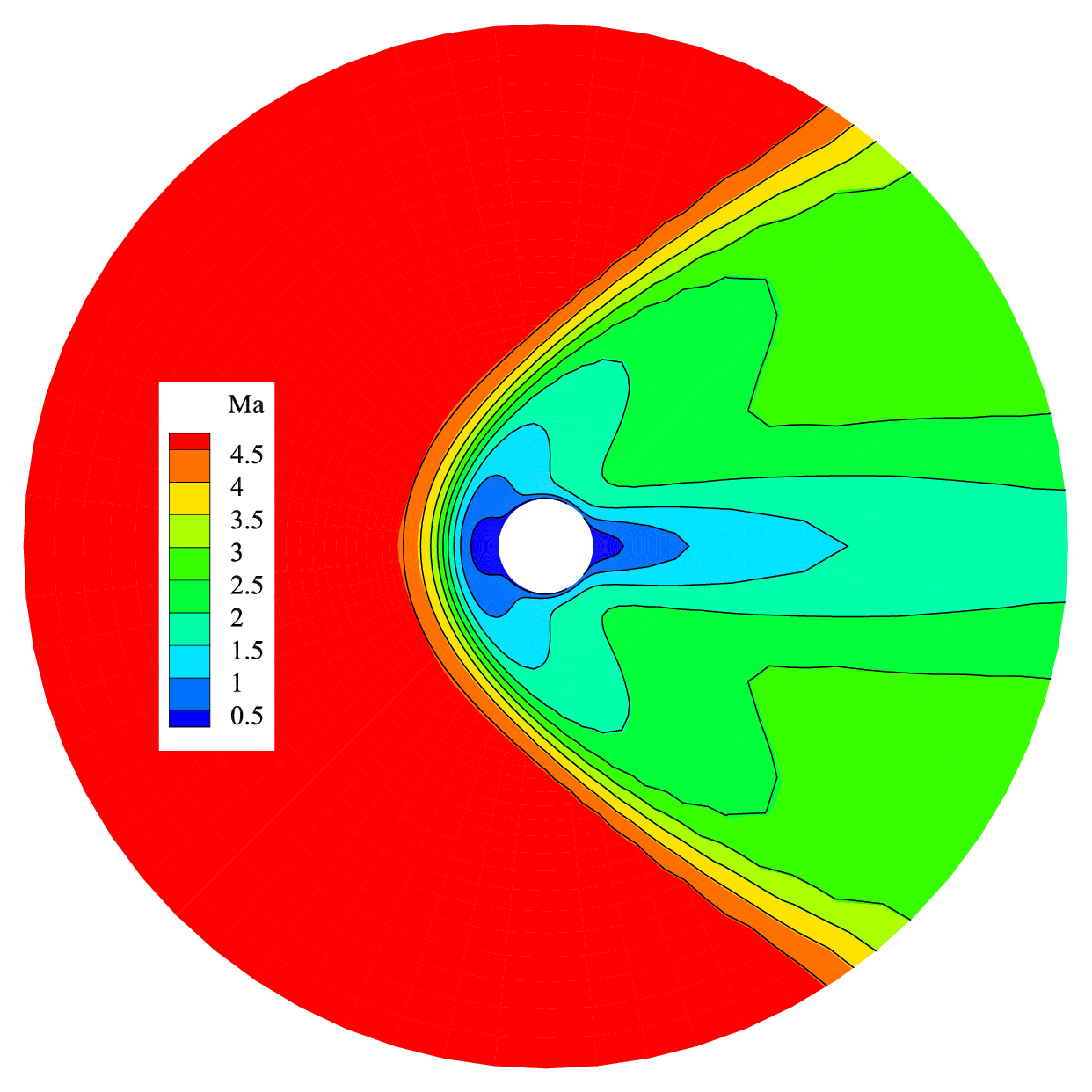}
		\caption{Mach number of $\mathrm{Kn} = 0.1$}
	\end{subfigure}
    \begin{subfigure}[t]{0.3\textwidth}
		\centering
		\includegraphics[width=\textwidth]{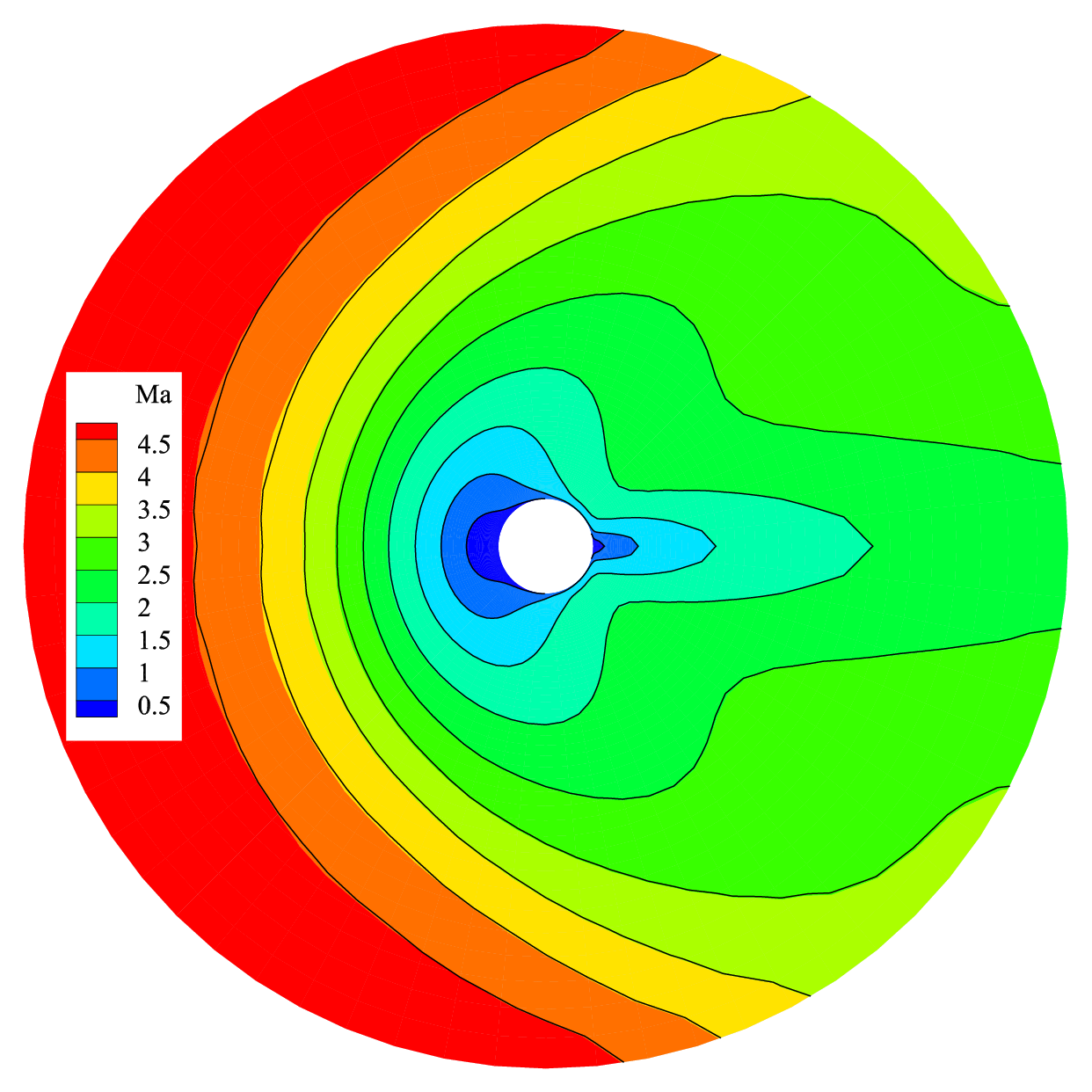}
		\caption{Mach number of $\mathrm{Kn} = 1$}
	\end{subfigure}

    \begin{subfigure}[t]{0.3\textwidth}
		\centering
		\includegraphics[width=\textwidth]{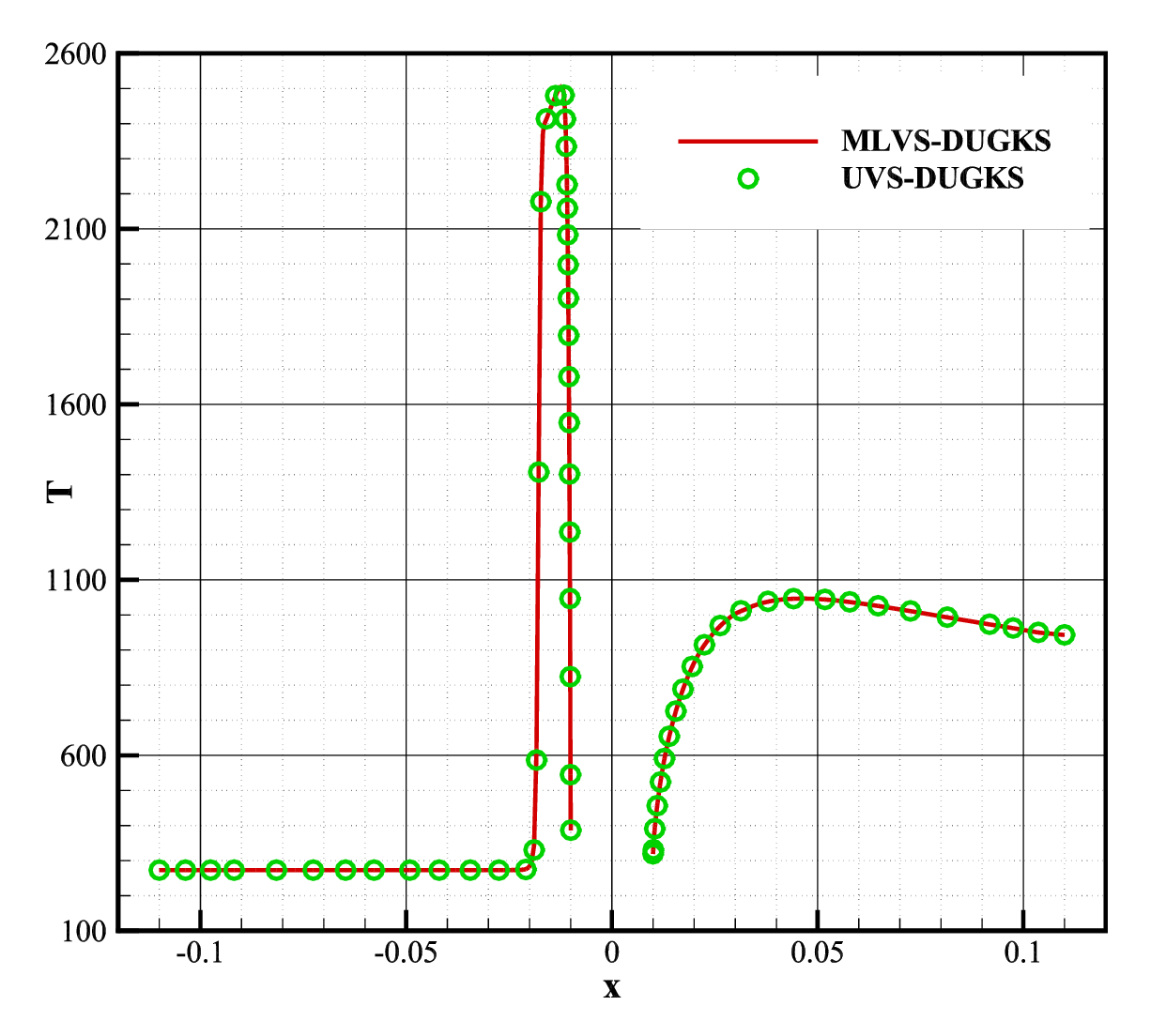}
		\caption{Temperature of $\mathrm{Kn} = 0.01$}
	\end{subfigure}
	\begin{subfigure}[t]{0.3\textwidth}
		\centering
		\includegraphics[width=\textwidth]{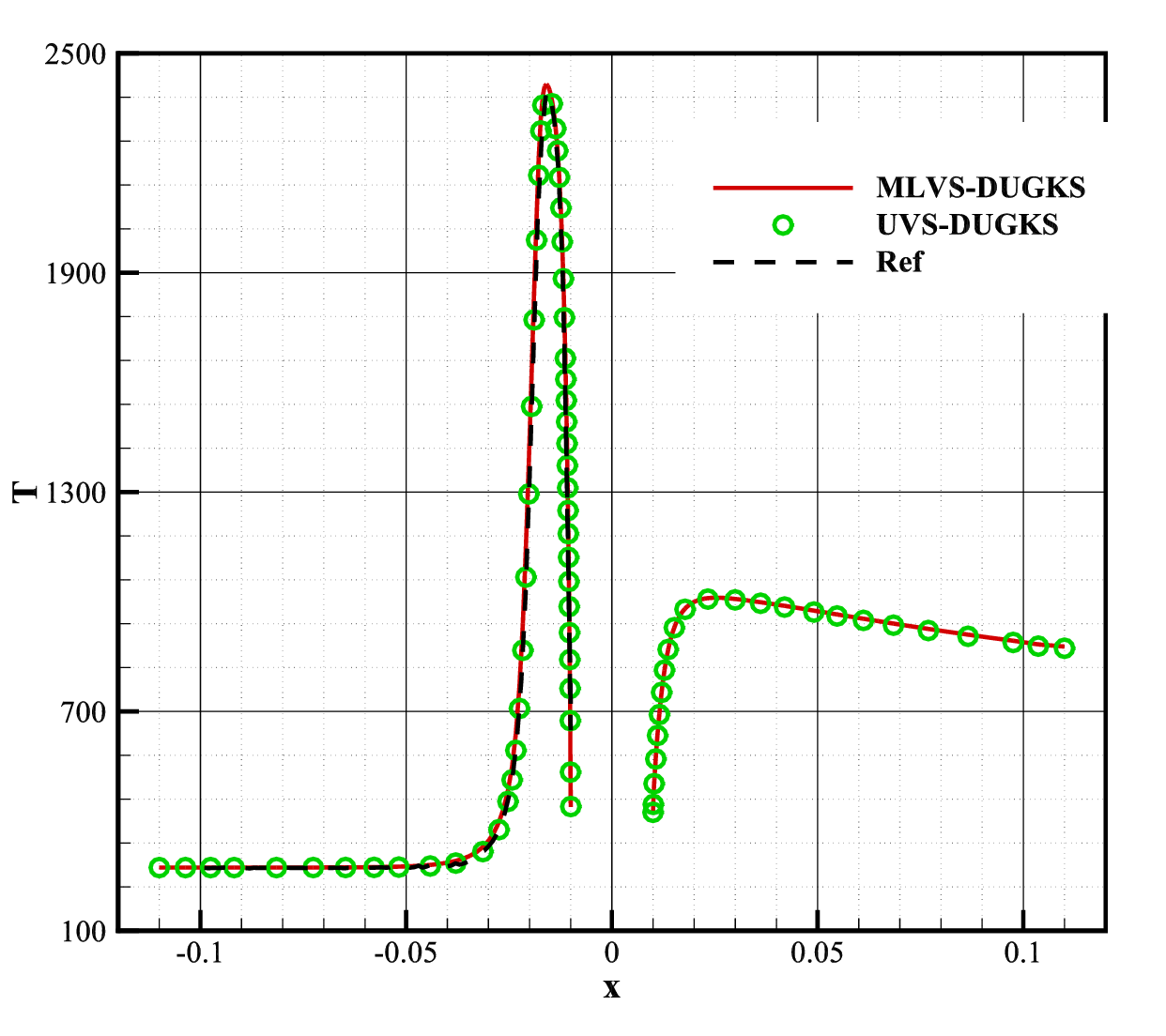}
		\caption{Temperature of $\mathrm{Kn} = 0.1$}
	\end{subfigure}
	\begin{subfigure}[t]{0.3\textwidth}
		\centering
		\includegraphics[width=\textwidth]{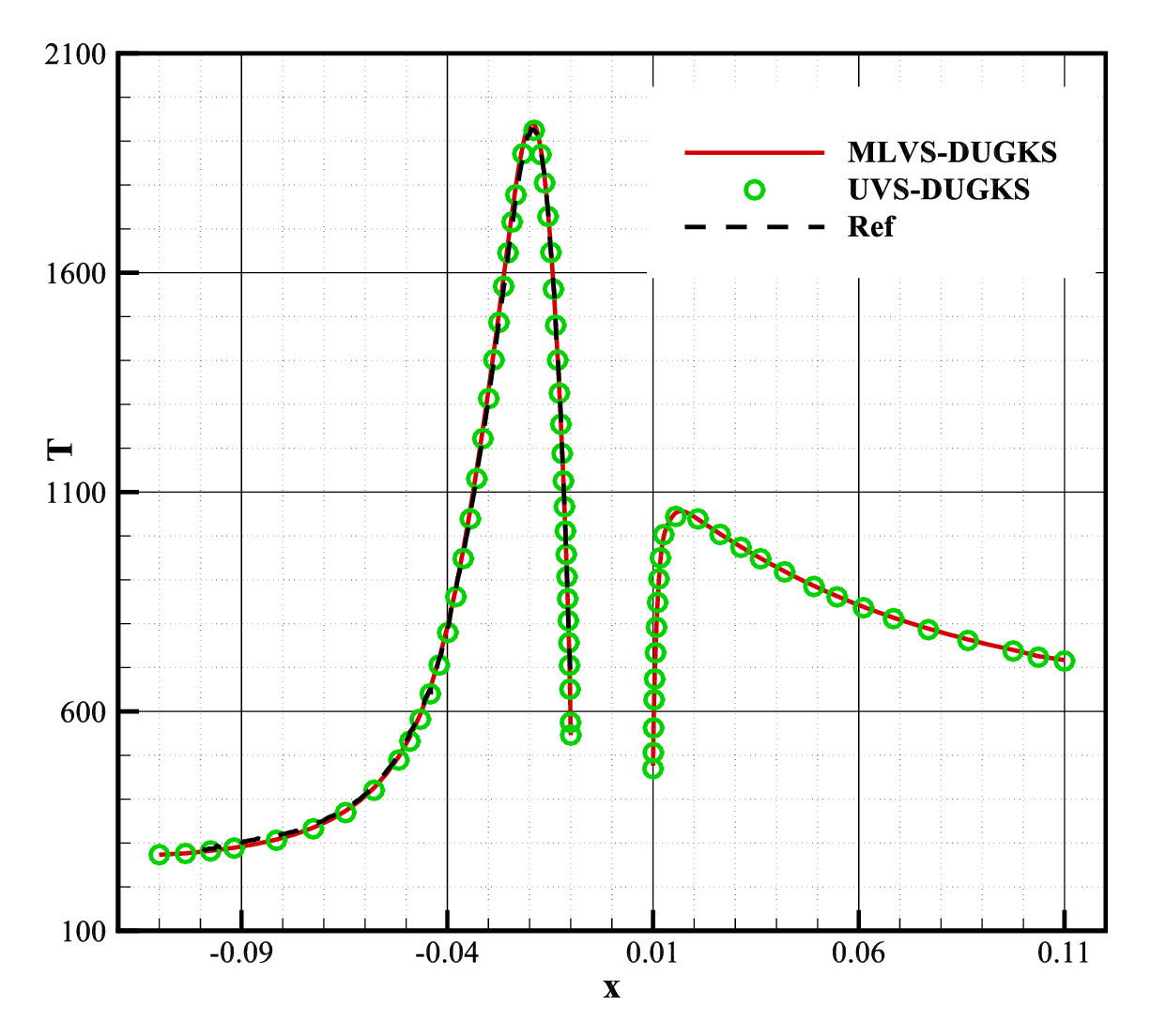}
		\caption{Temperature of $\mathrm{Kn} = 1$}
	\end{subfigure}
	\caption{Mach-number field and stagnation-line temperature for the Mach-5 cylinder flow. In the Mach-number field, the filled contour is computed by MLVS-DUGKS and the line is obtained by UVS-DUGKS.}
	\label{fig:cylinder_result}
\end{figure}

\begin{table}[htbp]
  \centering
  \small
  \caption{Representative velocity-space configurations for the Mach-5 cylinder flow. Each MLVS entry gives $N_\xi/N_{\mathrm{cell}}$, the number of velocity points and the number of physical cells assigned to it.}
  \label{tab:cylinder_velocity_config}
  \begin{tabular}{lccccccc}
    \toprule
    $\mathrm{Kn}$ & $\mathcal{V}_1$ & $\mathcal{V}_2$ & $\mathcal{V}_3$ & $\mathcal{V}_4$ & $\mathcal{V}_5$ & $N_{\mathrm{avg}}$ & Compression \\
    \midrule
    $0.01$ & 898\,/\,330    & 796\,/\,644    & 796\,/\,644    & 670\,/\,2,538 & 661\,/\,964    & 714.7 & 11.08 \\
    $0.1$  & 853\,/\,380    & 724\,/\,1,315 & 721\,/\,546   & 682\,/\,473    & 664\,/\,2,406 & 701.2 & 11.30 \\
    $1.0$  & 844\,/\,1{,}824 & 763\,/\,865   & 739\,/\,2,431 & --            & --             & 780.5 & 10.15 \\
    \bottomrule
  \end{tabular}
\end{table}

\begin{figure}[htbp]
    \centering
    \includegraphics[width=0.4\textwidth]{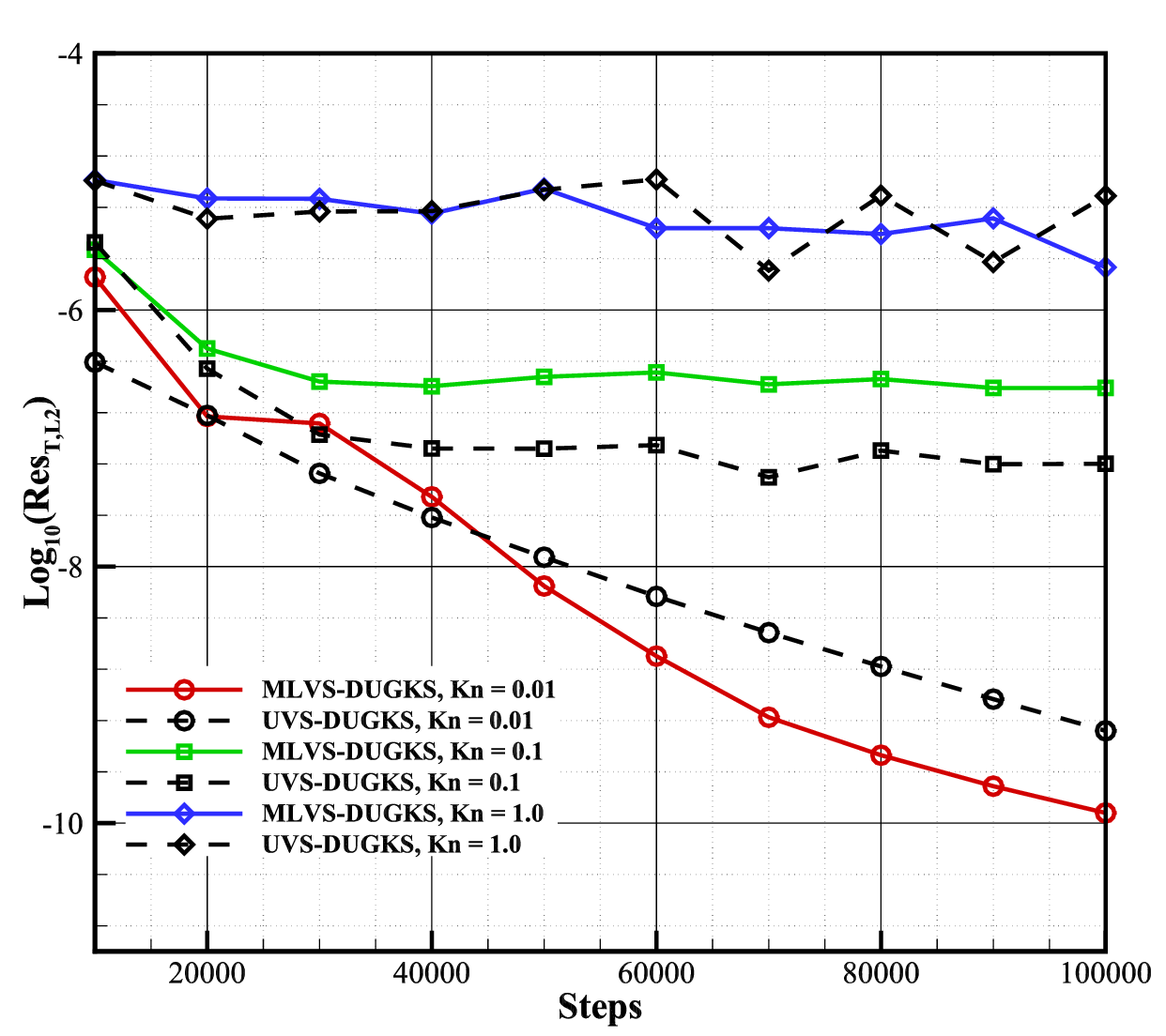}
    \caption{Comparison of the residual histories for Mach-5 flow past a circular cylinder.}
    \label{fig:cylinder_residual}
\end{figure}

Fig.~\ref{fig:cylinder_result} compares the Mach-number fields and stagnation-line temperature profiles obtained from the MLVS-DUGKS and the UVS-DUGKS. For $\mathrm{Kn}=0.1$ and $1.0$, the stagnation-line temperature profiles are also compared with the reference data reported by Zhu et al.~\cite{Zhu_Discrete_2016}. As shown, the MLVS-DUGKS and UVS-DUGKS solutions remain close for all three Knudsen numbers, and the stagnation-line temperature profiles at $\mathrm{Kn}=0.1$ and $1.0$ also agree well with the reference data. Quantitative differences between the MLVS-DUGKS and the UVS-DUGKS are listed in Table~\ref{tab:cylinder_field_error}. The relative errors in density, temperature, Mach number, and velocity remain below $1.22\%$ for all three Knudsen numbers. Heat-transfer accuracy is further assessed in the region of strongest compression by evaluating the normal heat flux near the stagnation point. It remains below $0.42\%$ over the entire Knudsen-number range and is $0.211\%$ at $\mathrm{Kn}=0.01$. These results demonstrate that the MLVS-DUGKS reproduces the principal flow fields and the stagnation-region heat transfer of the UVS-DUGKS with small discrepancies. Fig.~\ref{fig:cylinder_residual} compares the relative $L_2$ temperature-update residual histories of the two methods. At $\mathrm{Kn}=0.01$, both methods show sustained residual decay, reaching values below $10^{-9}$ by $10^5$ time steps. At $\mathrm{Kn}=0.1$, both histories approach residual plateaus after an initial decrease, and at $\mathrm{Kn}=1.0$, both display nonmonotonic variations. Although the absolute residual levels differ, the overall evolution patterns of the two methods agree at all three Knudsen numbers, indicating that the adaptive velocity space preserves the qualitative convergence behavior of the UVS-DUGKS.

\begin{table}[htbp]
	\centering
	\caption{$L_2$ differences in the macroscopic fields of the Mach-5 cylinder flow.}
	\label{tab:cylinder_field_error}
	\small
	\begin{tabular}{c c c c c c}
		\toprule
		$\mathrm{Kn}$ & $\varepsilon_{2,\rho}^{\mathrm{rel}}$ & $\varepsilon_{2,T}^{\mathrm{rel}}$ & $\varepsilon_{2,Ma}^{\mathrm{rel}}$ & $\varepsilon_{2,\boldsymbol{u}}^{\mathrm{rel}}$ & $\varepsilon_{q_{n,\mathrm{stag}}}^{\mathrm{rel}}$ \\
		\midrule
		0.01          & $1.36\times10^{-3}$                   & $2.30\times10^{-3}$                & $1.04\times10^{-3}$                 & $5.28\times10^{-4}$                             & $2.11\times10^{-3}$                                \\
		0.1           & $7.40\times10^{-3}$                   & $1.22\times10^{-2}$                & $5.08\times10^{-3}$                 & $2.44\times10^{-3}$                             & $4.18\times10^{-3}$                                \\
		1.0           & $3.13\times10^{-3}$                   & $8.12\times10^{-3}$                & $4.70\times10^{-3}$                 & $1.40\times10^{-3}$                             & $3.75\times10^{-3}$                                \\
		\bottomrule
	\end{tabular}
\end{table}

\begin{table}[htbp]
	\centering
	\caption{Computational performance of the Mach-5 cylinder flow}
	\label{tab:cylinder_performance}
	\small
	\begin{tabular}{l ccc ccc}
		\toprule
		& \multicolumn{3}{c}{Time (s)} & \multicolumn{3}{c}{GPU memory (MB)} \\
		\cmidrule(lr){2-4} \cmidrule(lr){5-7}
		$\mathrm{Kn}$ & UVS & MLVS & Speedup & UVS & MLVS & Ratio \\
		\midrule
		$0.01$ & 9786.18 & 1325.48 & 7.38 & 2528.44 & 337.44 & 7.49 \\
		$0.1$  & 9917.56 & 1250.76 & 7.93 & 2528.44 & 339.44 & 7.45 \\
		$1.0$  & 9917.29 & 1352.09 & 7.33 & 2528.44 & 375.44 & 6.73 \\
		\bottomrule
	\end{tabular}
\end{table}

The velocity-space compression yields significant computational savings. As shown in Table~\ref{tab:cylinder_performance}, the MLVS-DUGKS speeds up the simulation by a factor of $7.33$--$7.93$ and cuts GPU memory usage by $6.73$--$7.49$ times. However, these numbers fall short of the roughly tenfold reduction in average velocity points. This stems from two issues. First, the adaptive scheme introduces additional overheads (e.g., velocity-space updates, inter-level transfers, and conservative moment corrections). Second, the current GPU implementation focuses on validating the MLVS-DUGKS framework rather than hardware-level optimization. Future tuning of data layouts, memory access, kernel fusion, synchronization, and thread-block configurations can help close this gap. Nevertheless, these results demonstrate that the MLVS-DUGKS achieves substantial speedups without compromising accuracy.

\FloatBarrier

\subsection{Multiscale flow expansion between two connected cavities}
To evaluate the MLVS-DUGKS for transient multiscale flows, gas expansion between two connected cavities~\cite{Zhu_Discrete_2016} is considered. The geometry consists of two square cavities with side length $L=1\,\mathrm{m}$ connected by a channel of length $1.0\,\mathrm{m}$ and width $0.125\,\mathrm{m}$. The complete domain is discretized using 12,012 unstructured cells. Initially, the gas is stationary with the gas temperature and wall temperature set to $273\,\mathrm{K}$. The initial pressure states on the two sides of $x=0$ are
\begin{equation}
	p_L=6.436\,\mathrm{Pa},
	\quad
	p_R=10^{-3}p_L
	=6.436\times10^{-3}\,\mathrm{Pa}.
\end{equation}
Using the cavity side length as the reference length, the corresponding Knudsen numbers are $\mathrm{Kn}_L=0.001$ and $\mathrm{Kn}_R=1.0$. The UVS-DUGKS employs a $101\times101$ uniform velocity grid over $\left[-7\sqrt{2RT_w},7\sqrt{2RT_w}\right]^2$. The MLVS-DUGKS uses the same velocity-space range and constructs representative velocity spaces, which are updated every 10 steps using $C_{\mathrm{split}}=0.01$, $C_{\mathrm{merge}}=0.005$, $\varepsilon_{\mathrm{merge}}=2.0\times10^{-5}$, and a clustering threshold of $0.05$.

\begin{figure}[htbp]
	\centering
	\begin{subfigure}[t]{0.32\textwidth}
		\centering
		\includegraphics[width=\textwidth]{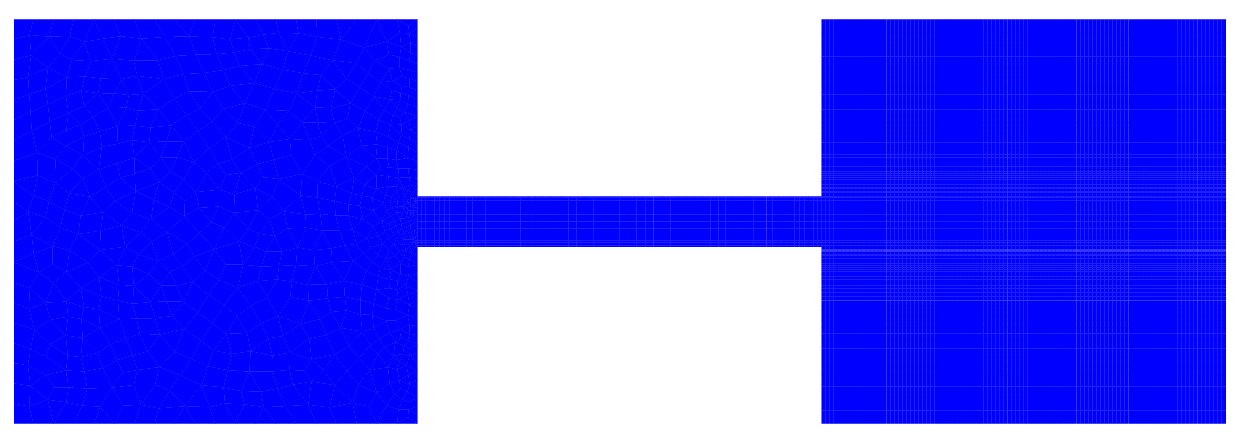}
		\caption{$t = 0.1$}
	\end{subfigure}
	\begin{subfigure}[t]{0.32\textwidth}
		\centering
		\includegraphics[width=\textwidth]{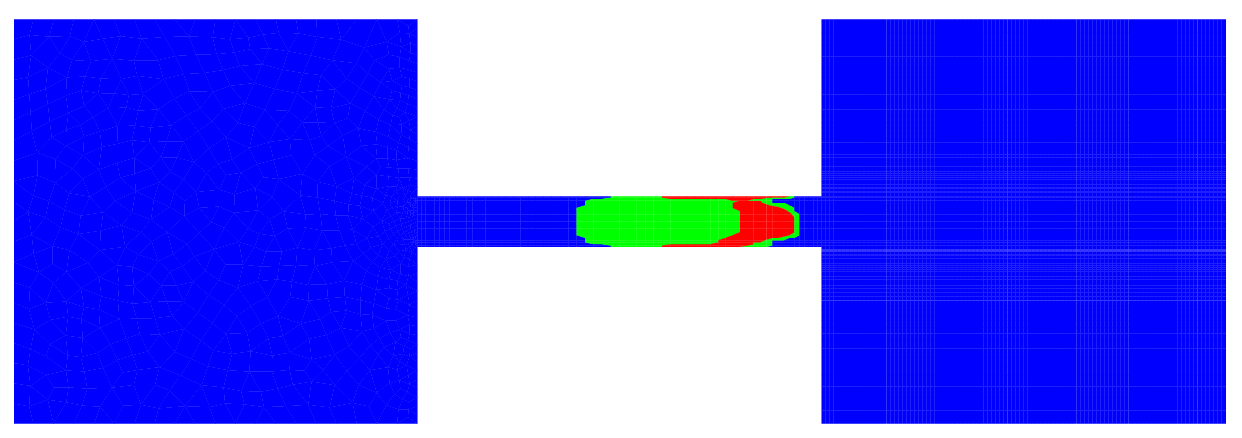}
		\caption{$t = 0.2$}
	\end{subfigure}
	\begin{subfigure}[t]{0.32\textwidth}
		\centering
		\includegraphics[width=\textwidth]{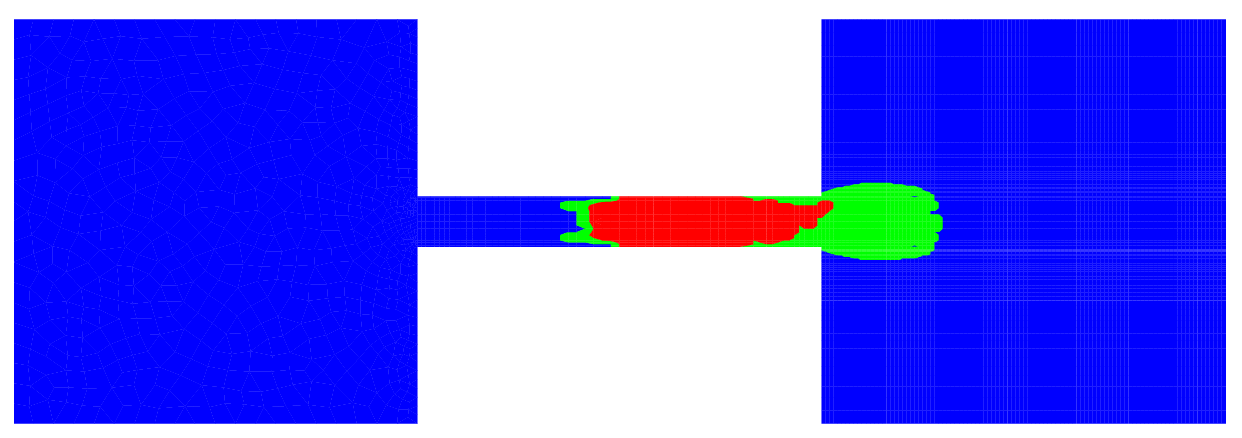}
		\caption{$t = 0.3$}
	\end{subfigure}

	\begin{subfigure}[t]{0.32\textwidth}
		\centering
		\includegraphics[width=\textwidth]{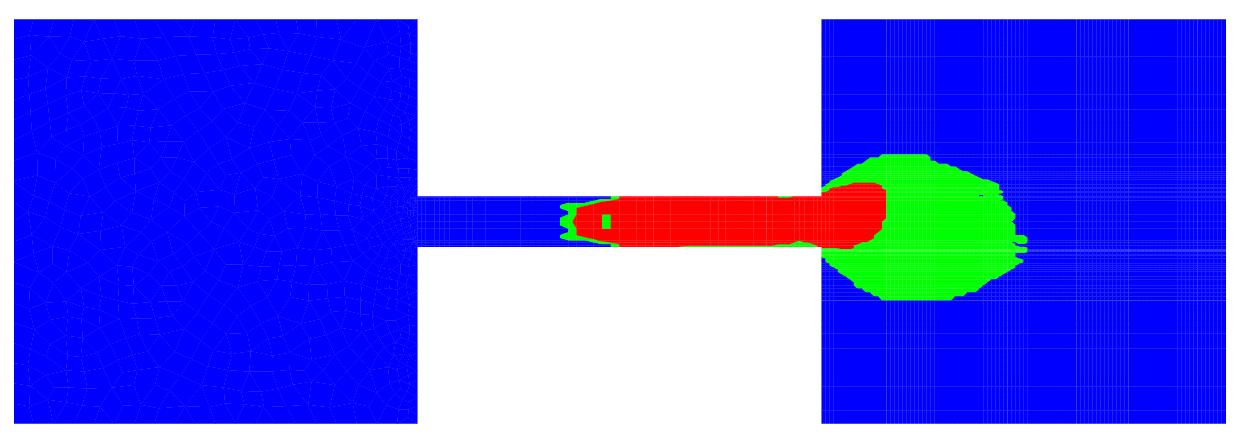}
		\caption{$t = 0.4$}
	\end{subfigure}
	\begin{subfigure}[t]{0.32\textwidth}
		\centering
		\includegraphics[width=\textwidth]{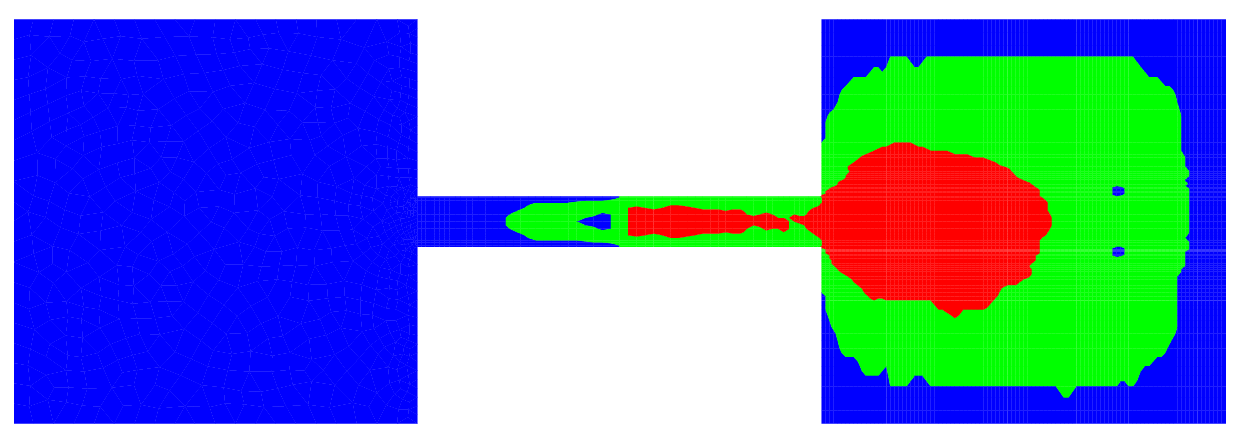}
		\caption{$t = 2.0$}
	\end{subfigure}
	\begin{subfigure}[t]{0.32\textwidth}
		\centering
		\includegraphics[width=\textwidth]{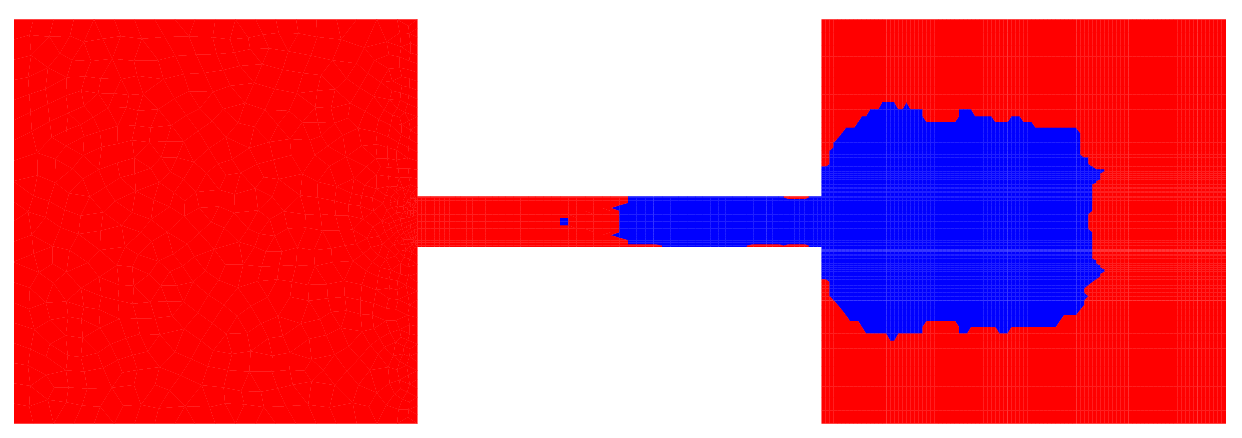}
		\caption{$t = 4.0$}
	\end{subfigure}

	\caption{Temporal evolution of the representative velocity-spaces.}
	\label{fig:twocavity_tag_evolution}
\end{figure}

\begin{figure}[htbp]
	\centering
	\begin{subfigure}[t]{0.5\textwidth}
		\centering
		\includegraphics[width=\textwidth]{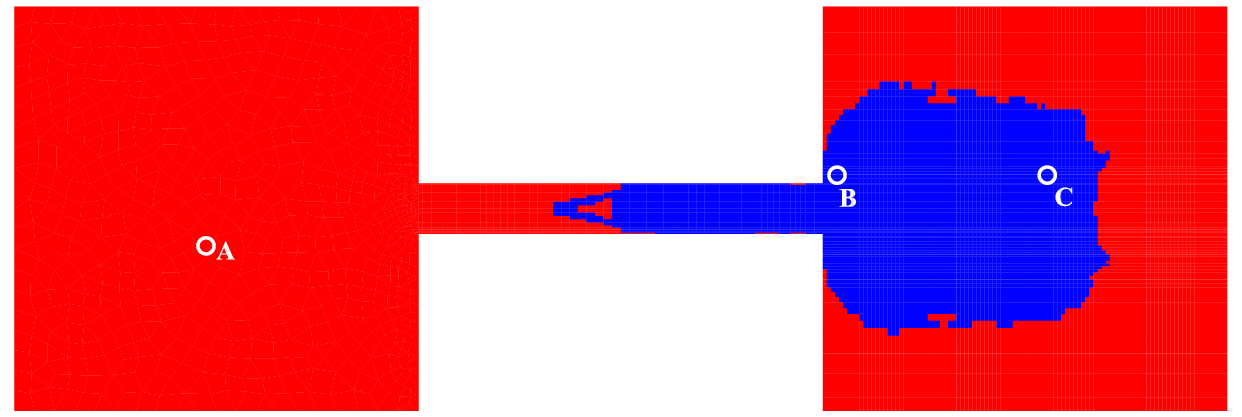}
		\caption{Sampled cell locations in the two-cavity configuration.}
		\label{fig:twocavity_sampling_points}
	\end{subfigure}

	\begin{subfigure}[t]{0.3\textwidth}
		\centering
		\includegraphics[width=0.9\textwidth]{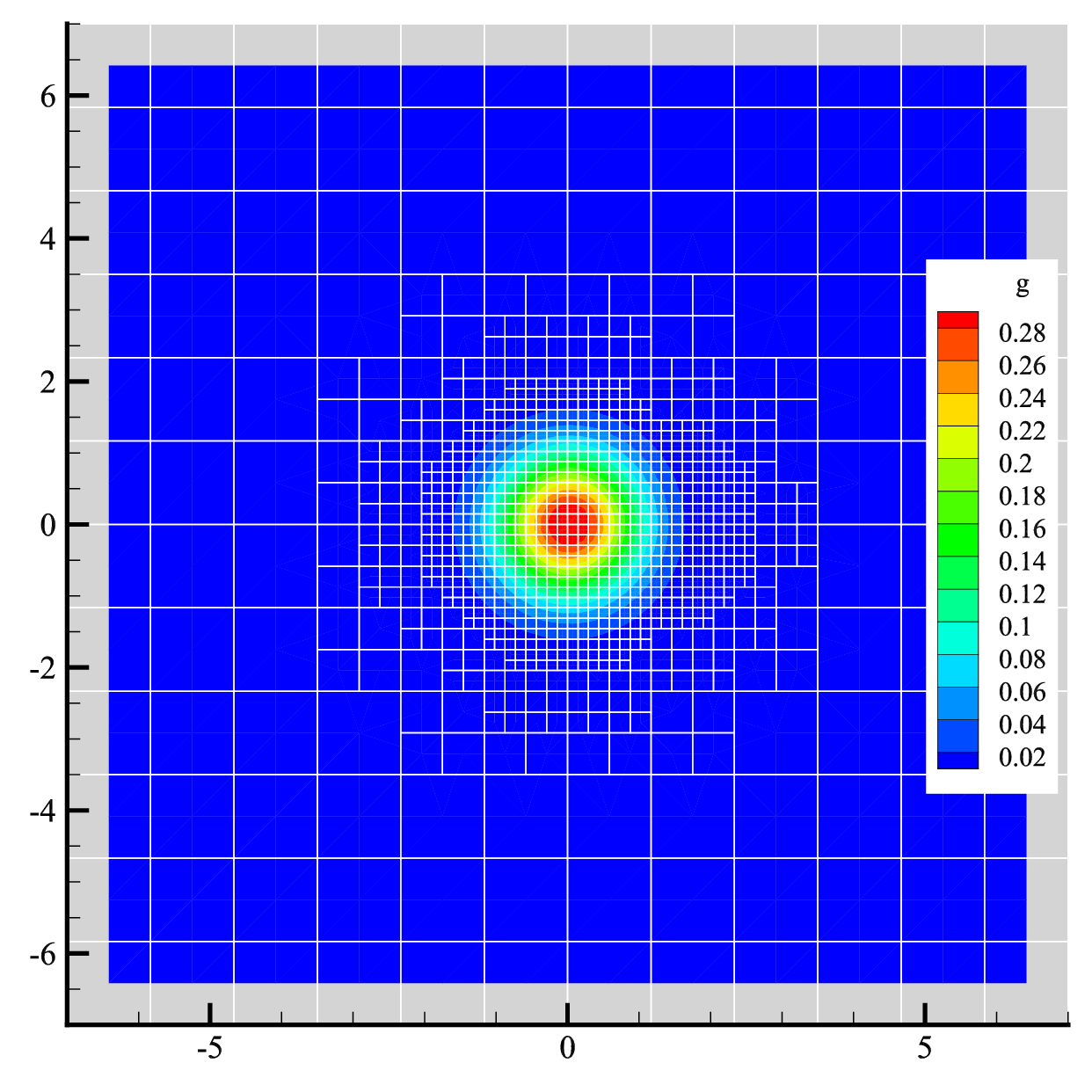}
		\caption{Point A.}
		\label{fig:twocavity_cell11208}
	\end{subfigure}
	\begin{subfigure}[t]{0.3\textwidth}
		\centering
		\includegraphics[width=0.9\textwidth]{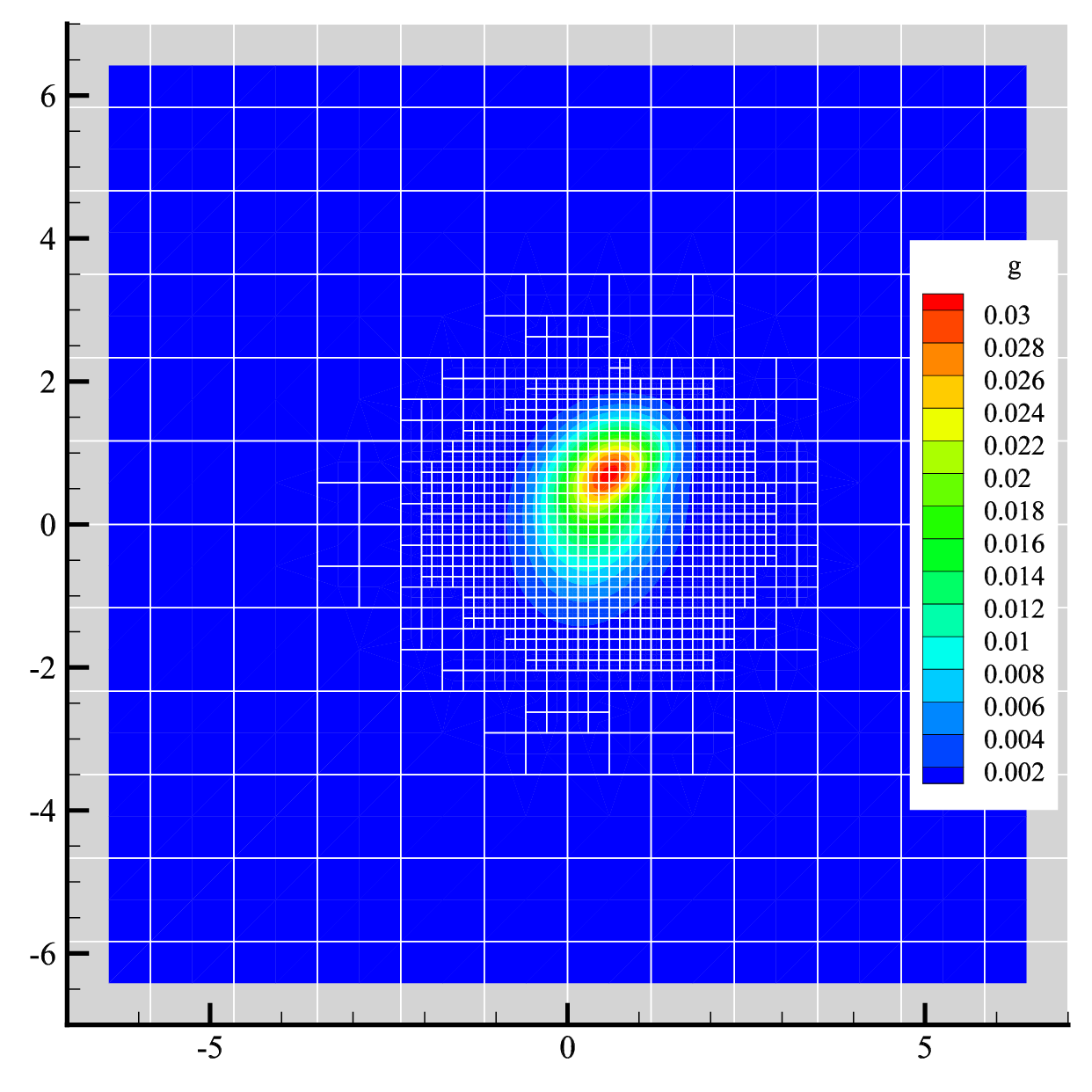}
		\caption{Point B.}
		\label{fig:twocavity_cell3303}
	\end{subfigure}
	\begin{subfigure}[t]{0.3\textwidth}
		\centering
		\includegraphics[width=0.9\textwidth]{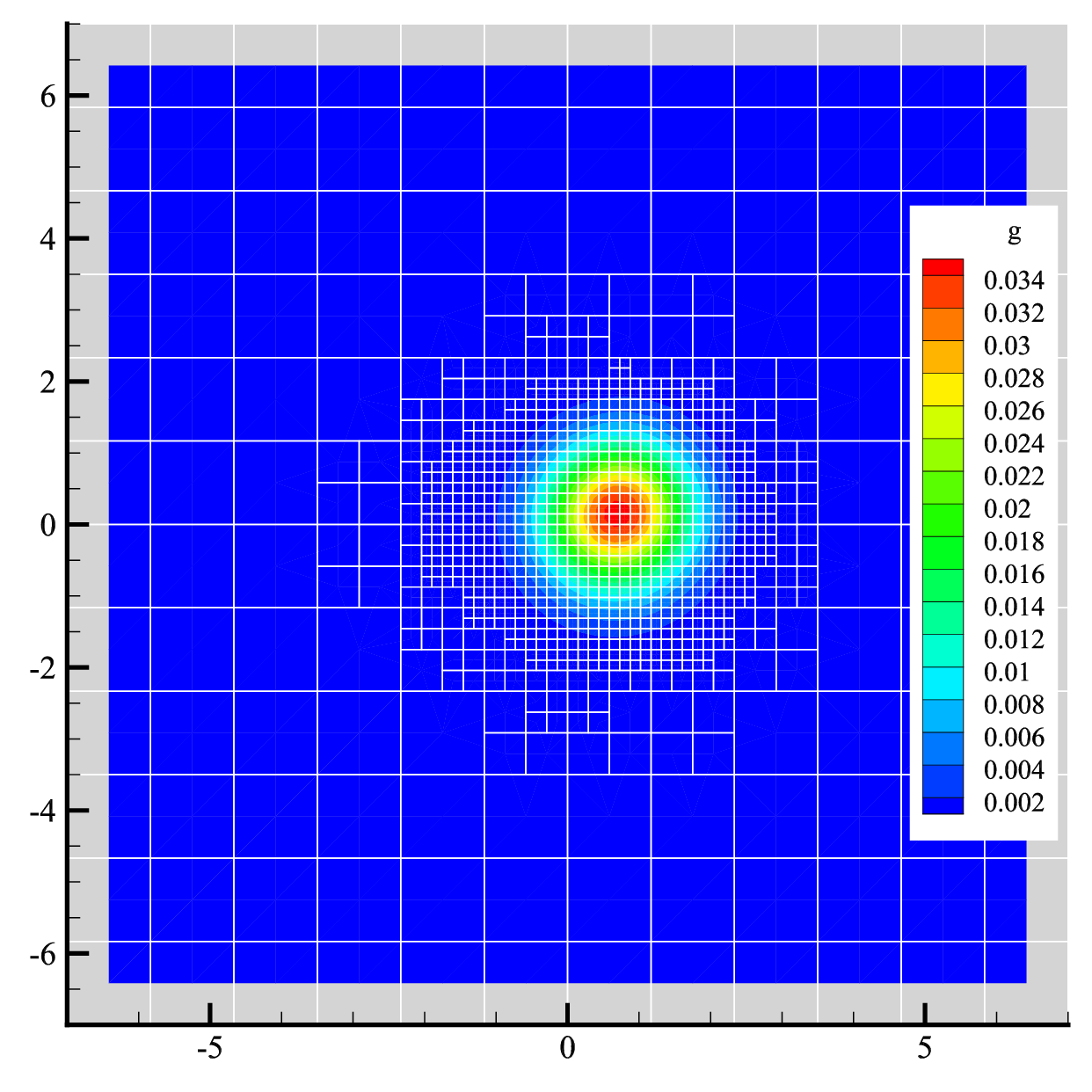}
		\caption{Point C.}
		\label{fig:twocavity_cell3355}
	\end{subfigure}

	\caption{Two-dimensional reduced distribution functions and corresponding adaptive velocity grids at $t=4$.}
	\label{fig:twocavity_distribution}
\end{figure}

The temporal evolution of the representative velocity-space assignments is illustrated in Fig.~\ref{fig:twocavity_tag_evolution}. Initially, a single representative space is shared by the entire domain. As the unsteady expansion develops, the local flow states in the connecting channel and the cavities diverge, dynamically triggering the generation of additional representative spaces. At later times, these spaces redistribute or merge as certain flow regions become more uniform. This demonstrates that the adaptive strategy responds to the temporal evolution of the unsteady flow by dynamically adjusting both the number and spatial assignments of the representative spaces, without imposing a uniformly detailed velocity grid throughout the physical domain. The three sampling points in Fig.~\ref{fig:twocavity_distribution} exhibit distinct local kinetic characteristics at $t=4$. At point A (near the left-cavity center), the distribution remains nearly isotropic and centered close to the velocity-space origin, consistent with a state close to local equilibrium. Near the channel exit, point B exhibits an obliquely shifted and strongly asymmetric distribution, indicating a substantial departure from local equilibrium. At point C (near the right-cavity center), the distribution is displaced primarily in the positive streamwise direction but is more nearly isotropic and closer to the local Maxwellian than at point B. Despite these differences, points B and C share the same representative velocity space, whereas point A uses another.

\begin{figure}[htbp]
	\centering
	\begin{subfigure}[t]{0.48\textwidth}
		\centering
		\includegraphics[width=\textwidth]{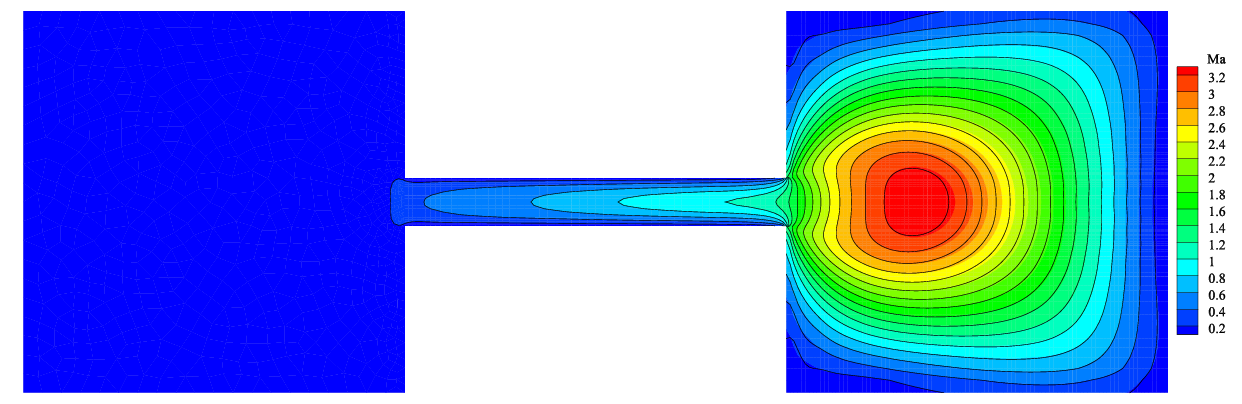}
		\caption{$t = 1$}
	\end{subfigure}
	\begin{subfigure}[t]{0.48\textwidth}
		\centering
		\includegraphics[width=\textwidth]{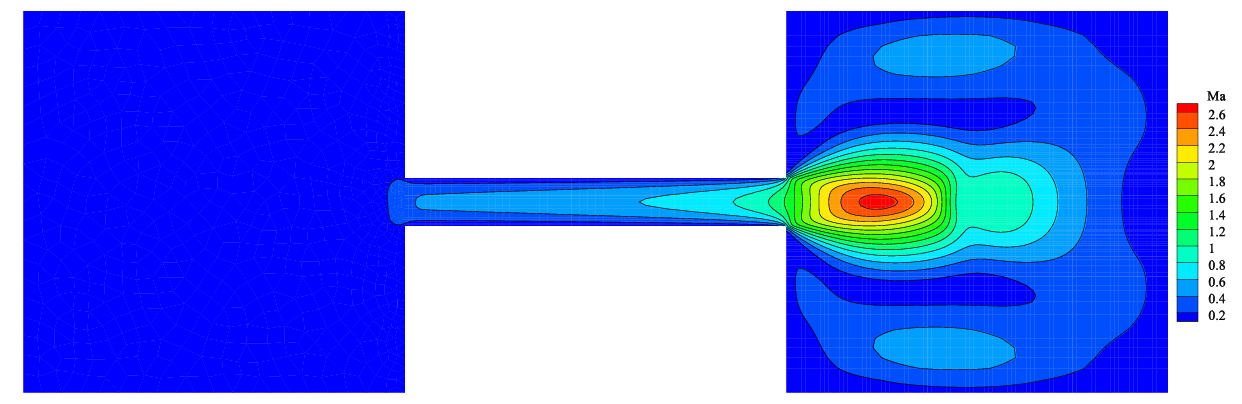}
		\caption{$t = 4$}
	\end{subfigure}
	\caption{Mach-number fields (filled contours: MLVS-DUGKS; lines: UVS-DUGKS).}
	\label{fig:twocavity_macontour}
\end{figure}

\begin{figure}[htbp]
	\centering
	\begin{subfigure}[t]{0.45\textwidth}
		\centering
		\includegraphics[width=0.75\textwidth]{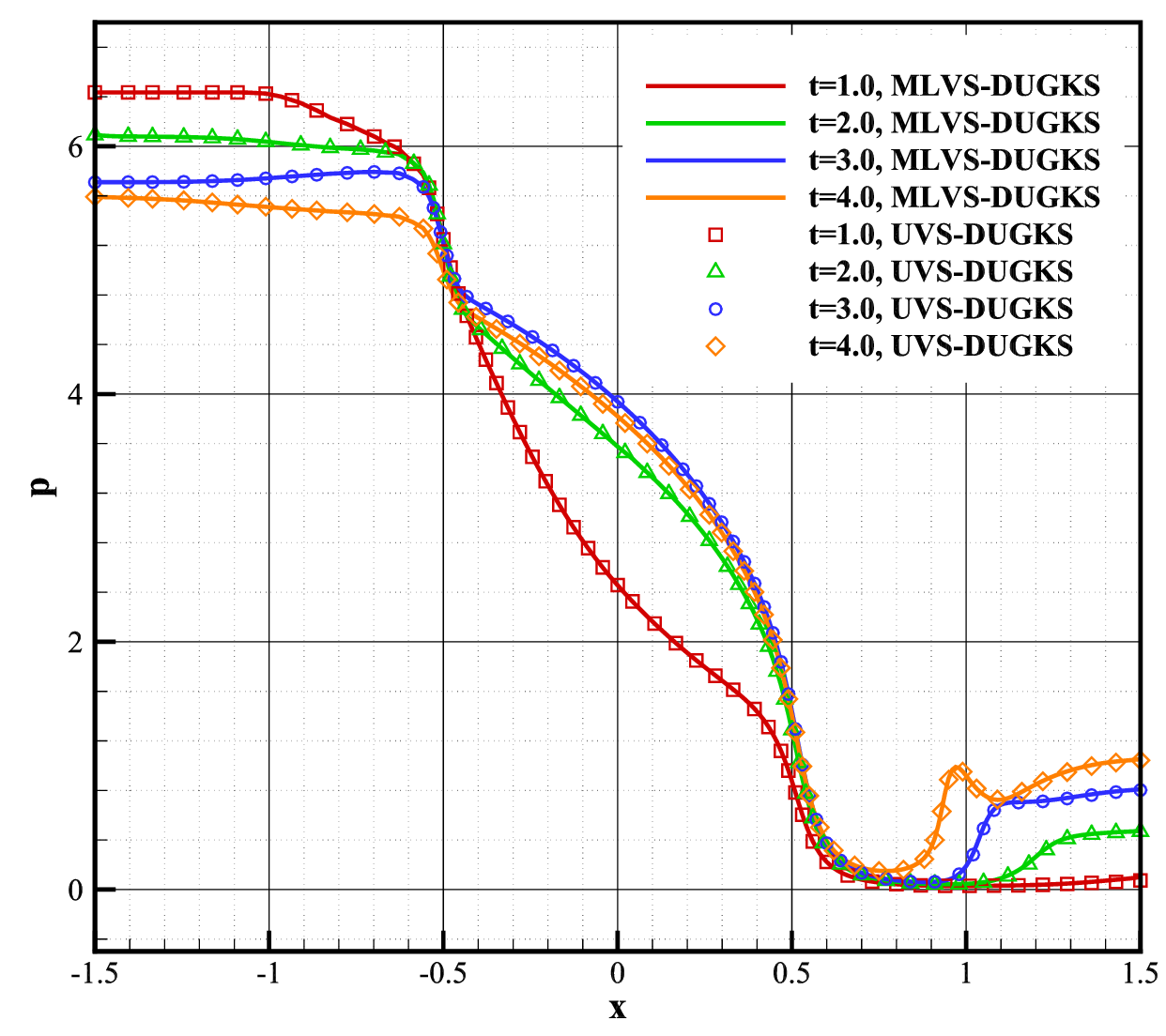}
		\caption{Pressure along $y=0$.}
	\end{subfigure}
	\begin{subfigure}[t]{0.45\textwidth}
		\centering
		\includegraphics[width=0.75\textwidth]{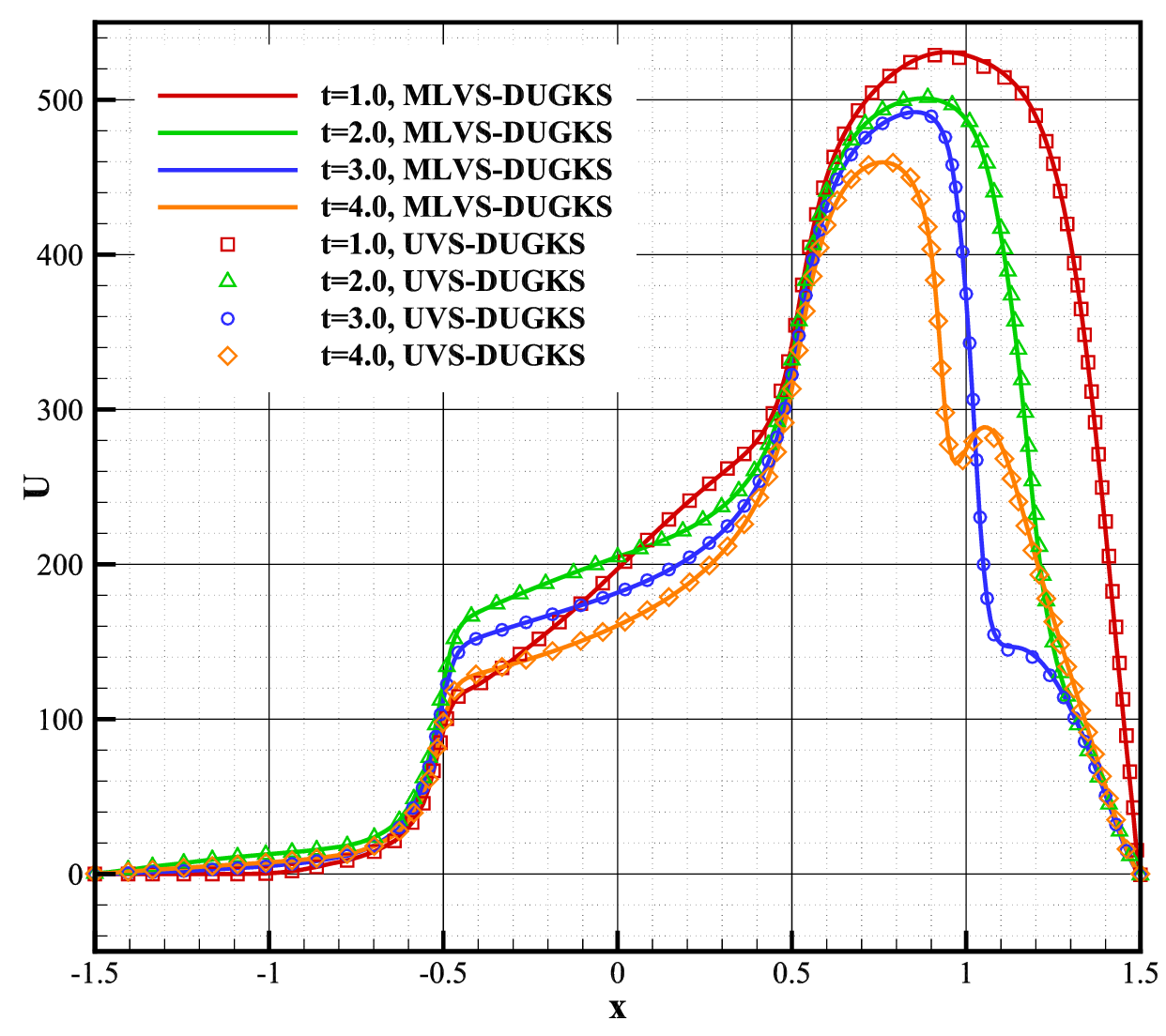}
		\caption{Streamwise velocity along $y=0$.}
	\end{subfigure}

	\begin{subfigure}[t]{0.45\textwidth}
		\centering
		\includegraphics[width=0.75\textwidth]{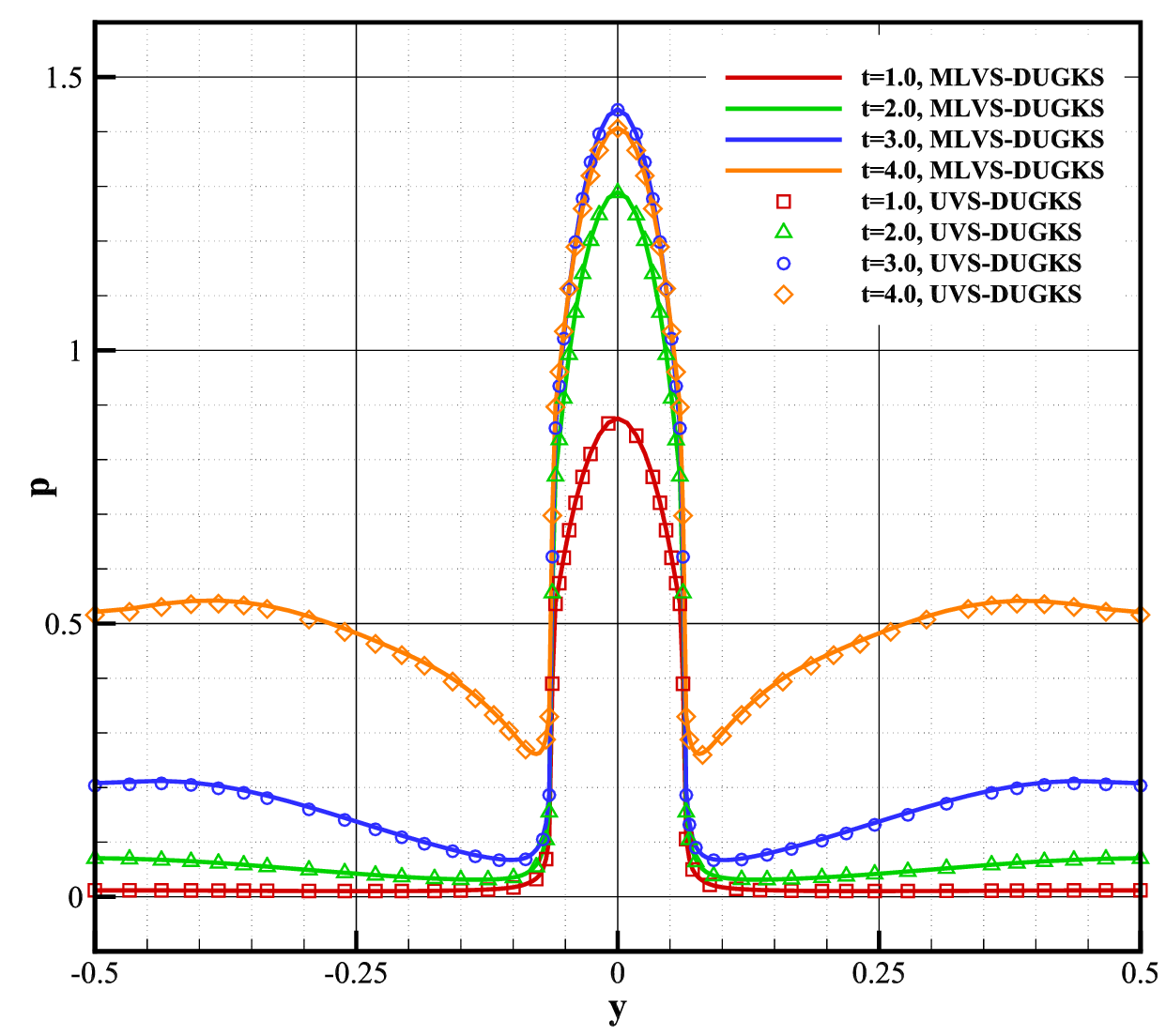}
		\caption{Pressure along $x=0.5$.}
	\end{subfigure}
	\begin{subfigure}[t]{0.45\textwidth}
		\centering
		\includegraphics[width=0.75\textwidth]{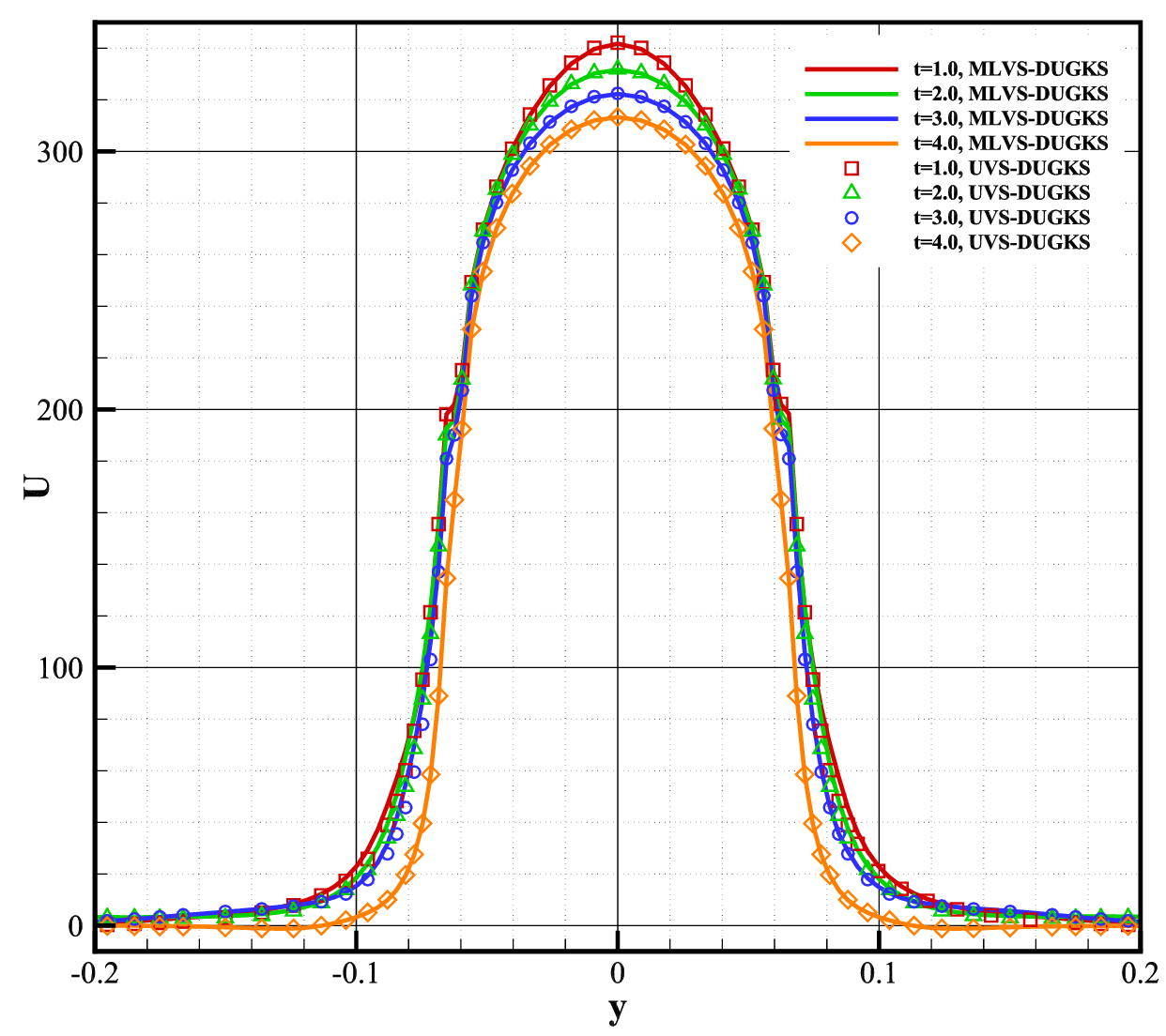}
		\caption{Streamwise velocity along $x=0.5$.}
	\end{subfigure}

	\caption{Comparisons of the pressure and streamwise-velocity profiles.}
	\label{fig:twocavity_line}
\end{figure}

\begin{table}[htbp]
	\centering
	\caption{$L_2$ differences between the MLVS-DUGKS and the UVS-DUGKS at $t=4$.}
	\label{tab:twocavity_field_error}
	\small
	\begin{tabular}{c c c c c}
		\toprule
		$\varepsilon_{2,\rho}^{\mathrm{rel}}$ & $\varepsilon_{2,p}^{\mathrm{rel}}$ & $\varepsilon_{2,\boldsymbol{u}}^{\mathrm{rel}}$ & $\varepsilon_{2,T}^{\mathrm{rel}}$ & $\varepsilon_{2,\boldsymbol{q}}^{\mathrm{rel}}$ \\
		\midrule
		$6.18\times10^{-4}$                   & $8.36\times10^{-4}$                & $6.66\times10^{-3}$                             & $2.26\times10^{-3}$                & $1.33\times10^{-2}$                             \\
		\bottomrule
	\end{tabular}
\end{table}

The transient macroscopic solutions obtained with the MLVS-DUGKS and the UVS-DUGKS are compared in Figs.~\ref{fig:twocavity_macontour} and~\ref{fig:twocavity_line}. The high-speed region develops from a broad expansion zone in the right cavity at $t=1$ into an elongated core downstream of the channel exit at $t=4$. Throughout this evolution, the Mach-number contours from the MLVS-DUGKS agree closely with those of the UVS-DUGKS. The pressure and streamwise-velocity profiles at $t=1$, $2$, $3$, and $4$ also show close agreement, including in the steep-gradient regions near the channel exit. For a quantitative assessment, Table~\ref{tab:twocavity_field_error} summarizes the volume-weighted relative $L_2$ differences at $t=4$. The relative differences in density and pressure are $6.18\times10^{-4}$ and $8.36\times10^{-4}$, respectively, while those in velocity and temperature are $6.66\times10^{-3}$ and $2.26\times10^{-3}$. Although the heat flux is not explicitly constrained during inter-level transfer, its relative difference is $1.33\%$ at this final time. These final-time differences, together with the preceding transient comparisons, show that the proposed strategy preserves the principal macroscopic fields and retains the nonequilibrium heat-transport information.

The representative velocity-space configurations at $t=4$ are summarized in Table~\ref{tab:twocavity_velocity_config}. Two active representative velocity spaces are assigned to the 12,012 physical cells in the MLVS-DUGKS. These spaces yield a cell-weighted average of 1,029.3 velocity points per physical cell. The resulting velocity-space compression factor is 9.91. As reported in Table~\ref{tab:twocavity_performance}, the MLVS-DUGKS achieves a speedup factor of $7.00$ and reduces GPU-memory consumption by a factor of $7.36$. Combined with the preceding flow-field comparisons, these results demonstrate that the MLVS-DUGKS captures the transient multiscale evolution while substantially reducing the computational and memory requirements.

\begin{table}[htbp]
	\centering
	\caption{Representative velocity-space configurations for gas expansion between two connected cavities at $t=4$.}
	\label{tab:twocavity_velocity_config}
	\small
	\begin{tabular}{lcccc}
		\toprule
		& $\mathcal{V}_1$ & $\mathcal{V}_2$ & $N_{\mathrm{avg}}$ & Compression \\
		\midrule
		$t=4$ & 1,044\,/\,6,129 & 1,014\,/\,5,883 & 1,029.3 & 9.91 \\
		\bottomrule
	\end{tabular}
\end{table}

\begin{table}[htbp]
	\centering
	\caption{Computational performance for gas expansion between two connected cavities.}
	\label{tab:twocavity_performance}
	\small
	\begin{tabular}{cccccc}
		\toprule
		\multicolumn{3}{c}{Time (s)} & \multicolumn{3}{c}{GPU memory (MB)} \\
		\cmidrule(lr){1-3} \cmidrule(lr){4-6}
		UVS & MLVS & Speedup & UVS & MLVS & Ratio \\
		\midrule
		20130.61 & 2876.88 & 7.00 & 7725 & 1049.75 & 7.36 \\
		\bottomrule
	\end{tabular}
\end{table}

\subsection{Analysis of clustering effects}
\label{sec:threshold_study}
The preceding benchmarks demonstrate the computational savings of the MLVS-DUGKS relative to the UVS-DUGKS. This section examines how the organization of adaptive velocity spaces affects computational efficiency. The clustering threshold is varied from 1 to 0.035 for the Mach-5 cylinder flow at $\mathrm{Kn}=0.01$, with the other settings identical to those of the preceding cylinder benchmark. Table~\ref{tab:threshold_performance} summarizes the number of representative velocity spaces $N_{\mathrm{rep}}$, the cell-weighted average number of velocity points $\bar{N}_{\xi}$, the number of cross-level interfaces $N_{\mathrm{cross}}$, the measured runtimes and GPU memory. As shown, decreasing $\tau_{\mathrm{clust}}$ increases the number of representative spaces from one to eight and reduces $\bar{N}_{\xi}$ from 1033.0 to 689.2, reflecting the ability of multiple representative spaces to accommodate differing local velocity-space requirements. At the same time, the cross-level interface fraction increases from zero to $9.46\%$, introducing additional work for inter-level distribution transfer, auxiliary-interface flux evaluation, and moment-constrained correction. The measured runtime does not decrease monotonically with $\bar{N}_{\xi}$. Among the tested configurations, $\tau_{\mathrm{clust}}=0.05$ gives the shortest measured runtime of $1325.48\,\mathrm{s}$. Further decreasing the threshold to 0.035 reduces $\bar{N}_{\xi}$ from 714.7 to 689.2, but increases the cross-level interface fraction from $5.28\%$ to $9.46\%$ and the runtime to $1429.89\,\mathrm{s}$. The measured GPU memory behaves similarly. As a result, the reduction in the average number of velocity points does not guarantee a monotonic improvement in computational performance, because the gains from velocity-space compression are counterbalanced by cross-level coupling overheads.

\begin{table}[htbp]
	\centering
	\caption{Effect of the clustering threshold.}
	\label{tab:threshold_performance}
	\renewcommand{\arraystretch}{1.15}
	\small
	\setlength{\tabcolsep}{5pt}
	\begin{tabular}{c c c c c c c}
		\toprule
		$\tau_{\mathrm{clust}}$
		      & $N_{\mathrm{rep}}$
		      & $\bar{N}_{\xi}$
		      & $N_{\mathrm{cross}}$
		      & $N_{\mathrm{cross}}/N_{\mathrm{int}}$
		      & Measured time (s)
			  & GPU memory (MB)\\
		\midrule
		1.000 & 1 & 1033.000 & 0   & $0.00\%$ & 1580.13 & 365.44 \\
		0.070 & 2 & 921.796  & 424 & $4.17\%$ & 1573.01 & 393.44 \\
		0.050 & 5 & 714.698  & 537 & $5.28\%$ & 1325.48 & 337.44 \\
		0.040 & 6 & 704.011  & 586 & $5.76\%$ & 1330.06 & 349.44 \\
		0.035 & 8 & 689.201  & 963 & $9.46\%$ & 1429.89 & 365.44 \\
		\bottomrule
	\end{tabular}
\end{table}

The $\tau_{\mathrm{clust}}=1$ configuration, in which all physical cells share a single adaptive velocity space, coincides with the organization of global adaptive approaches such as the strategy of Chen et al.~\cite{Chen_global_2024a}. It is constructed within the present solver framework rather than reproducing their method, so the following comparison isolates the effect of the multi-level organization itself rather than ranking against the published implementation. Compared with this global adaptive configuration, the $\tau_{\mathrm{clust}}=0.05$ configuration reduces $\bar{N}_{\xi}$ by $30.8\%$, yielding a speedup of 1.192 and a $7.7\%$ reduction in GPU memory usage.

Although the threshold $\tau_{\mathrm{clust}}=0.05$ yields the optimal performance for the present $\mathrm{Kn}=0.01$ case, it is not necessarily optimal for other flow conditions. To avoid case-specific tuning, this threshold is kept constant across all benchmarks, meaning the reported speedups represent conservative estimates rather than tuned optima.

\section{Conclusions}\label{sec: Conclusions}
A multi-level velocity-space adaptive discrete unified gas-kinetic scheme (MLVS-DUGKS) has been developed for the efficient simulation of multiscale nonequilibrium gas flows. Motivated by the observation that local flow states often take a limited number of characteristic forms, the method groups the velocity-space requirements of individual cells into a few shared representative velocity spaces. Because both velocity coarsening and data transfer between representative spaces can violate discrete moment consistency, a linear moment-constrained correction is introduced to enforce the prescribed moments during collision-model construction and inter-space mapping. A local linear collision analysis confirms the decay of the nonconservative modes, and small-perturbation and uniform-flow simulations support the stability and conservation of the full solver under the tested conditions.

Benchmark tests, including the cylinder and two-cavity problems, show that the principal macroscopic fields remain in close agreement with those of the UVS-DUGKS. Compared to the UVS-DUGKS, the method reduces the average number of velocity points by approximately one order of magnitude, yielding speedups of $7.00$--$7.93$ and GPU-memory reductions by factors of $6.73$--$7.49$. Furthermore, comparison with a single-space global adaptive organization shows that the multi-level arrangement yields an additional speedup of 1.192 and a $7.7\%$ reduction in GPU memory. As the gains from velocity-space compression are partly counterbalanced by cross-level coupling overheads, the reported speedups represent conservative estimates.

The overall efficiency of the MLVS-DUGKS depends on implementation details and the trade-off between cross-level overheads and velocity-space compression. The present GPU implementation prioritizes validating the framework over architectural optimization. Future work will focus on optimizing cross-level data transfer, memory layouts, and on exploring the distribution of representative velocity spaces across multiple GPUs to reduce global communication.



\section*{Acknowledgements}
The authors gratefully acknowledge Professor Zhaoli Guo of Huazhong University of Science and Technology for his support and guidance throughout this research.

This work was supported by the National Natural Science Foundation of China (Grant Nos. 12202490 and 12502275), the Hunan Province Natural Science Foundation (Grant No. 2024JJ6455), and the Hunan Provincial Innovation Foundation for Postgraduate (Grant No. CX20250007).


\appendix
\numberwithin{table}{section}
\renewcommand{\thetable}{\Alph{section}.\arabic{table}}
\renewcommand{\thesubsection}{\Alph{section}.\arabic{subsection}}

\section{Local linear stability of the corrected collision update}
\label{app:stability}
This appendix derives the local amplification factors in Eq.~\eqref{eq:local_collision_factors}. Fix the discrete velocity nodes and quadrature weights, and write $\boldsymbol g=(g_1,\ldots,g_{N_\xi})^T$ and $\boldsymbol h=(h_1,\ldots,h_{N_\xi})^T$. For $\mathrm{Pr}=2/3$, the corrected distributions satisfy 
\begin{equation}
	\left[
		\boldsymbol W(g^*,h^*), \boldsymbol q(g^*,h^*)
	\right]^T
	=
	\boldsymbol D_6
	\begin{pmatrix}
		\boldsymbol W, \boldsymbol q
	\end{pmatrix}^T,
	\quad
	\boldsymbol D_6
	=
	\operatorname{diag}
	\left(
	1,1,1,1,\tfrac13,\tfrac13
	\right),
	\label{eq:model_six_moment_identity}
\end{equation}
where the corrected distributions $(g_i^*,h_i^*)$ are computed from the current discrete moments $(\boldsymbol W,\boldsymbol q)$. Substituting the corrected distributions into Eq.~\eqref{eq:reduced_gh_eq}
\begin{equation}
	\frac{\mathrm d g_i}{\mathrm dt} =\Omega_{g,i} =-\frac{g_i-g_i^*}{\tau}, \qquad \frac{\mathrm d h_i}{\mathrm dt} =\Omega_{h,i} =-\frac{h_i-h_i^*}{\tau}, \label{eq:discrete_collision_operator}
\end{equation}
Let $\boldsymbol W_0$ denote the macroscopic variables of the uniform reference equilibrium state, and $\boldsymbol 0$ the zero heat flux, around which the linearization is performed. At this state, Eq.~\eqref{eq:model_six_moment_identity} gives the equilibrium $\boldsymbol g_0 = \boldsymbol g^*(\boldsymbol W_0,\boldsymbol 0), \; \boldsymbol h_0 = \boldsymbol h^*(\boldsymbol W_0,\boldsymbol 0)$, whose discrete moments are exactly $(\boldsymbol W_0,\boldsymbol 0)$, so evaluating the corrected model at these moments returns $(\boldsymbol g_0,\boldsymbol h_0)$. Assume that the correction is differentiable near this equilibrium, and define
\begin{equation}
	\begin{aligned} \boldsymbol L &= \left. \begin{pmatrix} \dfrac{\partial\boldsymbol W}{\partial\boldsymbol g} & \dfrac{\partial\boldsymbol W}{\partial\boldsymbol h} \\[6pt] \dfrac{\partial\boldsymbol q}{\partial\boldsymbol g} & \dfrac{\partial\boldsymbol q}{\partial\boldsymbol h} \end{pmatrix} \right|_{(\boldsymbol g_0,\boldsymbol h_0)} \in\mathbb R^{6\times 2N_\xi}, \quad \boldsymbol V = \left. \begin{pmatrix} \dfrac{\partial\boldsymbol g^*}{\partial\boldsymbol W} & \dfrac{\partial\boldsymbol g^*}{\partial\boldsymbol q} \\[6pt] \dfrac{\partial\boldsymbol h^*}{\partial\boldsymbol W} & \dfrac{\partial\boldsymbol h^*}{\partial\boldsymbol q} \end{pmatrix} \right|_{(\boldsymbol W_0,\boldsymbol 0)} \in\mathbb R^{2N_\xi\times6}. \end{aligned} \label{eq:stability_jacobians}
\end{equation}
The heat-flux derivatives include the dependence of the peculiar velocity on the conservative variables, so $\boldsymbol q$, unlike $\boldsymbol W$, is a nonlinear function of the distributions. Differentiating Eq.~\eqref{eq:model_six_moment_identity} with respect to the input moments and applying the chain rule gives
\begin{equation}
	\boldsymbol L\boldsymbol V=\boldsymbol D_6. \label{eq:stability_LV_identity}
\end{equation}
Let $\tau_0=\tau(\boldsymbol W_0)>0$. Linearizing Eq.~\eqref{eq:discrete_collision_operator} about $(\boldsymbol g_0,\boldsymbol h_0)$ yields
\begin{equation}
	\frac{\mathrm d}{\mathrm dt}
	\begin{pmatrix}
		\delta\boldsymbol g \\
		\delta\boldsymbol h
	\end{pmatrix}
	=
	\boldsymbol J_c
	\begin{pmatrix}
		\delta\boldsymbol g \\
		\delta\boldsymbol h
	\end{pmatrix},
	\quad
	\boldsymbol J_c
	=
	\frac{\boldsymbol V\boldsymbol L-\boldsymbol I}{\tau_0},
	\label{eq:stability_collision_jacobian}
\end{equation}
where $\boldsymbol I$ is the identity matrix in distribution space. Derivatives of the state-dependent relaxation time multiply the vanishing collision residual at equilibrium and contribute nothing to $\boldsymbol J_c$. Partitioning the two Jacobians into conservative and heat-flux blocks,
\begin{equation}
	\boldsymbol L
	=
	\begin{pmatrix}
		\boldsymbol L_W \\
		\boldsymbol L_q
	\end{pmatrix},
	\quad
	\boldsymbol V
	=
	\begin{pmatrix}
		\boldsymbol V_W & \boldsymbol V_q
	\end{pmatrix},
	\label{eq:stability_blocks}
\end{equation}
Eq.~\eqref{eq:stability_LV_identity} gives
\begin{equation}
	\begin{aligned}
		\boldsymbol L_W\boldsymbol V_W =\boldsymbol I_4,\quad
		\boldsymbol L_W\boldsymbol V_q =\boldsymbol0,\quad
		\boldsymbol L_q\boldsymbol V_W =\boldsymbol0,\quad
		\boldsymbol L_q\boldsymbol V_q =\tfrac13\boldsymbol I_2.
	\end{aligned}
	\label{eq:stability_block_relations}
\end{equation}
Since $\boldsymbol D_6$ is invertible, $\boldsymbol V$ has six independent columns and $\boldsymbol L$ has full row rank, so every distribution perturbation has the unique decomposition
\begin{equation}
	\begin{pmatrix}
		\delta\boldsymbol g \\
		\delta\boldsymbol h
	\end{pmatrix}
	=
	\boldsymbol V_W\delta\boldsymbol W
	+3\boldsymbol V_q\delta\boldsymbol q
	+
	\begin{pmatrix}
		\delta\boldsymbol g_\perp \\
		\delta\boldsymbol h_\perp
	\end{pmatrix},
	\quad
	\boldsymbol L
	\begin{pmatrix}
		\delta\boldsymbol g_\perp \\
		\delta\boldsymbol h_\perp
	\end{pmatrix}
	=\boldsymbol0,
	\label{eq:stability_mode_decomposition}
\end{equation}
where $(\delta\boldsymbol W^T,\delta\boldsymbol q^T)^T =\boldsymbol L(\delta\boldsymbol g^T,\delta\boldsymbol h^T)^T$. The three terms belong to the conservative, heat-flux, and moment-free subspaces of dimensions $4$, $2$, and $2N_\xi-6$. The subscript $\perp$ denotes the moment-free component and does not imply orthogonality in the Euclidean inner product. Combining Eqs.~\eqref{eq:stability_collision_jacobian} and~\eqref{eq:stability_block_relations} gives
\begin{equation}
	\begin{aligned}
		\boldsymbol J_c\boldsymbol V_W =\boldsymbol0,   \quad
		\boldsymbol J_c\boldsymbol V_q
		                               =-\frac{2}{3\tau_0}\boldsymbol V_q, \quad
		\boldsymbol J_c
		\begin{pmatrix}
			\delta\boldsymbol g_\perp ,\delta\boldsymbol h_\perp
		\end{pmatrix}^T
		                                =
		-\frac{1}{\tau_0}
		\begin{pmatrix}
			\delta\boldsymbol g_\perp ,\delta\boldsymbol h_\perp
		\end{pmatrix}^T,
	\end{aligned}
	\label{eq:stability_spectral_decomposition}
\end{equation}
so the four conservative modes have collision eigenvalue $0$, the two heat-flux modes $-2/(3\tau_0)$, and the remaining modes $-1/\tau_0$. Applying the DUGKS trapezoidal collision update to the linearized equation gives
\begin{equation}
	\left(
	\boldsymbol I-\frac{\Delta t}{2}\boldsymbol J_c
	\right)
	\begin{pmatrix}
		\delta\boldsymbol g^{n+1} \\
		\delta\boldsymbol h^{n+1}
	\end{pmatrix}
	=
	\left(
	\boldsymbol I+\frac{\Delta t}{2}\boldsymbol J_c
	\right)
	\begin{pmatrix}
		\delta\boldsymbol g^n \\
		\delta\boldsymbol h^n
	\end{pmatrix}.
	\label{eq:stability_trapezoidal}
\end{equation}
For a collision eigenvalue $\lambda$, the amplification factor is $(1+\Delta t\lambda/2)/(1-\Delta t\lambda/2)$. Substituting the three eigenvalues above gives the conservative factor $1$ and the nonconservative factors $r_q$ and $r_\perp$ in Eq.~\eqref{eq:local_collision_factors}. The denominators are positive for every $\Delta t>0$ and $\tau_0>0$, and $|(1-a)/(1+a)|<1$ for every $a>0$, so the update is well defined with $|r_q|,|r_\perp|<1$. The full local update thus has spectral radius $1$, while on the invariant subspace of perturbations with zero conservative moments it is $r_b=\max\{|r_q|,|r_\perp|\}<1$. For an initial perturbation with $\delta\boldsymbol W^0=\boldsymbol 0$, repeated local updates give
\begin{equation}
	\begin{pmatrix}
		\delta\boldsymbol g^n \\
		\delta\boldsymbol h^n
	\end{pmatrix}
	=
	3r_q^n\boldsymbol V_q\delta\boldsymbol q^0
	+
	r_\perp^n
	\begin{pmatrix}
		\delta\boldsymbol g_\perp^0 \\
		\delta\boldsymbol h_\perp^0
	\end{pmatrix}
	\longrightarrow\boldsymbol0,
	\label{eq:stability_repeated_decay}
\end{equation}
and, in any fixed vector norm,
\begin{equation}
	\left\|
	\begin{pmatrix}
		\delta\boldsymbol g^n \\
		\delta\boldsymbol h^n
	\end{pmatrix}
	\right\|
	\leq
	r_b^n
	\left[
		3\|\boldsymbol V_q\delta\boldsymbol q^0\|
		+
		\left\|
		\begin{pmatrix}
			\delta\boldsymbol g_\perp^0 \\
			\delta\boldsymbol h_\perp^0
		\end{pmatrix}
		\right\|
		\right].
	\label{eq:stability_norm_bound}
\end{equation}
This establishes the asymptotic decay of every zero-conservative-moment perturbation.

\section{One-step operator and implementation checks}
\label{app:response_protocol}
To distinguish the local model spectrum from its implementation, the complete one-step linear operator is derived and checked on a uniform layout at a homogeneous discrete equilibrium. Let $\boldsymbol J_c$ be the local collision Jacobian defined in Appendix~\ref{app:stability}, and set $s=\Delta t/2$. The collision-related factors are
\begin{equation}
	\begin{aligned}
		\boldsymbol A_c
		 & =(\boldsymbol I+s\boldsymbol J_c)(\boldsymbol I-s\boldsymbol J_c)^{-1},   \\
		\boldsymbol B_c
		 & =(\boldsymbol I+s\boldsymbol J_c/2)(\boldsymbol I-s\boldsymbol J_c)^{-1}, \\
		\boldsymbol H_c
		 & =(\boldsymbol I-s\boldsymbol J_c/2)^{-1}.
	\end{aligned}
	\label{eq:one_step_collision_factors}
\end{equation}
For a Fourier mode $\theta$ along $x$, the diagonal factor for centered reconstruction and half-step characteristic tracing has entries
\begin{equation}
	r_i(\theta)=
	\begin{cases}
		1+\mathrm i(\tfrac12-\xi_{x,i}s/\Delta x)\sin\theta,
		                                 & \xi_{x,i}>0, \\[2pt]
		e^{\mathrm i\theta}
			[1+\mathrm i(-\tfrac12-\xi_{x,i}s/\Delta x)\sin\theta],
		                                 & \xi_{x,i}<0, \\[2pt]
		\tfrac12(1+e^{\mathrm i\theta}), & \xi_{x,i}=0,
	\end{cases}
	\label{eq:one_step_reconstruction}
\end{equation}
and the diagonal matrix $\boldsymbol R(\theta)=\operatorname{diag}(r_1(\theta),\ldots)$ repeats these entries for the $g$ and $h$. Writing $\boldsymbol\Xi_x$ for the corresponding repeated diagonal matrix of $x$-velocities, the one-step symbol acting on the cell-centered transformed-distribution perturbation is
\begin{equation}
	\boldsymbol G(\theta)=\boldsymbol A_c-
	\frac{\Delta t}{\Delta x}(1-e^{-\mathrm i\theta})
	\boldsymbol\Xi_x\boldsymbol H_c\boldsymbol R(\theta)\boldsymbol B_c.
	\label{eq:analytic_one_step}
\end{equation}
At $\theta=0$, the flux difference vanishes and $\boldsymbol G(0)=\boldsymbol A_c$, agreeing with the homogeneous trapezoidal collision update. For nonzero wavenumbers, transport and reconstruction are present. Since the nontrivial part of $\boldsymbol J_c=(\boldsymbol V\boldsymbol L-\boldsymbol I)/\tau_0$ (cf. Eq.~\eqref{eq:stability_collision_jacobian}) has rank at most six, inversion can use
\begin{equation}
	(a\boldsymbol I-b\boldsymbol V\boldsymbol L)^{-1}
	=\frac1a\boldsymbol I+
	\frac ba\boldsymbol V(a\boldsymbol I-b\boldsymbol L\boldsymbol V)^{-1}
	\boldsymbol L.
	\label{eq:one_step_low_rank_inverse}
\end{equation}
This reduces inversion in distribution space to a six-dimensional moment-space solve. To verify the production implementation against this symbol, the check uses a $16\times2$ periodic physical mesh, $17\times17$ uniform velocity nodes, $\Delta x=1/16$, $\Delta t=0.005$, and a single direction with $\theta=2\pi\times3/16$. The centered finite-difference directional derivative of the production one-step map is compared with $\boldsymbol G(\theta)\boldsymbol v$ for an arbitrary direction $\boldsymbol v$ and step sizes $h_{\mathrm{FD}}=10^{-5}$ and $3\times10^{-6}$. Table~\ref{tab:one_step_check} reports the relative directional errors and the residual of one uniform-state update. The same checks give $\max|\boldsymbol L\boldsymbol V-\boldsymbol D_6| =1.22125\times10^{-15}$, consistent with Eq.~\eqref{eq:stability_LV_identity}. The reported results verify agreement of the production one-step map with the analytic operator for the specified wavenumber and direction when the Venkatakrishnan limiter is enabled.

\begin{table}[htbp]
	\centering
	\caption{Uniform-state one-step residuals and relative errors.}
	\label{tab:one_step_check}
	\small
	\begin{tabular}{ccc}
		\toprule
		Kn        & \makecell{Uniform-state residual}
		          & \makecell{Relative directional error}\\
		\midrule
		$10^{-3}$ & $4.796\times10^{-16}$   & $2.209\times10^{-10}$ \\
		$0.1$     & $7.604\times10^{-17}$   & $1.602\times10^{-11}$ \\
		$10$      & $7.385\times10^{-17}$   & $1.436\times10^{-11}$ \\
		\bottomrule
	\end{tabular}
\end{table}

\bibliographystyle{unsrt}
\bibliography{mybibfile}

\end{document}